\documentclass[12pt]{article}
\pdfoutput=1
\usepackage[nosort]{cite}

\usepackage{xcolor}

\usepackage{epsfig}
\usepackage{amsfonts}
\usepackage{amscd}
\usepackage{latexsym}
\usepackage{amsmath,amssymb}
\usepackage{verbatim}
\usepackage{setspace}
\usepackage{color}
\usepackage{fancyhdr}
\usepackage{cite}
\usepackage{hyperref}
\usepackage{tikz}
\usepackage{slashed}
\usepackage{multirow}
\usetikzlibrary{calc}
\usetikzlibrary{topaths}
\usetikzlibrary{decorations}
\usetikzlibrary{decorations.pathmorphing}

\usepackage{draft}
\usepackage{hyperref}
\usepackage{graphicx,color,subfig}
\usepackage{cite}
\usepackage{mciteplus}
\usepackage{skak}

\numberwithin{equation}{section}

\def\({\left(}
\def\){\right)}
\def\d{\partial}

\begin{document}

\begin{titlepage}
	
	\begin{center}
		
		\hfill YITP-SB-2022-20, MIT-CTP/5392\\
		\hfill \\
		\vskip 1cm

		\title{Global Dipole Symmetry, Compact Lifshitz Theory, Tensor Gauge Theory, and Fractons}
		
		\author{Pranay Gorantla$^{1}$, Ho Tat Lam$^{2}$, Nathan Seiberg$^{3}$ and Shu-Heng Shao$^{4}$}
		
		\address{${}^{1}$Physics Department, Princeton University}
				\address{${}^{2}$Center for Theoretical Physics, Massachusetts Institute of Technology}
		\address{${}^{3}$School of Natural Sciences, Institute for Advanced Study}
				\address{${}^{4}$C.\ N.\ Yang Institute for Theoretical Physics, Stony Brook University}

	\end{center}
	
	\begin{abstract}\noindent
We study field theories with global dipole symmetries and gauge dipole symmetries. The famous Lifshitz theory is an example of a theory with a global dipole  symmetry.  We study in detail its 1+1d version with a compact field.  When this global symmetry is promoted to a $U(1)$ dipole gauge symmetry, the corresponding gauge field is a tensor gauge field. This theory is known to lead to fractons.  In order to resolve various subtleties in the precise meaning of these global or gauge symmetries, we place these 1+1d theories on a lattice and then take the continuum limit.  Interestingly, the continuum limit is not unique.  Different limits lead to different continuum theories, whose operators, defects, global symmetries, etc. are different.  We also consider a lattice gauge theory with a ${\mathbb Z}_N$ dipole gauge group. Surprisingly, several physical observables, such as the ground state degeneracy and the mobility of defects depend sensitively on the number of sites in the lattice.

Our analysis forces us to think carefully about global symmetries that do not act on the standard Hilbert space of the theory, but only on the Hilbert space in the presence of defects.  We refer to them as time-like global symmetries and discuss them in detail.  These time-like global symmetries allow us to phrase the mobility restrictions of defects (including those of fractons) as a consequence of a global symmetry.

	\end{abstract}
	
	\vfill
	
\end{titlepage}

\eject
	
\tableofcontents

\section{Introduction}

Symmetric tensor gauge theories \cite{Gu:2006vw,Xu:2006,Pankov:2007,Xu2008,Gu:2009jh,rasmussen,Slagle:2018kqf,Qi:2020jrf,Du:2021pbc} are non-relativistic field theories, which have been studied extensively in recent years due to their association with fracton models \cite{Pretko:2016kxt,Pretko:2016lgv,Pretko:2018jbi}. (See \cite{Nandkishore:2018sel,Pretko:2020cko,Grosvenor:2021hkn} for a review on this subject.) The simplest theory in this class, commonly known as the ``rank-2 scalar charge theory'' \cite{Pretko:2016kxt,Pretko:2016lgv}, involves $U(1)$ gauge fields $(A_\tau,A_{ij})$ with gauge symmetry
\ie\label{scalarchargetheory-gaugesym}
A_\tau \sim A_\tau + \partial_\tau \alpha~, \quad A_{ij} \sim A_{ij} + \partial_i \partial_j \alpha~.
\fe
Here, $\alpha$ is the gauge parameter, and $A_{ij}$ is symmetric in the spatial indices $i,j$, i.e., $A_{ij} = A_{ji}$.\footnote{A variant of the above theory has only off-diagonal components of the gauge field $A_{ij}$, i.e., $A_{ij}=0$ for $i =j$ \cite{Xu2008,Slagle:2017wrc,Bulmash:2018lid,Ma:2018nhd,Oh2021,You:2018zhj,You:2019cvs,You:2019bvu,
Radicevic:2019vyb,paper1,paper2,paper3,Dubinkin:2020kxo}.  Its properties and dynamics are quite different than the theory we will study here.}

Most of the discussion in this note with be in Euclidean signature spacetime and we will denote the Euclidean time direction by $\tau$.  In the few places, where we will rotate to Lorentzian signature, we will denote the Lorentzian signature time by $t$.   We will often abuse the terminology and use the phrase ``time-like'' to mean ``along the Euclidean time direction.''

When this gauge theory is coupled to a matter theory, the gauge field $(A_\tau,A_{ij})$ couples to Noether current $(J_\tau,J^{ij})$.  The gauge symmetry \eqref{scalarchargetheory-gaugesym} shows that the Noether current $(J_\tau,J^{ij})$ must satisfy a dipole current conservation equation
\ie\label{Noecn}
\partial_\tau J_\tau = \partial_i \partial_j J^{ij}~, \qquad J^{ij} = J^{ji}.
\fe
This global symmetry and the current conservation have been studied in \cite{Griffin:2013dfa,Griffin:2014bta,Pretko:2016lgv,Pretko:2018jbi,Gromov:2018nbv,Seiberg:2019vrp,Shenoy:2019wng,
Gromov:2020rtl,Gromov:2020yoc,Du:2021pbc,Stahl:2021sgi}. A matter theory containing the Noether current $(J_\tau,J^{ij})$ has a \emph{dipole} global symmetry generated by the conserved scalar and dipole charges\footnote{The variant of the tensor gauge theory with only off-diagonal terms is coupled to a matter theory whose Noether current $(J_\tau,J^{ij})$ satisfies \eqref{Noecn}, but it has only off-diagonal components, i.e., $J^{ij}=0$ for $i=j$. As for the gauge theory, this matter theory is quite different than the theory with diagonal elements in $J^{ij}$.  In particular, its Noether current leads to a \emph{subsystem} global symmetry generated by the charges
\ie
Q_i(x^i) = \int_{\text{fixed}~x^i} J_\tau~,
\fe
where the integral is over the subspace with fixed $x^i$. This symmetry is significantly larger than the dipole symmetry \eqref{dipolechargem}. Examples of such theories were analyzed in  \cite{PhysRevB.66.054526,You:2018zhj,You:2019bvu,paper1,paper2}.}
\ie\label{dipolechargem}
Q = \int_\text{space} J_\tau~, \qquad Q^i = \int_\text{space} x^i J_\tau~.
\fe
As we will discuss below, such global symmetries should be handled with care.  The factor of $x^i$ in the charge is not well defined in compact space.  And even on $\mathbb R^d$ it grows at infinity and the action of this $Q^i$ might take us out of the allowed space of fields.

If the matter theory is invariant under spatial translations, there is also a conserved momentum operator $P_i$. Together, they satisfy
\ie\label{PQcommu}
[P_j,Q^i] = -i\delta^i_j Q~.
\fe
This mixture between the global symmetry and translations will be important below.

A typical matter theory in $d$ spatial dimensions with the conservation equation \eqref{Noecn} is the Lifshitz theory (see \cite{Chen:2009ka}, and references therein) with the action
\ie\label{intro-dipphi-action1t}
S =  \oint d\tau d^dx~ \left[ \frac{\mu_0}{2} (\partial_\tau \phi)^2 + \frac{1}{2\mu} \left(\sum_i\partial_i^2 \phi\right)^2 \right]~.
\fe
In this case, the conserved current \eqref{Noecn} is\footnote{Here we follow the conventions in \cite{Gorantla:2021svj} where when we analytically continue to Lorentzian signature, $J_\tau$ does not get another factor of $i$ due to its subscript.\label{conventionsf}}
\ie\label{dipphi-momcuri}
&J_\tau = i\mu_0 \partial_\tau \phi~,\qquad J^{ij} = \frac{i}{\mu} \partial^i\partial^j \phi~,
\\
&\partial_\tau J_\tau = \partial_i\partial_j J^{ij}~
\fe
and the conserved charges are the scalar and dipole charges \eqref{dipolechargem} implementing
\ie\label{phishifi}
\phi \to \phi +c + c_{i} x^i~,
\fe
with constants $c$ and $c_i$.  As we commented after \eqref{dipolechargem}, such a transformation is subtle.  If we work in compact space, it is not well defined. And if we work on $\mathbb R^d$, then this shift changes the behavior of $\phi$ at infinity and takes us out of the allowed space of fields.\footnote{Actually, in this case, the equation of motion $\mu_0\partial_\tau^2 \phi ={1\over \mu} \partial_i\partial^i \partial_j\partial^j\phi$ suggests that there are additional conserved charges -- multipole charges, e.g., $Q^{ij}=\int_\text{space} x^ix^j J_\tau$ and $Q^{ijk}=\int_\text{space} x^ix^jx^k J_\tau$, implementing the transformations $\phi \to \phi + c_{ij} x^ix^j$ and $\phi \to \phi + c_{ijk} x^ix^jx^k$, respectively.  These transformations are even more singular than the shift \eqref{phishifi} and might not even leave the action \eqref{intro-dipphi-action1t} invariant.}

In most papers on the Lifshitz theory, the scalar field $\phi$ is noncompact.
Instead, following the discussion of the 2+1d case in \cite{Henley1997,
Moessner2001,Vishwanath:2004,Fradkin:2004,Ardonne:2003wa,2005,2018PhRvB..98l5105M,Yuan:2019geh,Lake:2022ico}, we will be interested in the case where the scalar is compact, i.e., $\phi \sim \phi + 2\pi$.
This compactness will have important consequences below.  Among other things, the theory \eqref{intro-dipphi-action1t} has more global symmetries in addition to \eqref{dipphi-momcuri}.

We can also go in the reverse direction.  Given a matter theory containing the Noether current $(J_\tau,J^{ij})$ satisfying \eqref{Noecn}, we can gauge the dipole global symmetry by coupling the current to the gauge field $(A_\tau,A_{ij})$ as
\ie
iA_\tau J_\tau + iA_{ij} J^{ij}~.
\fe
We can also add kinetic terms for the gauge fields, such as $E_{ij}^2$ and $B_{[ij]k}^2$, where
\ie
&E_{ij} = \partial_\tau A_{ij} - \partial_i \partial_j A_\tau~,\\
&B_{[ij]k} = \partial_i A_{jk} - \partial_j A_{ik}
\fe
are the electric and magnetic fields. Then, we can study the pure gauge theory of $(A_\tau,A_{ij})$ without matter.

There are some important questions and subtleties in both the matter and gauge theories mentioned above:
\begin{itemize}
\item It is common to analyze a theory in finite volume by placing it on a compact space, such as a flat spatial torus with periodic boundary conditions. However, if we place the matter theory with a dipole global symmetry on a compact space, the dipole charge $Q^i$ is not well defined even if the scalar charge $Q$ vanishes \cite{Seiberg:2019vrp}. See also the comment after \eqref{dipolechargem}.
\item The pure gauge theory of $(A_\tau,A_{ij})$ famously has fracton defects, i.e., defects that describe world-lines of immobile particles, or fractons. The immobility of fractons is usually attributed to conservation of scalar and dipole charges \cite{Pretko:2016kxt}.   However, in a gauge theory, the notion of ``conservation of charge'' does not make sense in compact space because the global symmetry generated by that charge is gauged.\footnote{When the theory is placed on a space with a boundary, the notion of gauge charge depends on the boundary conditions, and in a noncompact space, we can discuss the total gauge charge measured at infinity.} Here, ``charge'' refers to both scalar and dipole charges.
\item What is the geometric setup for such tensor gauge theories? What are the allowed gauge transformations and transition functions?  What are the nontrivial bundles and how are they characterized?  What replaces the notion of holonomies?
\end{itemize}
The goal of this paper is to address these subtleties, and make the statement of immobility of fractons in theories with dipole global symmetries more precise. Following \cite{Seiberg:2019vrp,paper1,paper2,paper3,Gorantla:2020xap,Gorantla:2020jpy,Rudelius:2020kta,Gorantla:2021svj,
Gorantla:2021bda,Burnell:2021reh}, we will focus on the global symmetries and their consequences and then we will study the corresponding gauge theory.

For simplicity, let us consider the 1+1d continuum theory described by the action
\ie\label{intro-dipphi-action1}
S =  \oint d\tau dx~ \left[ \frac{\mu_0}{2} (\partial_\tau \phi)^2 + \frac{1}{2\mu} (\partial_x^2 \phi)^2 \right]~,
\fe
where $\phi \sim \phi + 2\pi$ is a compact scalar, and $\mu_0$ and $\mu$ are coupling constants with mass dimensions $0$ and $2$. Due to the mass dimension and periodicity of $\phi$, the local operators $e^{i\phi}$ and $\partial_x \phi$ exist in this continuum theory.  The obvious fact that since $\phi$ is dimensionless, $e^{i\partial_x \phi}$ does not exist, will have important consequences below.

The theory has a  dipole global symmetry that shifts $\phi$ as
\ie
\phi(\tau,x) \rightarrow \phi(\tau,x) + c + c_x x~.
\fe
We will comment on the global properties of $c$ and $c_x$ momentarily.
This is the simplest scalar field theory with a dipole global symmetry, while  more general ones with multipole global symmetries have been discussed extensively in \cite{Griffin:2013dfa,Griffin:2014bta,Pretko:2018jbi,Gromov:2018nbv,Shenoy:2019wng,Gromov:2020rtl,Gromov:2020yoc,Stahl:2021sgi}.

We now turn to the global aspects of the above dipole global symmetry.
The parameter $c\sim c+2\pi$ is a circle-valued constant, which generates an ordinary $U(1)$ symmetry.
Following the standard terminology in string theory,  we will refer  to this symmetry  as the $U(1)$ momentum global symmetry.\footnote{Here by ``momentum" we mean the momentum in the target space, rather than on the worldsheet. In the condensed matter literature, this symmetry is referred to as the ``particle number symmetry."}
On the other hand,  $c_x$ is a real constant with mass dimension $1$, which generates a momentum dipole symmetry.
On noncompact space, the symmetry group of the momentum dipole symmetry is the noncompact group of real numbers $\mathbb{R}$ (rather than the compact group $U(1)$).
If we place the theory on a spatial circle of length $\ell_x$, the shift $c_x x$ is not well defined unless $c_x \in \frac{2\pi}{\ell_x} \mathbb Z$.
So, on a compact space, the symmetry group generated by $c_x$ is actually the discrete group of integers $\mathbb Z$.

The action \eqref{intro-dipphi-action1} is also invariant under spatial translations. Denote the $U(1)$ charge of momentum symmetry by $Q$, the generator of $\mathbb Z$ momentum dipole symmetry by $U$, and the generator of spatial translations by $P$. They satisfy
\ie
\, [P,U] = \frac{2\pi}{\ell_x} QU~.
\fe
It differs from \eqref{PQcommu} because on a compact space the dipole symmetry is $\mathbb Z$ rather than $\mathbb R$.

Interestingly, the continuum theory \eqref{intro-dipphi-action1} has an infinite ground state degeneracy.\footnote{A similar phenomenon has been noted in the 2+1d quantum dimer model at the Rokhsar-Kivelson point \cite{Rokhsar1988} and in the 2+1d quantum Lifshitz model \cite{Henley1997,Moessner2001}.}   This can be understood as a result of the symmetries of the model as follows.

In addition to the $U(1)$ momentum and the $\mathbb{Z}$ momentum dipole symmetries discussed above, the continuum theory \eqref{intro-dipphi-action1} has a $U(1)$ winding dipole global symmetry.  We will discuss it in detail below.
Denoting the $U(1)$ charge of the winding dipole symmetry by $\tilde{\mathcal Q}$, we have
\ie
\, [\tilde{\mathcal{Q}},U] = U ~,
\fe
or, in terms of the more general group elements $U_m=U^m$ and $e^{i\theta \tilde{\mathcal{Q}}}$,
\ie
U_m e^{i\theta \tilde{\mathcal{Q}}} = e^{-im\theta}e^{i\theta \tilde{\mathcal{Q}}}U_m~.
\fe
This lack of commutativity between the group elements of the momentum and winding dipole symmetries means that the Hilbert space realizes this symmetry projectively.  And as a result, the ground state is infinitely degenerate.  More abstractly, this can be described as an 't Hooft anomaly between these symmetries.

In Section \ref{sec:dipphi}, we will analyze the theory \eqref{intro-dipphi-action1} in more detail.  In order to regularize the infinite ground state degeneracy, we will formulate it on a finite Euclidean lattice.  And then, in order to preserve the symmetries of the continuum theory, we will study its modified Villain version following \cite{Sulejmanpasic:2019ytl,Gorantla:2021svj}.
On a lattice with $L_x$ sites, the modified Villain model has $L_x$ ground states.  It becomes infinite in the continuum limit.
Curiously, there are at least three natural continuum limits of this lattice model. They have the same action \eqref{intro-dipphi-action1}, but differ in the identifications on the scalar field.

It turns out that this lattice model is the same as the modified Villain version of the 2+1d $\phi$-theory\footnote{The continuum 2+1 $\phi$-theory is described by the action
\ie\label{2+1contphi}
S=\oint d\tau dx dy\left[\frac{\mu_0}{2}(\partial_\tau\phi)^2+\frac{1}{2\mu}(\partial_x\partial_y\phi)^2\right]~.
\fe} of \cite{Gorantla:2021svj} on a slanted spatial 2-torus (as in \cite{Rudelius:2020kta}) with identifications
\ie
(\hat x, \hat y) \sim (\hat x+L_x,\hat y) \sim (\hat x+1, \hat y-1)~,
\fe
where the integers $(\hat x,\hat y)$ label the sites of the spatial lattice,\footnote{Note that $\hat x $ and $\hat y$ are not unit vectors, but integers labeling the sites.} and $L_x$ is the number of sites in the $x$ direction.
This equivalence between the 1+1d theory and the 2+1d theory on a slanted torus exists even for the $U(1)$ and $\mathbb Z_N$ dipole gauge theories described below.  The agreement between the analyses of the 1+1d theory here and the 2+1d theory in \cite{Rudelius:2020kta} provides an interesting perspective and a good check on these two independent discussions.

In Section \ref{sec:dipA}, we will study the pure gauge theory of the $U(1)$ dipole gauge fields $(A_\tau,A_{xx})$ that couple to the $U(1)$ dipole global symmetry of the 1+1d dipole $\phi$-theory. It is the 1+1d version of the pure gauge theory of $(A_\tau,A_{ij})$ mentioned around \eqref{scalarchargetheory-gaugesym}. The gauge theory has defects that describe world-lines of fractons. Which defects should be considered and their properties depend on the second and third subtleties above.

A crucial element in the analysis of gauge theories is their electric and magnetic global symmetries \cite{Gaiotto:2014kfa}. The electric global symmetries are associated with shifts of the gauge fields that leave the action invariant, but are not gauge transformations.

In a pure gauge theory like ours, the system does not have charged dynamical fields and the objects charged under the global symmetry are various line operators and defects.  In a relativistic system, people often abuse the terminology and do not distinguish between line operators, which acts at a given time, and line defects, which are supported on a time-like line.\footnote{As we said above, our discussion will be mostly in Euclidean spacetime.  Then, when we say that the defect is supported on a time-like line, we mean that it is supported on a line along the Euclidean time direction.}   The latter represent the world-line of a probe massive particles.

However, in our case, which is not relativistic, the distinction between these two notions is important.  We refer to symmetries that act on operators as ordinary or \emph{space-like global symmetries}, and to symmetries that act on defects, but not on operators, as \emph{time-like global symmetries}.  See Appendix \ref{app:timelike} for a more detailed discussion of time-like global symmetries.

Let us return to the $U(1)$ dipole gauge theory.  It was originally argued in \cite{Pretko:2016kxt,Pretko:2016lgv,Pretko:2018jbi} that when it is coupled to matter fields, the matter fields are immobile. We will study the theory without dynamical matter fields.  Instead, the theory has defects
\ie\label{fractondefi}
\exp\left( i \oint d\tau~A_\tau(\tau,x) \right)~,
\fe
which represent the world-lines of probe charged particles.  As in the discussion in  \cite{paper1,paper2,paper3} of a different theory, it is easy to see, using the gauge transformation laws of the gauge field \eqref{scalarchargetheory-gaugesym} that these defects are immobile -- these defects are fractons.

Below, we will derive the immobility of the defect \eqref{fractondefi} as a consequence of a global symmetry.  The theory
has a time-like global symmetry that acts on this fracton defect as
\ie\label{timelike-dip-sym}
\exp\left( i \oint d\tau~A_\tau(\tau,x) \right) \rightarrow  \exp\left(ic_\tau + 2\pi i m \frac{x}{\ell_x}\right) \exp\left( i \oint d\tau~A_\tau(\tau,x) \right)~,
\fe
where $c_\tau \sim c_\tau + 2\pi$ is circle-valued, and $m$ is an integer. Fracton defects at different positions carry different time-like dipole charges, so they cannot be deformed to each other without violating the time-like global symmetry.
This explains the restricted mobility of fractons using global symmetry, rather than gauge symmetry. It also gives a more precise explanation of the intuitive ``dipole moment conservation'' discussed in \cite{Pretko:2016kxt,Pretko:2016lgv,Pretko:2018jbi}.

Curiously, the operator   $\oint dx~ A_{xx}$, which is a line observable  acting at a fixed time,  does not need exponentiation for gauge invariance.  This will be discussed in detail below.

We will see that there are other consistent continuum tensor gauge theories with the same Lagrangian, but with different global properties of the fields where the fractons \eqref{fractondefi} are absent.

In Section \ref{sec:dipZN}, we will study the 1+1d $\mathbb Z_N$ dipole gauge theory. Following \cite{Gorantla:2021svj}, we will consider a BF version of the theory on a Euclidean lattice. It has a ground state degeneracy of $N \gcd(N,L_x)$, where $L_x$ is the number of sites in the $x$ direction.  This is a consequence of the mixed 't Hooft anomaly in the space-like symmetry of the model.

Surprisingly, unlike the $U(1)$ theory, the $\mathbb Z_N$ dipole gauge theory has no fractons on the lattice. (This was pointed out in a closely related model in \cite{Bulmash:2018lid,Ma:2018nhd,Oh2021}.)    First, a particle can hop by $N$ sites. In addition, on a finite lattice with $L_x$ sites, a particle can move around the whole space a number of times and end up hopping by $\gcd(N,L_x)$ sites. Once again, the relaxed restriction on the mobility is explained by the time-like global symmetry of the model.

To summarize, the ground state degeneracy of these models follows from their space-like global symmetries and the restricted mobility of their defects is controlled by their time-like global symmetries.

Observe that both the ground state degeneracy and the mobility of defects depend on the number theoretic properties of $L_x$. Consequently, this theory does not have a smooth $L_x\to \infty$ continuum limit, which is a manifestation of the UV/IR mixing of such theories \cite{paper1,Gorantla:2021bda}. Related phenomena have also been observed in various models, for example, \cite{Haah:2011drr,Yoshida:2013sqa,Meng,Manoj:2020wwy,Rudelius:2020kta}.  Our example is presumably the simplest setup exhibiting this phenomenon.\footnote{This phenomenon in our 1+1d model is perhaps not as surprising as the UV/IR mixing in other exotic models in higher spacetime dimensions. Indeed, it is common for standard lattice systems, such systems with frustration, to exhibit a ground state degeneracy that depends sensitively on the number of lattice sites. It would be nice to understand whether a similar interpretation of the degeneracy exists in our example.}

In Appendix \ref{app:timelike}, we will introduce and explain the notion of time-like global symmetry in various well-known theories. In an ordinary gauge theory, it is part of the one-form global symmetry. In exotic gauge theories with subsystem symmetries, including models containing fractons, it explains the restricted mobility of fractons, lineons, etc. In particular, using time-like global symmetries, we will show that the mobility of fractons and lineons in the X-cube model, which is na\"ively a local property, can depend sensitively on the global geometry of the lattice. In all these examples, the time-like symmetry is a consequence of Gauss law in the presence of defects. However, gauge fields are not essential for the existence of time-like symmetry, as we will demonstrate in the case of 2+1d compact boson.

\section{1+1d dipole $\phi$-theory}\label{sec:dipphi}

In this section, we will study a compact scalar field theory in  1+1d with dipole global symmetries.  We will place the modified Villain version of this theory on a finite Euclidean space-time lattice with $L_x$ sites in the $x$-direction, and $L_\tau$ sites in the $\tau$-direction, and impose periodic boundary conditions. This will lead us to explore various different continuum limits.

\subsection{First look at the continuum theory -- compact Lifshitz theory}\label{sec:dipphi-contfirst}

Consider the continuum action
\ie\label{dipphi-action1}
S =  \oint d\tau dx~ \left[ \frac{\mu_0}{2} (\partial_\tau \phi)^2 + \frac{1}{2\mu} (\partial_x^2 \phi)^2 \right]~,
\fe
where $\phi$ is a dimensionless compact scalar with the identification $\phi(\tau,x) \sim \phi(\tau,x) +2\pi$.
We will refer to this theory as the 1+1d dipole $\phi$-theory.

The action \eqref{dipphi-action1} is the same as that of a 1+1d version of Lifshitz scalar field theory (see \cite{Chen:2009ka}, and references therein). However, it differs from the conventional Lifshitz theory in that our scalar field is compact. Hence the term ``compact Lifshitz theory."
It is also similar to that of a 1+1d version of the 2+1d quantum Lifshitz model \cite{Henley1997,
Moessner2001,Vishwanath:2004,Fradkin:2004,Ardonne:2003wa,2005,2018PhRvB..98l5105M,Yuan:2019geh,Lake:2022ico}, which has a compact scalar field. The relation to Lifshitz theory will be reviewed in Section \ref{sec:dipphi-Lifshitz}.

Let us place this continuum system on the circle of length $\ell_x$ and analyze its global symmetries:
\begin{itemize}
\item
 A $U(1)$ momentum symmetry shifts $\phi \rightarrow \phi + c$, where $c$ is a real constant. The periodicity of $\phi$ makes $c$ circle-valued. The Noether current is\footnote{Recall our conventions, as discussed in footnote \ref{conventionsf}.  They guarantee that in Lorentzian signature the charge operator is hermitian.}
\ie\label{dipphi-momcur}
&J_\tau = i\mu_0 \partial_\tau \phi~,\qquad J_{xx} = \frac{i}{\mu} \partial_x^2 \phi~,
\\
&\partial_\tau J_\tau = \partial_x^2 J_{xx}~.
\fe
The conserved charge is
\ie\label{dipphi-momcharge}
Q = \oint dx ~ J_\tau~.
\fe
The operators charged under this symmetry are $e^{in\phi}$ with integer $n$.
\item
A $U(1)$ winding dipole symmetry has Noether current\footnote{We refer to this $U(1)$ symmetry as a dipole symmetry for reasons that will become clear in Section \ref{Villain}.}
\ie\label{dipphi-windcur}
&\tilde{\mathcal J}_\tau = -\frac{1}{2\pi} \partial_x \phi~,\qquad \tilde{\mathcal J}_x = -\frac{1}{2\pi} \partial_\tau \phi~,
\\
& \partial_\tau \tilde{\mathcal J}_\tau = \partial_x \tilde{\mathcal J}_x~.
\fe
The conserved charge is
\ie\label{dipphi-windcharge}
\tilde{\mathcal Q} = \oint dx~ \tilde{\mathcal J}_\tau~.
\fe
A configuration that carries a nontrivial $U(1)$ winding dipole charge $p\in\mathbb Z$ is
\ie
\phi(\tau,x) = -2\pi p \frac{x}{\ell_x}~.
\fe
Since this charge is quantized, this symmetry is $U(1)$, rather than $\mathbb R$.

\item
A $\mathbb Z$ momentum dipole symmetry acts on $\phi$ as
\ie\label{dipphi-momdip}
U_m:\quad \phi(\tau,x) \rightarrow \phi(\tau,x) + 2\pi m \frac{x}{\ell_x}~,
\fe
for some integer $m$. Here we use $U_m$ to denote the corresponding symmetry operator. This symmetry acts on the local operator $\partial_x \phi$ inhomogeneously. Importantly, $e^{i\partial_x \phi}$ is not a well-defined local operator in this continuum theory because $\partial_x \phi$ has mass-dimension $1$.  Note that because of the compact space and the compact target space, this dipole symmetry action does not suffer from the subtlety mentioned after \eqref{dipolechargem}.
\end{itemize}

The $\mathbb Z$ momentum dipole symmetry does not commute with the $U(1)$ winding dipole symmetry
\ie
U_m e^{i\theta \tilde{\mathcal{Q}}} = e^{-im\theta}e^{i\theta \tilde{\mathcal{Q}}}U_m~.
\fe
The minimal representation of this algebra is infinite dimensional with states $|p\rangle$, $p\in\mathbb{Z}$
\ie
&\tilde{\mathcal{Q}} |p\rangle = p|p\rangle~,
\\
&U_m |p\rangle = |p+m\rangle~.
\fe
It implies that all energy levels, in particular the lowest energy level, are infinitely degenerate and the states transform in a projective representation of the two symmetries. This signals a mixed 't Hooft anomaly.

It is easy to see that the theory is not invariant under any scale transformation.  For example, under the scale transformation of the Lifshitz theory with a noncompact $\phi$,
\ie\label{xtauscal}
x\rightarrow \lambda x ~ , \qquad \tau \rightarrow \lambda^2\tau~,
\fe
the couplings scale as
\ie
\mu_0 \rightarrow \frac{\mu_0}{\lambda}~,
\quad \mu\rightarrow \lambda \mu~.
\fe
Alternatively, we can keep the coupling constants unchanged, but scale $\phi$.  This has the effect of changing the periodicity of $\phi$ from $2\pi$ to $2\pi/\sqrt{\lambda}$.  Either way, we see that the dipole $\phi$-theory \eqref{dipphi-action1} is not scale invariant.

Since the continuum theory is very singular, we would like to regularize it on a lattice, while preserving its global symmetries. Below we will discuss the modified Villain model of \eqref{dipphi-action1}, which provides an unambiguous regularization of the continuum theory.

\subsection{Modified Villain formulation}\label{Villain}

In the Villain form, the continuum theory is associated with the lattice action\footnote{In Section \ref{sec:dipphi-robustness}, we will comment on the relation between them.}
\ie\label{dipphi-Vill-action}
S = \frac{\beta_0}{2} \sum_{\tau\text{-link}} (\Delta_\tau \phi - 2\pi n_\tau)^2 + \frac{\beta}{2} \sum_\text{site} (\Delta_x^2 \phi - 2\pi n_{xx})^2 ~,
\fe
where $\Delta_x^2 f(\hat x) = f(\hat x+1) + f(\hat x-1) - 2f(\hat x)$.  Here, $\phi$ is a real-valued scalar field and $(n_\tau, n_{xx})$ are integer-valued gauge fields with gauge symmetry
\ie\label{dipphi-modVill-gaugesym}
&\phi \sim \phi + 2\pi k~,
\\
&n_\tau \sim n_\tau + \Delta_\tau k~,
\\
&n_{xx} \sim n_{xx} + \Delta_x^2 k~,
\fe
where $k$ are integer gauge parameters.  This $\mathbb Z$ gauge symmetry effectively makes $\phi$ compact.

We can further deform the action \eqref{dipphi-Vill-action} to the modified Villain version:
\ie\label{dipphi-modVill-action}
S = \frac{\beta_0}{2} \sum_{\tau\text{-link}} (\Delta_\tau \phi - 2\pi n_\tau)^2 + \frac{\beta}{2} \sum_\text{site} (\Delta_x^2 \phi - 2\pi n_{xx})^2 + i\sum_{\tau\text{-link}}\tilde \phi (\Delta_\tau n_{xx} - \Delta_x^2 n_\tau)~,
\fe
where $\tilde \phi$ is a Lagrange multiplier that makes the integer gauge fields $(n_\tau, n_{xx})$ flat. It has a gauge symmetry
\ie\label{dipphi-modVill-gaugesym2}
\tilde \phi \sim \tilde \phi + 2\pi \tilde k~,
\fe
where $\tilde k$ are integer gauge parameters. Physically, the above deformation suppresses all topological excitations, or vortices.

In the rest of this subsection, we will analyze this modified Villain model \eqref{dipphi-modVill-action} following similar steps in \cite{Gorantla:2021svj}.

\subsubsection{Relation to 2+1d $\phi$-theory}\label{sec:dipphi-compactify}

First, we will provide an alternative interpretation of  the 1+1d action \eqref{dipphi-modVill-action}.
 We will show that it arises from the modified Villain version of the 2+1d $\phi$-theory\footnote{The continuum limit of this modified Villain lattice model is the 2+1d $\phi$-theory of  \cite{PhysRevB.66.054526,paper1} with Lagrangian \eqref{2+1contphi}.  See also \cite{Tay_2011,You:2019cvs,You:2019bvu,Karch:2020yuy,You:2021tmm,Gorantla:2021bda,Burnell:2021reh,Distler:2021qzc,Distler:2021bop} for related discussions.} \cite{Gorantla:2021svj}
\ie\label{2+1dphi}
S = \frac{\beta_0}{2} \sum_{\tau\text{-link}} (\Delta_\tau \phi - 2\pi n_\tau)^2 + \frac{\beta}{2} \sum_{xy\text{-plaq}} (\Delta_x\Delta_y \phi - 2\pi n_{xy})^2 + i\sum_\text{cube} \phi^{xy} (\Delta_\tau n_{xy} - \Delta_x\Delta_y n_\tau)
\fe
on a special torus.
This relation can be viewed roughly as a dimension reduction, but we emphasize that this is an exact equivalence with no approximation involved.

We place this 2+1d lattice model \eqref{2+1dphi} on a slanted spatial torus with identifications
\ie\label{slanted}
(\hat x, \hat y) \sim (\hat x+L_x,\hat y) \sim (\hat x+1, \hat y-1)~.
\fe
On this slanted torus, we have the following relations:
\ie
&\Delta_y \phi(\tau, \hat x, \hat y) = \Delta_x \phi(\tau, \hat x, \hat y) \,,\\
&\Delta_x \Delta_y \phi(\tau, \hat x,\hat y) = \Delta_x^2 \phi(\tau, \hat x+1,\hat y)~.
\fe
In \cite{Rudelius:2020kta}, the continuum 2+1d $\phi$-theory was studied on a    torus with more general complex structure.

Next, we treat the $y$-direction as the compactified direction, and view the resulting system as 1+1 dimensional.
More specifically, we can always use the identification \eqref{slanted} to bring any field to $\hat y=0$.
We thus replace the fields of the 2+1d model by the fields in 1+1d:
\ie
&\phi(\tau,\hat x, 0) = \phi(\tau, \hat x)\,,&&\qquad \Delta_x \Delta_y \phi(\tau, \hat x-1,0) = \Delta_x^2 \phi(\tau, \hat x)\,,\\
&n_\tau (\tau, \hat x,0)= n_\tau (\hat \tau ,\hat x)\,,&&\qquad n_{xy}(\tau, \hat x-1, 0) = n_{xx}(\tau, \hat x) \,,\\
&  \phi^{xy}(\tau, \hat x-1,0)= \tilde \phi(\tau, \hat x)\,.
\fe
Under this replacement, the 2+1d action \eqref{2+1dphi} on this elongated torus \eqref{slanted} is exactly equivalent  to the 1+1d model \eqref{dipphi-modVill-action}.
From this exact equivalence, all the analysis in the rest of this section follows from the  2+1d $\phi$-theory on a slanted torus  \cite{Rudelius:2020kta}.

\subsubsection{Global symmetry}\label{sec:villainsymmetry}

The global symmetry of the modified Villain model \eqref{dipphi-modVill-action} includes:

\begin{itemize}
\item
The $U(1)$ momentum symmetry acts as $\phi \rightarrow \phi + c$, where $c$ is a real constant. The Noether current is
\ie\label{dipphi-modVill-momcur}
&J_\tau = i\beta_0 (\Delta_\tau \phi - 2\pi n_\tau)~,\qquad J_{xx} = i\beta (\Delta_x^2 \phi - 2\pi n_{xx})~,
\\
&\Delta_\tau J_\tau = \Delta_x^2 J_{xx}~,
\fe
which follows from the equation of motion of $\phi$. The charge is
\ie\label{U1momentumVillain}
Q = \sum_{\tau\text{-link: fixed }\hat \tau} J_\tau~,
\fe
and the charged operators are $e^{i n\phi}$ with charge $n\in \mathbb Z$.  The symmetry transformations with $c$ and with $c+2\pi$  are related by a gauge transformation.  Therefore, this symmetry group is $U(1)$ rather than $\mathbb R$.

\item
The $\mathbb Z_{L_x}$ momentum dipole symmetry acts as
\ie\label{dipphi-modVill-momdip}
&\phi \rightarrow \phi + 2\pi m \frac{\hat x}{L_x}~, &&\quad\text{for}\quad 0 \le \hat x < L_x~,
\\
&n_{xx} \rightarrow n_{xx} + m \left( \delta_{\hat x,0} - \delta_{\hat x,L_x-1} \right)~,
\fe
where $m=0,1,\ldots, L_x-1$, and $\delta_{\hat x,\hat x_0}$ is the Kronecker delta function. It is a $\mathbb Z_{L_x}$ rather than a $\mathbb Z$ symmetry, because the shift corresponding to $m \in L_x \mathbb Z$ is a gauge transformation \eqref{dipphi-modVill-gaugesym}. The symmetry operator is\footnote{The current $J_\tau$ is imaginary in Euclidean signature.  Following the comment in footnote \ref{conventionsf}, it is hermitian in Lorentzian signature and consequently, $U_m$ is unitary.}
\ie\label{dipphi-modVill-momdip-symop}
U_m = \exp\left( \frac{2\pi i m}{L_x} \sum_{\tau\text{-link: fixed }\hat \tau} \hat x J_\tau - i m \left[ \tilde \phi(\hat \tau, 0) - \tilde \phi(\hat \tau,L_x-1) \right] \right)~.
\fe
Here the sum is restricted to the fundamental domain $0\leq \hat x<L_x$. It can be understood in a simple way: the first and second terms shift $\phi$ and $n_{xx}$, respectively, as in \eqref{dipphi-modVill-momdip}. The charged operators are $e^{i\phi}$ and $e^{i p \Delta_x \phi}$ with $p \in \mathbb Z$. The operator $e^{i p \Delta_x \phi}$ has $\mathbb{Z}_{L_x}$ charge $p$ mod $L_x$.  See below how the symmetry acts on these charged operators.

\item
There is a $U(1)$ winding symmetry that shifts $\tilde \phi \rightarrow \tilde \phi + \tilde c$, where $\tilde c$ is a circle-valued real constant. The Noether current is
\ie\label{dipphi-modVill-windcur}
&\tilde J_\tau = \frac{1}{2\pi} (\Delta_x^2 \phi - 2\pi n_{xx})~, \qquad \tilde J_{xx} = \frac{1}{2\pi} (\Delta_\tau \phi - 2\pi n_\tau)~,
\\
&\Delta_\tau \tilde J_\tau = \Delta_x^2 \tilde J_{xx}~,
\fe
which follows from the equation of motion of $\tilde \phi$. The charge is
\ie\label{dipphi-modVill-windcharge}
\tilde Q = \sum_{\text{site: fixed }\hat \tau} \tilde J_\tau = - \sum_{\text{site: fixed }\hat \tau} n_{xx}~.
\fe
The charged operators are $e^{i\tilde n \tilde \phi}$ with charge $\tilde n \in \mathbb Z$.

\item
Finally, there is a $\mathbb Z_{L_x}$ winding dipole symmetry that shifts
\ie\label{dipphi-modVill-winddip}
\tilde \phi \rightarrow \tilde \phi + 2\pi \tilde m \frac{\hat x}{L_x}~,\quad \text{for}\quad 0 \le \hat x < L_x~,
\fe
where $\tilde m \in \mathbb Z$. It is a $\mathbb Z_{L_x}$ rather than a $\mathbb Z$ symmetry because the shift corresponding to $\tilde m \in L_x \mathbb Z$ is a gauge transformation \eqref{dipphi-modVill-gaugesym}. The symmetry operator is
\ie\label{dipphi-modVill-winddipo}
\tilde U_{\tilde m} = \exp\left(-\frac{2\pi i \tilde m}{L_x} \sum_{\text{site: fixed }\hat \tau} \hat x n_{xx} \right)~.
\fe
The charged operators are $e^{i\tilde \phi}$ and $e^{i \tilde p \Delta_x \tilde \phi}$ with $\tilde p \in \mathbb Z$. The operator $e^{i \tilde p \Delta_x \tilde \phi}$ has $\mathbb{Z}_{L_x}$ charge $\tilde p$ mod $L_x$.  See below how the symmetry acts on these charged operators.
\end{itemize}

The operator $e^{in\phi}$ ($e^{i\tilde n\tilde \phi}$), which is charged under the $U(1)$ momentum (winding) symmetry does not transform simply under the $\mathbb Z_{L_x}$ momentum (winding) dipole symmetry. The reason is that the $\mathbb Z_{L_x}$ spatial translation symmetry does not commute with the dipole symmetries. Let $T$ be the generator of the lattice translation $\hat x \rightarrow \hat x + 1$. Then,
\ie
&TU_mT^{-1} = e^{\frac{2\pi i}{L_x} mQ} U_m~,
\\
&T\tilde U_{\tilde m}T^{-1} = e^{\frac{2\pi i}{L_x} \tilde m \tilde Q} \tilde U_{\tilde m}~,
\fe
Note that this lack of commutativity is not a central extension of the symmetries generated by the dipole symmetries and the translations.  It is not an anomaly.

On the other hand, the non-commutativity of the two $\mathbb Z_{L_x}$ dipole symmetries,
\ie\label{noncommute}
U_m \tilde U_{\tilde m} = e^{-\frac{2\pi i}{L_x} m \tilde m}\tilde U_{\tilde m} U_m~,
\fe
does signal a mixed 't Hooft anomaly between them.  (This follows from using \eqref{dipphi-modVill-momdip} and \eqref{dipphi-modVill-winddip} in the operators \eqref{dipphi-modVill-winddipo} and \eqref{dipphi-modVill-momdip-symop}.) As a result, every energy level is $L_x$-fold degenerate. In particular, there is a large ground state degeneracy, which depends on the number of lattice sites $L_x$.  As in \cite{Gorantla:2021bda}, this degeneracy, which depends on the number of sites is a manifestation of UV/IR mixing.

As discussed in Section \ref{sec:dipphi-compactify}, the 1+1d modified Villain model \eqref{dipphi-modVill-action} is equivalent to the 2+1d $\phi$-theory on a slanted torus \eqref{slanted}.
Indeed, the algebra \eqref{noncommute} arises from the projective $\mathbb{Z}_M\times \mathbb{Z}_M$ symmetry discussed in Section 3 of \cite{Rudelius:2020kta}, with $M=L_x$ in that reference for the torus \eqref{slanted}.

\subsubsection{Self-duality}\label{sec:villainduality}

Using Poisson resummation of the integers $(n_\tau,n_{xx})$, the modified Villain model \eqref{dipphi-modVill-action} is self-dual with $\phi \leftrightarrow \tilde \phi$ and $\beta_0 \leftrightarrow \frac{1}{(2\pi)^2\beta}$. The dual action is
\ie\label{dipphi-modVill-dualaction}
S = \frac{1}{2(2\pi)^2\beta} \sum_\text{site} (\Delta_\tau \tilde \phi - 2\pi \tilde n_\tau)^2 + \frac{1}{2(2\pi)^2\beta_0} \sum_{\tau\text{-link}} (\Delta_x^2 \tilde \phi - 2\pi \tilde n_{xx})^2 - i\sum_\text{site} \phi (\Delta_\tau \tilde n_{xx} - \Delta_x^2 \tilde n_\tau)~,
\fe
where $(\tilde n_\tau,\tilde n_{xx})$ are integer gauge fields that make $\tilde \phi$ compact. Under the gauge symmetry \eqref{dipphi-modVill-gaugesym2}, they transform as
\ie
\tilde n_\tau \sim \tilde n_\tau + \Delta_\tau \tilde k~, \qquad \tilde n_{xx} \sim \tilde n_{xx} + \Delta_x^2 \tilde k~.
\fe

\subsubsection{Gauge fixing the integers}\label{sec:dipphi-modVill-gaugefix}

Following the same procedure as in \cite{Gorantla:2021svj},  after integrating out $\tilde \phi$ and gauge fixing the integer gauge fields, the action \eqref{dipphi-modVill-action} can be written in terms of a new field $\bar \phi$ as
\ie\label{dipphi-modVill-gaugefixaction}
S = \frac{\beta_0}{2} \sum_{\tau\text{-link}} (\Delta_\tau \bar \phi)^2 + \frac{\beta}{2} \sum_\text{site} (\Delta_x^2 \bar \phi)^2~.
\fe
The new field is defined as
\ie
&\bar \phi(0,0) = \phi(0,0)~,\qquad \Delta_x \bar\phi(0,-1) = \Delta_x \phi(0,-1)~,
\\
&\Delta_x^2 \bar \phi = \Delta_x^2 \phi - 2\pi n_{xx}~,
\\
&\Delta_\tau \bar \phi = \Delta_\tau \phi - 2\pi n_\tau~.
\fe
The integer gauge fields are gauge fixed to zero except that
\ie
&n_\tau(L_\tau-1,\hat x)=-\bar n_\tau~,
\\
&n_{xx}(0,0)=-\bar n_{xx}-\bar p_{xx}~,
\\
&n_{xx}(0,L_x-1)=\bar p_{xx}~,
\fe
where $\bar n_\tau,\bar n_{xx},\bar p_{xx}\in\mathbb{Z}$.
The residual gauge symmetry acts on $\bar\phi$ as
\ie\label{dipphi-modVill-gaugesym-barphi}
\bar \phi(\hat\tau,\hat x,\hat y)\sim \bar\phi(\hat\tau,\hat x,\hat y) + 2\pi k_0 + 2\pi k_x \hat x~,\quad k_0,k_x\in\mathbb{Z}~.
\fe
The remaining gauge parameters $k_0,k_x$ are constant on the lattice.

Unlike $\phi$, the field $\bar \phi$ need not be single-valued. Instead, it can wind with the boundary condition
\ie
&\bar \phi(\hat \tau+L_\tau, \hat x)=\bar \phi(\hat \tau, \hat x)+2\pi \bar n_\tau~,
\\
&\bar \phi(\hat \tau, \hat x+L_x)=\bar \phi(\hat \tau, \hat x)+2\pi \bar n_{xx} \hat x +2\pi \bar p_{xx}~.
\fe
Because of the gauge symmetry \eqref{dipphi-modVill-gaugesym-barphi}, $\tilde p_{xx}\sim\tilde p_{xx}+L_x$. One configuration that winds in the $x$-direction is
\ie\label{dipphi-modVill-windconfig}
\bar \phi(\hat \tau, \hat x) = 2\pi \bar n_{xx} \frac{\hat x(\hat x- L_x)}{2L_x} +2\pi \bar p_{xx} \frac{\hat x}{L_x}~.
\fe

\subsubsection{Spectrum}\label{spectrum}

We will now determine the spectrum of the theory.  We will work with a continuous Lorentzian time, denoted by $t$, while keeping the space discrete. We do this by introducing a lattice spacing $a_\tau$ in the $\tau$-direction, taking the limit $a_\tau\rightarrow 0$, while keeping $\beta_0'=\beta_0 a_\tau$ and $\beta'=\beta/a_\tau$ fixed, and then Wick rotating from Euclidean time to Lorentzian time.

The spectrum of the modified Villain model \eqref{dipphi-modVill-action} includes plane waves with nonzero spatial momentum and states charged under the $U(1)$ momentum and winding symmetries. The dispersion relation for the plane waves is
\ie
\omega_{n_x} = 4\sqrt{\frac{\beta'}{\beta_0'}} \sin^2\left( \frac{\pi n_x}{L_x} \right)~.
\fe

The winding configuration \eqref{dipphi-modVill-windconfig} has the minimal energy with those charges:
\ie
H = \frac{\beta'}{2 } \sum_{\text{site: fixed }\tau} (\Delta_x^2 \bar\phi)^2 = \frac{(2\pi)^2\beta'}{2  L_x} \tilde n^2~.
\fe
Note that the energy does not depend on $\tilde p$. This is related to the fact that the two $\mathbb Z_{L_x}$ dipole symmetries have a mixed 't Hooft anomaly resulting in a degeneracy in the spectrum.

Similarly, the minimal energy of a state with $U(1)$ momentum charge $n$ is
\ie
H = \frac{n^2}{2 \beta_0' L_x}~.
\fe

For fixed lattice parameters (recall that we have taken $a_\tau\to 0$ and rotated to Lorentzian signature), the energies of the three kinds of states scale with $L_x$ as
\ie\label{fixedbetas}
E_\text{wave} \sim \frac{1}{L_x^2}~,\qquad E_\text{mom} \sim \frac{1}{L_x}~,\qquad E_\text{wind} \sim \frac{1}{L_x}~,
\fe
i.e., for large $L_x$, the states charged under the $U(1)$ momentum or winding symmetry are parametrically heavier than the plane waves.

Finally, recall that each state appears $L_x$ times forming a projective representation of the two ${\mathbb Z}_{L_x}$ momentum and winding symmetries.

The above degeneracy is lifted if we  impose only the momentum symmetries.
Indeed, the deformation of the modified Villain model \eqref{dipphi-modVill-action} by the winding dipole operator $e^{i\Delta_x \tilde \phi}$ breaks the $\mathbb Z_{L_x}$ winding dipole symmetry explicitly and lifts the ground state degeneracy.

\subsection{Continuum limits}\label{sec:dipphi-cont}

Now that we understand the modified Villain model \eqref{dipphi-modVill-action}, we can explore its continuum limit.  Surprisingly, there are three possible continuum limits.  All of them have the same continuum Lagrangian, but their fields have different properties.  Consequently, the three different continuum theories have different global symmetries (see Table \ref{tbl:lat-cont-sym}) and different spectra (see Table  \ref{tbl:lat-cont-energies}). One of these theories corresponds to the continuum theory \eqref{dipphi-action1}.

In all these limits, we introduce the spatial lattice spacing $a$, and take the limit $a,a_\tau \rightarrow 0$ and $L_x,L_\tau \rightarrow \infty$ such that $\ell_x = a L_x$ and $\ell_\tau = a_\tau L_\tau$ are fixed.

Before analyzing the system in detail, let us discuss the limit of the algebra of dipole symmetries \eqref{noncommute}
\ie\label{noncommutea}
U_m \tilde U_{\tilde m} = e^{-\frac{2\pi i}{L_x} m \tilde m}\tilde U_{\tilde m} U_m~.
\fe
As we take $L_x\to \infty$, we can focus on different elements of this algebra to find different limits.  Here are some options.
\begin{itemize}
\item We can focus on $U_m$ and $\tilde U_{\tilde m}$ with finite $m $ and $\tilde m$.  In this limit, these operators lead to two commuting copies of $\mathbb Z$.
\item We can focus on $U_m$ with finite $m$ and $\tilde U_{\tilde m}$ with $\tilde m \to \infty$ and finite ${\tilde m\over L_x}\to \tilde r$.  (Clearly, $\tilde r$ is circle-valued.)  In this limit, the operators $U_m$ lead to $\mathbb Z$, the operators $\tilde U_{\tilde m}\to \tilde U_{\tilde r}$ lead to $U(1)$ and they do not commute
    \ie
    U_m \tilde U_{\tilde r} = e^{-2\pi i m \tilde r}\tilde U_{\tilde r} U_m~.
    \fe
\item We can exchange $U\leftrightarrow \tilde U$ in the previous limit.
\item We can focus on $U_m$ and $\tilde U_{\tilde m}$ with $m, \tilde m \to \infty$, with fixed $m \over \sqrt{L_x}$, $\tilde m\over \sqrt{L_x}$.  In this limit, we can write $U_m\to \exp(i{m\over \sqrt{L_x}}{\cal U})$ and $\tilde U_{\tilde m} \to \exp(i{\tilde m\over \sqrt{L_x}}\tilde {\cal U})$. $\cal U$ and $\tilde {\cal U}$ generate two copies of $\mathbb R$, which do not commute
    \ie
   \,  [{\cal U},\tilde {\cal U}]=2\pi i~.
    \fe
\end{itemize}
Below we will see these algebras (except the first one) in various limits of the lattice system.

\subsubsection{1+1d dipole $\phi$-theory} \label{sec:dipphi-cont2}

To obtain the continuum dipole $\phi$-theory \eqref{dipphi-action1}, we scale the lattice coupling constants with $a,a_\tau$ as
\ie\label{dipphi-scaling}
\beta_0 = \frac{\mu_0 a}{a_\tau}~, \qquad \beta = \frac{a_\tau}{\mu a^3}~,
\fe
where $\mu_0$ and $\mu$ are fixed continuum coupling constants with mass dimensions $0$ and $2$ respectively.

In this continuum limit, the global symmetries of the modified Villain model reduce to the ones discussed in Section \ref{sec:dipphi-contfirst}. This is the second option in the list following \eqref{noncommutea}. See Table \ref{tbl:lat-cont-sym}, for the relation between the global symmetries in this continuum limit and on the lattice. In particular, the $U(1)$ winding symmetry of the modified Villain lattice model does not act in the continuum theory. Since $\partial_x \phi$ is a well-defined operator, the $U(1)$ winding charge associated with \eqref{dipphi-modVill-windcur} vanishes\footnote{We see here an interesting analogy with the $\phi$-theories with subsystem symmetry in 2+1d.  There, the momentum and winding subsystem symmetry currents exist in the continuum limit.  But the continuum theory has no charged finite energy states  \cite{paper1,Gorantla:2021svj,Gorantla:2021bda}.\label{analowss}}
\ie
\tilde Q = \frac{1}{2\pi}\oint dx~ \partial_x^2\phi = 0.
\fe

Relatedly, the lattice operator $e^{i\Delta_x \phi}$ becomes neutral under the momentum dipole symmetry.  It does not lead to exponential operators, but to operators of the form $\partial_x \phi$, which transforms under the $\mathbb Z$ momentum dipole symmetry inhomogeneously. In contrast, $e^{i\phi}$ is a well-defined local operator charged under the $U(1)$ momentum symmetry. See Table \ref{tbl:lat-cont-chargedop} for a comparison of charged operators on the lattice and in the continuum theory.

\renewcommand{\arraystretch}{1.5}
\begin{table}[t]
\begin{center}
\begin{tabular}{|c|c|c|}
\hline
Lattice & Continuum dipole $\phi$-theory & Continuum dipole $\hat \phi$-theory
\tabularnewline
\hline
$U(1)$ momentum & $U(1)$ momentum & does not act
\tabularnewline
\hline
$U(1)$ winding & does not act & does not act
\tabularnewline
\hline
$\mathbb Z_{L_x}$ momentum  dipole & $\mathbb Z$ momentum dipole & $\mathbb R$ momentum dipole
\tabularnewline
\hline
$\mathbb Z_{L_x}$ winding dipole & $ U(1)$ winding dipole & $\mathbb R$ winding dipole
\tabularnewline
\hline
\end{tabular}
\caption{Relation between the global symmetries on the lattice and in various continuum limits. The global symmetries of the continuum dipole $\Phi$-theory are the same as in the second column after swapping ``momentum'' and ``winding.''}\label{tbl:lat-cont-sym}
\end{center}
\end{table}

\renewcommand{\arraystretch}{1.6}
\begin{table}[t]
\begin{center}
\begin{tabular}{|c|c|c|c|}
\hline
Theory & $E_\text{wave}$ & $E_\text{mom}$ & $E_\text{wind}$
\tabularnewline
\hline
Modified Villain model & $\frac{1}{L_x^2}$ & $\frac{1}{L_x}$ & $\frac{1}{L_x}$
\tabularnewline
\hline
Continuum dipole $\phi$-theory & $\frac{1}{\ell_x^2}$ & $\frac{1}{\ell_x}$ & $\frac{1}{a^2\ell_x}$
\tabularnewline
\hline
Continuum dipole $\hat \phi$-theory & $\frac{1}{\ell_x^2}$ & $\frac{1}{a\ell_x}$ & $\frac{1}{a\ell_x}$
\tabularnewline
\hline
\end{tabular}
\caption{Energies of the three kinds of states on the lattice and in various continuum limits. The energies of the continuum dipole $\Phi$-theory are the same as in the third row after swapping ``momentum'' and ``winding.'' The fact that the energy of the winding states of the $\phi$-theory diverge in the continuum limit is compatible with the lack of local winding operators in this theory (Table \ref{tbl:lat-cont-chargedop}). A similar comment applies to the momentum and winding states in the $\hat\phi$-theory.}\label{tbl:lat-cont-energies}
\end{center}
\end{table}

\begin{table}[t]
\begin{center}
\begin{tabular}{|c|c|c|c|c|}
\hline
Symmetry & Lattice & $\phi$-theory & $\Phi$-theory & $\hat \phi$-theory
\tabularnewline
\hline
Momentum & $e^{i\phi}$ & $e^{i\phi}$ & --  & --
\tabularnewline
\hline
Winding & $e^{i\tilde \phi}$ & -- & $e^{i\tilde \phi}$ & --
\tabularnewline
\hline
Momentum dipole & $e^{i\Delta_x\phi}$ & $\partial_x\phi$ & $e^{i\partial_x \Phi}$ & $\partial_x \hat \phi$
\tabularnewline
\hline
Winding dipole & $e^{i\Delta_x\tilde \phi}$ & $e^{i\partial_x \tilde \Phi}$ & $\partial_x \tilde \phi$ & $\partial_x \check \phi$
\tabularnewline
\hline
\end{tabular}
\caption{Some local operators that transform under various symmetries on the lattice and in the three continuum limits. The exponentiated operators shown here transform linearly under their respective symmetry transformations, whereas the others transform inhomogeneously. Here, $\phi$ and $\tilde \phi$ have mass dimension $0$, $\Phi$ and $\tilde \Phi$ have mass dimension $-1$, and $\hat \phi$ and $\check\phi$ have mass dimension $-\frac12$. Note that the exponent is always dimensionless, and its coefficient is always an integer. Consequently, no continuum theory has local operators of the form $e^{i\phi}$ and $e^{i\partial_x \phi}$ at the same time. The continuum fields $\Phi$ and $\tilde \phi$ are dual to each other as discussed in Section \ref{sec:dipphi-cont3}. Similarly, the continuum fields $\phi$ and $\tilde \Phi$ are dual to each other. The continuum fields $\hat \phi$ and $\check \phi$ are dual to each other, as discussed in Section \ref{sec:dipphi-cont1}.}\label{tbl:lat-cont-chargedop}
\end{center}
\end{table}

As mentioned around \eqref{xtauscal}, the dipole $\phi$-theory is not scale invariant under any scaling of $x$ and $\tau$  because the periodicity of $\phi$ is not preserved under this scaling.

The energies of the three kinds of states in this limit are (see Table \ref{tbl:lat-cont-energies})
\ie\label{momeli}
E_\text{wave} \sim \frac{1}{\sqrt{\mu_0 \mu}}\frac{1}{\ell_x^2}~,\qquad E_\text{mom} \sim \frac{1}{\mu_0\ell_x}~,\qquad E_\text{wind} \sim \frac{1}{\mu a^2 \ell_x}~.
\fe
We see that the plane waves and momentum states have finite energy, but the winding states are infinitely heavy. This is consistent with the fact the $U(1)$ winding symmetry of the lattice model does not exist in this continuum limit. Here we see that the dipole $\phi$-theory is not self-dual. We will study the dual theory in subsection \ref{sec:dipphi-cont3}.

\subsubsection{1+1d dipole $\Phi$-theory}\label{sec:dipphi-cont3}

We consider a different continuum limit of \eqref{dipphi-modVill-action} by scaling
\ie\label{dipphi-dual-scaling}
\beta_0 = \frac{M_0 a^3}{a_\tau}~, \qquad \beta = \frac{a_\tau}{M a}~,
\fe
where $M_0$ and $M$ are fixed continuum coupling constants with mass dimensions $2$ and $0$,  respectively.
At the same time, we define the  continuum field $\Phi$ as
\ie
\Phi   \equiv   a\bar\phi\,.
\fe
Recall that $\bar\phi$ is the gauge-fixed version of $\phi$ on the modified Villain lattice model.
The action of this continuum limit is
\ie\label{dipphi-dualaction}
S = \oint d\tau dx~ \left[ \frac{M_0}{2} (\partial_\tau \Phi)^2 + \frac{1}{2M} (\partial_x^2 \Phi)^2 \right]~.
\fe

This action is very similar to that of the dipole $\phi$-theory \eqref{dipphi-action1}, but $\Phi$ has a different mass dimension of $-1$, and a different identification
\ie\label{tildecx}
\Phi(\tau,x) \sim \Phi(\tau,x) +  c + 2\pi x~,
\fe
where $ c$ is an arbitrary constant. We will refer to this theory as the 1+1d dipole $\Phi$-theory.

Using the standard duality transformation in the continuum, we find that the $\Phi$-theory is dual to the 1+1d dipole $\phi$-theory \eqref{dipphi-action1}:
\ie
S =  \oint d\tau dx~ \left[ \frac{\tilde \mu_0}{2} (\partial_\tau \tilde \phi)^2 + \frac{1}{2\tilde \mu} (\partial_x^2 \tilde \phi)^2 \right]~,
\fe
 with the following identification of the couplings:
\ie
&M_0  = {\tilde \mu \over (2\pi)^2}\,,~~~M = (2\pi)^2 \tilde \mu_0\,.
\fe
Here, $\tilde \phi$ has mass-dimension $0$.  It is subject to the identification $\tilde\phi\sim\tilde\phi+2\pi$ and has the same global properties as $\phi$ of Section \ref{sec:dipphi-cont2}. The currents from the two dual descriptions are mapped to each other as follows:
\ie\label{dipphi-duality}
iM_0 \partial_\tau \Phi = \frac{1}{2\pi} \partial_x^2 \tilde \phi ~, \qquad \frac{i}{M} \partial_x^2 \Phi = \frac{1}{2\pi} \partial_\tau \tilde \phi~.
\fe
This is the continuum version of the duality in the modified Villain lattice model discussed in Section \ref{sec:villainduality}.

We now discuss the global symmetries in this continuum limit.
This theory corresponds to the third option in the list following \eqref{noncommutea}.
\begin{itemize}
\item
Since the constant shift of the continuum field $\Phi$ is part of the gauge symmetry \eqref{tildecx},  the $U(1)$ momentum charge \eqref{U1momentumVillain} vanishes:
\ie
 Q = i M_0 \oint dx~ \partial_\tau\Phi = 0\,.
\fe
We conclude that the $U(1)$ momentum symmetry on the lattice does not act in the continuum $\Phi$-theory. This is consistent with the fact that the would-be charged local operator $e^{i\Phi}$ does not exist in the continuum theory because $\Phi$ has mass dimension $-1$ and has the gauge symmetry \eqref{tildecx}. The analogy mentioned in footnote \ref{analowss} is applicable also here.

\item The momentum dipole symmetry \eqref{dipphi-modVill-momdip} shifts $\Phi\rightarrow \Phi+\theta x$ with $\theta\sim\theta+2\pi$.
The charged operator on the  lattice is $e^{i\Delta_x \phi}$, which becomes a non-trivial local operator $e^{i \partial_x \Phi}$ in the continuum.  The $\mathbb{Z}_{L_x}$ momentum dipole symmetry becomes  a $U(1)$ symmetry in the continuum.

\item The symmetry group of the   winding symmetry \eqref{dipphi-modVill-windcharge} is still $U(1)$ in the continuum $\Phi$-theory. The $U(1)$ winding charge $\tilde Q$ is
\ie
\tilde Q = {1\over 2\pi} \oint dx  \,\partial_x^2\Phi\,.
\fe
The minimally charged configuration is
\ie
\Phi = \frac{2\pi x (x-\ell_x)}{2\ell_x}\,.
\fe
The charged local operator is $e^{i\tilde \phi}$, where $\tilde \phi$ is the dimensionless dual field of $\Phi$.

\item The $\mathbb{Z}_{L_x}$ winding dipole symmetry operator \eqref{dipphi-modVill-winddipo}  on the lattice becomes a $\mathbb{Z}$ symmetry in the continuum $\Phi$-theory. The symmetry operator in the continuum is
\ie\label{dipphi-dipolesymop}
\tilde U_{\tilde m} = \exp\left[ -\frac{i\tilde m}{\ell_x} \left(\int_{x_0}^{x_0+\ell_x} dx~\partial_x \Phi - 2\pi x_0\tilde Q\right) \right]~,\qquad \text{for}\qquad \tilde m\in \mathbb Z~.
\fe
Although $\partial_x \Phi$ is not a well-defined operator, the symmetry operator $\tilde U_{\tilde m}$ is well defined. Moreover, it is independent of $x_0$. A configuration that carries a nontrivial $\mathbb Z$ dipole charge is $\Phi = \theta x$ (where we identify $\theta\sim\theta +2\pi$). The charged local operator is $\partial_x \tilde \phi$ which realizes the $\mathbb Z$ winding dipole symmetry inhomogeneously.
\end{itemize}

See Table \ref{tbl:lat-cont-sym} for the relation between the global symmetries in this continuum limit and on the lattice.

As in the discussion of the $\phi$-theory in Section \ref{sec:dipphi-cont2}, because of the identification \eqref{tildecx}, this theory is also not scale invariant under any scaling of $x$ and $\tau$.

The energies of the three kinds of states in this limit are
\ie
E_\text{wave} \sim \frac{1}{\sqrt{M_0M}}\frac{1}{\ell_x^2}~,\qquad E_\text{mom} \sim \frac{1}{M_0a^2 \ell_x}~,\qquad E_\text{wind} \sim \frac{1}{M  \ell_x}~.
\fe

\subsubsection{1+1d dipole $\hat \phi$-theory}\label{sec:dipphi-cont1}
We can also study the low-energy limit with fixed lattice couplings \eqref{fixedbetas} and focus on the lightest states, the plane waves, ignoring the momentum and winding states.   (In addition, each state appears an infinite number of times because of the momentum and winding dipole symmetry).  This leads to a self-dual spectrum.

We scale the lattice coupling constants as
\ie
\beta_0 = \frac{\hat \mu_0 a^2}{a_\tau}~,\qquad \beta = \frac{a_\tau}{\hat \mu a^2}~,
\fe
where $\hat \mu_0$ and $\hat \mu$ are fixed continuum coupling constants with mass dimension $1$.\footnote{We will soon see that this theory is scale invariant, so fixing the lattice coupling constants $\beta_0,\beta$ is equivalent to fixing the continuum coupling constants $\hat \mu_0,\hat \mu$.} We also define a new continuum field,
\ie
\hat \phi = \sqrt{a} \,\bar\phi\,,
\fe
with mass dimension $-\frac12$. Then the action \eqref{dipphi-modVill-gaugefixaction} becomes
\ie\label{dipphi-action2}
S = \oint d\tau dx~ \left[ \frac{\hat \mu_0}{2} (\partial_\tau \hat \phi)^2 + \frac{1}{2\hat \mu} (\partial_x^2 \hat \phi)^2 \right]~.
\fe
The field $\hat \phi$ has gauge symmetry
\ie\label{dipphi-gaugesym2}
\hat \phi(\tau,x) \sim \hat \phi(\tau,x) + \hat c~,
\fe
where $\hat c$ is a real constant. In other words, the zero mode of $\hat \phi$ is removed. This means that $\partial_x \hat \phi$ is a well-defined operator. We will refer to this theory as the 1+1d dipole $\hat \phi$-theory.

The dipole $\hat \phi$-theory is self-dual with $\hat \mu_0 \leftrightarrow \frac{\hat \mu}{(2\pi)^2}$. The field $\hat \phi$ and its dual field $\check\phi$  are related by the duality map
\ie\label{dipphi-duality2}
i\hat \mu_0 \partial_\tau \hat \phi = \frac{1}{2\pi} \partial_x^2 \check \phi ~, \qquad \frac{i}{\hat \mu} \partial_x^2 \hat \phi = \frac{1}{2\pi} \partial_\tau \check \phi~.
\fe
$\check \phi$ has the same gauge symmetry \eqref{dipphi-gaugesym2} as $\hat \phi$.

Let us study the fate of various global symmetries of the modified Villain model in this continuum limit.
This theory corresponds to the fourth option in the list following \eqref{noncommutea}.
\begin{itemize}
\item
The $U(1)$ momentum symmetry $\hat\phi\rightarrow \hat\phi+\hat c$ does not act in the dipole $\hat \phi$-theory because it is part of the gauge symmetry \eqref{dipphi-gaugesym2}.

\item
The $U(1)$ winding symmetry does not act in the dipole $\hat \phi$-theory because $\d_x\hat\phi$ is a well-defined operator and the winding charge vanishes,
\ie
\tilde Q = \frac{1}{2\pi}\oint dx~ \partial_x^2\hat\phi = 0.
\fe
This is consistent with the self-duality of $\hat \phi$-theory.

\item
The $\mathbb Z_{L_x}$ momentum dipole symmetry becomes an $\mathbb R$ momentum dipole symmetry, which acts as
\ie
\hat \phi(\tau,x) \rightarrow \hat \phi(\tau,x) + c \frac{x}{\ell_x}~,
\fe
where $c$ is a real constant.  This action seems inconsistent with the periodic boundary conditions in space.  However, because of the gauge symmetry \eqref{dipphi-gaugesym2} it maps $\hat\phi$ between different twisted sectors of the same theory and therefore it is an allowed transformation.

\item
Similarly, the $\mathbb Z_{L_x}$ winding dipole symmetry becomes an $\mathbb R$ winding dipole symmetry. A nontrivial charged configuration with charge $q \in \mathbb R$ is
\ie
\hat \phi(\tau,x) = -q \frac{x}{\ell_x}~.
\fe
Again, because of the gauge symmetry \eqref{dipphi-gaugesym2}, this is a valid configuration.
\end{itemize}

The self-duality exchanges the two $\mathbb R$ dipole symmetries. Moreover, they do not commute with each other, resulting in infinite ground state degeneracy.

See Table \ref{tbl:lat-cont-sym} for the relation between the global symmetries in this continuum limit and on the lattice.

Under the scale transformation, $x\rightarrow \lambda x$, $\tau \rightarrow \lambda^2\tau$, we can scale the field $\hat \phi$ as
\ie
\hat \phi \rightarrow \sqrt{\lambda} \hat \phi~,
\fe
which leaves the action \eqref{dipphi-action2} invariant. It does not change the identification \eqref{dipphi-gaugesym2}. Therefore, the dipole $\hat \phi$-theory is scale invariant.

The energies of the three kinds of states in this limit are
\ie\label{momeli}
E_\text{wave} \sim \frac{1}{\sqrt{\hat \mu_0 \hat \mu}}\frac{1}{\ell_x^2}~,\qquad E_\text{mom} \sim \frac{1}{\hat \mu_0 a \ell_x}~,\qquad E_\text{wind} \sim \frac{1}{\hat \mu a \ell_x}~.
\fe
We see that the plane waves have finite energy, but the momentum and winding states are infinitely heavy. This is consistent with the facts that the $U(1)$ momentum and winding symmetries of the lattice model do not act in this continuum limit, and the dipole $\hat \phi$-theory is scale invariant.

Again, the analogy mentioned in footnote \ref{analowss} is applicable also here.  In fact, here the analogy is even better because there are no finite energy states charged under either the momentum or the winding symmetry.

\subsection{More comments}\label{sec:dipphi-morecomments}

\subsubsection{Local operators in different continuum limits}

Let us compare the local operators that transform under the global symmetries  in the modified Villain lattice model and its three continuum limits.

The modified Villain lattice model \eqref{dipphi-modVill-action} has two dimensionless compact scalar fields, $\phi$ and $\tilde \phi$.
The local operators include $e^{i\phi}, e^{i\Delta_x \phi}, e^{i\tilde\phi}, e^{i\Delta_x\tilde \phi}$, which are charged under the four global symmetries discussed in Section \ref{sec:villainsymmetry}.  $e^{i\phi}$ is charged under the two momentum symmetries, while $e^{i\Delta_x \phi}$ is invariant under the $U(1)$ momentum symmetry, but transforms under the ${\mathbb Z}_{L_x}$ momentum dipole symmetry.  Similarly, $e^{i\tilde\phi}$ is charged under the two winding symmetries, while $e^{i\Delta_x \tilde \phi}$ is invariant under the $U(1)$ winding symmetry, but transforms under the ${\mathbb Z}_{L_x}$ dipole winding symmetry.

In the continuum $\phi$-theory, we have the local operators $e^{i\phi}$, and $\partial_x \phi$, but not $e^{i\partial_x \phi}$ because $\phi$ has mass dimension 0.\footnote{Note that $\partial_x \phi$ is invariant under the $U(1)$ momentum symmetry, but is not invariant under the $\mathbb Z$ momentum dipole symmetry.  It transforms inhomogenously under it.}
On the other hand, in the continuum $\Phi$-theory, we have the local operators $e^{i\partial_x \Phi}$, but not $e^{i\Phi}$ because $\Phi$ has mass dimension $-1$. . Finally, in the continuum $\hat \phi$-theory, we have the local operators $\partial_x \hat \phi$, but not $e^{i\hat \phi}$ because $\hat\phi$ has mass dimension $-\frac12$.

There are other local operators that cannot be written in terms of the fundamental fields in the Lagrangian, but rather in terms of their dual fields.
The dual fields of $\phi$, $\Phi$ and $\hat \phi$ are $\tilde \Phi$, $\tilde \phi$ and $\check \phi$ respectively. They have mass-dimensions $-1$, $0$, and $-\frac12$ respectively, and have the same identifications as $\Phi$, $\phi$, and $\hat \phi$ respectively.
See Section \ref{sec:dipphi-cont3} and \ref{sec:dipphi-cont1} for the duality transformations.

In terms of the dual field, the continuum $\phi$-theory has an additional local operator $e^{i\partial_x \tilde \Phi}$, but not $e^{i\tilde \Phi}$. On the other hand, the continuum $\Phi$-theory has additional local operators $e^{i\tilde \phi}$, and $\partial_x \tilde \phi$, but not $e^{i\partial_x \tilde \phi}$. Finally, the continuum $\hat \phi$-theory has additional local operator $\partial_x \check \phi$, but not $e^{i\check \phi}$.

Importantly, none of the three continuum theories has local operators of the form $e^{i\phi}$ and $e^{i\partial_x\phi}$ at the same time. These operators are summarized in Table \ref{tbl:lat-cont-chargedop}.

\subsubsection{Robustness} \label{sec:dipphi-robustness}

The Lifshitz theory \eqref{intro-dipphi-action1t} and specifically its 1+1d version \eqref{dipphi-action1} is natural in the high energy physics sense.  The absence of potential terms and two-derivative terms for $\phi$ is natural because such terms violate a global symmetry -- the two momentum symmetries.  Furthermore, this continuum theory also has the winding symmetries and therefore, it is natural to set the coefficients of all the winding violating operators to zero.

However, this theory might not be robust. If we start at short distances with a UV theory without some of these symmetries, some level of fine tuning might be needed in order to end up at long distances with this continuum theory.  (See \cite{paper1}, for a review of naturalness vs. robustness in high energy physics and in condensed matter physics.)

Let us study a concrete example.
Consider a lattice action with $U(1)$ variables $e^{i\varphi}$ at each site and the action
\ie\label{dipphi-lat-actionc}
 -{\beta_0}\sum_{\tau\text{-link}} \cos (\Delta_\tau \varphi) -{\beta} \sum_\text{site} \cos(\Delta_x^2 \varphi)~.
\fe
For $\beta_0,\beta \gg 1$, it is similar to the Villain theory \eqref{dipphi-Vill-action}
\ie\label{dipphi-Vill-actionc}
 \frac{\beta_0}{2} \sum_{\tau\text{-link}} (\Delta_\tau \phi - 2\pi n_\tau)^2 + \frac{\beta}{2} \sum_\text{site} (\Delta_x^2 \phi - 2\pi n_{xx})^2 ~.
\fe
These two theories preserves the $U(1)$ momentum and $\mathbb Z_{L_x}$ momentum dipole symmetries, but they do not have the winding symmetries of the modified Villain action \eqref{dipphi-modVill-action}
\ie\label{dipphi-modVill-actionc}
\frac{\beta_0}{2} \sum_{\tau\text{-link}} (\Delta_\tau \phi - 2\pi n_\tau)^2 + \frac{\beta}{2} \sum_\text{site} (\Delta_x^2 \phi - 2\pi n_{xx})^2 + i\sum_{\tau\text{-link}}\tilde \phi (\Delta_\tau n_{xx} - \Delta_x^2 n_\tau)~,
\fe
or the continuum theory.

Following \cite{Gorantla:2021svj}, we can explore the relation between the theory \eqref{dipphi-lat-actionc} (or \eqref{dipphi-Vill-actionc}) and \eqref{dipphi-modVill-actionc} by perturbing the latter by the winding dipole violating operator
\ie\label{winding-violating}
\cos(\Delta_x\tilde \phi)~.
\fe
Starting with \eqref{dipphi-modVill-actionc}, we flow in the IR to the continuum theory.  Then, we deform it by \eqref{winding-violating} to check whether the IR behavior changes.
If this deformation is irrelevant, then the modified Villain theory \eqref{dipphi-modVill-actionc} is robust and the continuum theory captures the long distance behavior of \eqref{dipphi-lat-actionc} and \eqref{dipphi-Vill-actionc}.  If, however, it is relevant, then the modified Villain theory \eqref{dipphi-modVill-actionc} is not robust and the lattice actions \eqref{dipphi-lat-actionc} and \eqref{dipphi-Vill-actionc} do not flow in the IR to the theory described by the continuum model.

In our case, it is easy to see that the deformation \eqref{winding-violating} is relevant.  The operator \eqref{winding-violating} carries dipole winding charge and therefore when it acts on a state with a given dipole charge, it changes this charge.  In particular, it acts nontrivially in the space of ground states.  As a result, if we deform the modified Villain model by this operator, the ground state degeneracy is removed.\footnote{Note that the discussion of the spectrum in Section \ref{spectrum} and, in particular, the ground state degeneracy of states charged under the two dipole symmetries is unlike the situation in \cite{paper1,Gorantla:2021bda}, where the charged states are heavier than the plane waves.  Consequently, the theory discussed in \cite{paper1,Gorantla:2021bda} is robust.}

We could reach the same conclusion if we deformed the action by $\cos(\Delta_x\phi)$ instead of \eqref{winding-violating}.  This would violate the momentum symmetries, but preserve the winding symmetries.

A closely related question is whether the infinite volume limit of our system exhibits spontaneous symmetry breaking (see \cite{Stahl:2021sgi,Lake:2022ico} for a recent discussion). Na\"ively, the answer is yes.  We have an infinite number of ground states carrying various charges under the dipole symmetries and as we take the volume to infinity, the Hilbert space of the theory could split into separate superselection sectors and lead to spontaneous symmetry breaking.  However, because of the singular nature of these states, we do not have a coherent picture of this phenomenon.  It would be nice to understand this issue better.

\subsubsection{Relation to Lifshitz theory}\label{sec:dipphi-Lifshitz}

The theory \eqref{dipphi-action1} can be viewed as the 1+1d version of the Lifshitz theory
\ie
S =  \oint d\tau d^dx~ \left[ \frac{\mu_0}{2} (\partial_\tau \phi)^2 + \frac{1}{2\mu} \left(\sum_i\partial_i^2 \phi\right)^2 \right]~.
\fe
In most of the literature, the scalar field $\phi$ is taken to be noncompact and the theory has a Lifshitz scale symmetry
\ie\label{eq:lif_scale}
\tau\rightarrow \lambda^2\tau~,\quad x^i\rightarrow \lambda x^i~,\quad \phi\rightarrow \lambda^{2-d}\phi~.
\fe
In this section, we have considered various versions of the 1+1d Lifshitz theory with different identifications on $\phi$. Typically, imposing identification on $\phi$ breaks the Lifshitz scale symmetry. For example, the identifications $\phi\sim\phi+2\pi$ in Section \ref{sec:dipphi-cont2} and $\Phi\sim\Phi+ c+2\pi x$ in Section \ref{sec:dipphi-cont3} make the theory incompatible with the Lifshitz scale symmetry.

The situation in the $\hat \phi$-theory in Section \ref{sec:dipphi-cont1} is different.  Here, the identification $\hat\phi\sim \hat\phi+\hat c$ removes the zero mode of the field and it is compatible with the scale symmetry.  Indeed, it describes the low-energy limit of the 1+1d modified Villain lattice model \eqref{dipphi-modVill-action} with fixed coupling.

In 2+1d, the Lifshitz scale transformation does not act on the scalar field $\phi$, so it is natural to consider a compact version of the Lifshitz theory with identification $\phi\sim\phi+2\pi$ \cite{Henley1997, Moessner2001, Vishwanath:2004, Fradkin:2004, Ardonne:2003wa,2005,2018PhRvB..98l5105M,Yuan:2019geh,Lake:2022ico}. Such a theory arises naturally  in the study of quantum dimer models  \cite{Rokhsar1988,Henley1997,Moessner2001} and the dipolar Bose-Hubbard model \cite{Lake:2022ico}. Most of our discussions about 1+1d compact Lifshitz theories, including their global symmetries and infinite ground state degeneracy, are applicable in 2+1d. In particular, the infinite ground state degeneracy due to different winding sectors has been noticed in the quantum dimer model \cite{Rokhsar1988} and its effective description in terms of the compact Lifshitz theory in \cite{Henley1997,Moessner2001}.

Unlike the 1+1d theory, the winding symmetry of the 2+1d theory is actually robust. In 2+1d, the states charged under the winding dipole symmetry are extended in space.  They are not created by point-like operators, but by line operators.  Consequently, the theory is robust under adding operators violating this winding symmetry.  This is similar to the fact that the standard 2+1d $U(1)$ gauge theory is not robust under deformations breaking its magnetic symmetry, which is the famous Polyakov mechanism, while the similar 3+1d theory is robust.

\section{1+1d $U(1)$ dipole gauge theory}\label{sec:dipA}

\subsection{First look at the continuum theory}\label{sec:dipAcontinuum}

We can gauge the momentum global symmetries of the dipole $\phi$-theory by coupling it to the gauge fields $(A_\tau, A_{xx})$ of mass dimensions $1$ and $2$, respectively. The gauge symmetry is
\ie\label{dipA-cont2-gaugesym}
A_\tau \sim A_\tau + \partial_\tau \alpha~, \qquad A_{xx} \sim A_{xx} + \partial_x^2 \alpha~,
\fe
where $\alpha$ is the gauge parameter with mass dimension $0$. The global properties of $\alpha$ are the same as those of $\phi$ in Section \ref{sec:dipphi-cont2}.
The continuum action of the pure gauge theory is
\ie\label{dipA-conti-action}
S = \oint d\tau dx~ \frac{1}{2g^2} E_{xx}^2 ~.
\fe
where
\ie
E_{xx} = \partial_\tau A_{xx} - \partial_x^2 A_\tau
\fe
is the electric field with mass dimension $3$.
Here, $g$ is a fixed continuum coupling of mass dimension 2.
We will refer to this continuum action as the 1+1d $U(1)$ dipole $A$-theory.
Below we will discuss some unusual subtleties of this continuum theory.

On a Euclidean torus, we can consider the following large gauge transformation
\ie
\alpha = \frac{2\pi n_\tau \tau}{\ell_\tau}+\frac{2\pi n_x x}{\ell_x}~,\quad n_\tau, n_x\in\mathbb{Z}~.
\fe
It shifts the gauge fields by
\ie\label{eq:large_gauge}
(A_\tau,A_{xx})\rightarrow \left(A_\tau+\frac{2\pi n_\tau}{\ell_\tau},A_{xx}\right)~.
\fe
Note that the gauge transformation associated with $n_x$ acts trivially on the gauge fields.

The theory has gauge invariant line defects
\ie\label{eq:defect}
\exp\left( in \oint d\tau~ A_\tau(\tau,x) \right)~,
\fe
with the integer $n$ quantized by the large gauge transformation \eqref{eq:large_gauge}.

The Lorentzian signature version of \eqref{eq:defect}
\ie\label{eq:defectL}
\exp\left( in \int_{-\infty}^{+\infty} dt~ A_t(t,x) \right)~,
\fe
represents the world-line of a charged particle at $x$.  Because of the gauge symmetry, this particle cannot move continuously.  Hence, it is a fracton.  (Below, we will discuss it in more detail.)

However, for $n\ne 1$, the particle is not completely immobile.  It can hop from $x$ to $x+{k\ell_x\over n} $ with any integer $k$.  One way to see that is to consider the defect
\ie\label{eq:defectLm}
\exp\left( in \int_{t}^{+\infty} dt'~ A_t(t',x') \right)\mathcal O(t,x,x')\exp\left( in \int_{-\infty}^{t} dt'~ A_t(t',x) \right)~.
\fe
Here, the first and the last factors represent the motion of the particle.  And the operator $\mathcal O(t,x,x')$ acts at time $t$ and moves the particle from $x$ to $x'$.  Gauge invariance restricts the hop to satisfy $x-x'\in {\ell_x\over n}\mathbb{Z}$.  Specifically, the shortest hop is implemented using
\ie
\mathcal O\left(t,x,x+{\ell_x\over n}\right)=\exp\left(-i\sum_{r=1}^{n-1}\int_x^{x+{\ell_x\over n}} dy \int_y^{y+ {r\ell_x\over n}} dy'~ A_{xx}(t,y')\right)~.
\fe
It is easy to check that with this operator, the combination \eqref{eq:defectLm} is gauge invariant. A crucial point is that the operator
$\mathcal O\left(t,x,x+{\ell_x\over n}\right)$ is supported over the whole space.  The motion of the particle from $x$ to $x+{\ell_x\over n}$ takes place by acting on the entire system.  In this sense this is not a local operation.

We also have another observable
\ie\label{dipA-dipdef}
\oint_{\mathcal C} \left[d\tau~ \partial_x A_\tau(\tau,x) + dx~ A_{xx}(\tau,x) \right] ~,
\fe
where $\mathcal C$ is a closed curve in the spacetime.
When $\mathcal C$ is purely space-like, \eqref{dipA-dipdef} simplifies to a gauge invariant operator
\ie\label{eq:operator}
\oint dx\, A_{xx} ~
\fe
at a fixed time.  Both the general observable \eqref{dipA-dipdef} and the special case \eqref{eq:operator} are gauge invariant, including under the large gauge transformation \eqref{eq:large_gauge} and do not need to be exponentiated.  In fact, they have dimension +1 and therefore, it makes no sense to exponentiate them.

Instead, the integrated version of \eqref{dipA-dipdef}
\ie\label{dipA-dipdefecti}
 \oint_{\mathcal C} \left[d\tau~ (A_\tau(\tau,x+x_0) - A_\tau(\tau,x)) + dx~ \int_x^{x+x_0}dx'~ A_{xx}(\tau,x') \right] ~,
\fe
with fixed $x_0$ is dimensionless and can be exponentiated to the defect
\ie\label{dipA-dipdefectid}
 \exp\left(ir\oint_{\mathcal C} \left[d\tau~ (A_\tau(\tau,x+x_0) - A_\tau(\tau,x)) + dx~ \int_x^{x+x_0}dx'~ A_{xx}(\tau,x') \right] \right)~,
\fe
with any real $r$.  In the special case where $r$ is an integer, this can be interpreted as a dipole of fractons \eqref{eq:defect} with opposite charges $\pm r$ separated by $x_0$.  More generally, for real $r$ it is a dipole of fractional charged sources.

We have seen that the fracton defect \eqref{eq:defect} cannot move continuously -- all it can do is to hop as in \eqref{eq:defectLm}.  This is to be contrasted with the dipole defect \eqref{dipA-dipdefectid}, which is mobile.  Below, in Section \ref{sec:dipA-cont}, we will discuss this fact in more detail.

We see that the line defects and the line operators are very different. The  defects \eqref{eq:defect} are exponentials with quantized coefficients, while the observables  \eqref{dipA-dipdef} and their special cases, the operators \eqref{eq:operator}, do not have to be exponentiated. To understand this better, we will regularize the continuum theory using a Villain lattice model.

\subsection{Villain formulation}\label{sec:dipA-modVill}

The Villain version of the continuum $U(1)$ dipole gauge theory is described by the lattice action
\ie\label{dipA-modVill-action}
S = \frac{\Gamma}{2} \sum_{\tau\text{-link}} (\Delta_\tau \mathcal A_{xx}-\Delta_x^2 \mathcal A_\tau - 2\pi n_{\tau xx})^2 = \frac{\Gamma}{2} \sum_{\tau\text{-link}} \mathcal E_{xx}^2~,
\fe
where $\Gamma$ is a coupling constant, $n_{\tau xx}$ is an integer-valued gauge field and $\mathcal A_\tau,\mathcal A_{xx}$ are real-valued gauge fields. Here, the electric field,
\ie\label{calEdef}
\mathcal E_{xx} = \Delta_\tau \mathcal A_{xx} - \Delta_x^2 \mathcal A_\tau - 2\pi n_{\tau xx}~,
\fe
is the only gauge invariant field strength under the gauge symmetry
\ie\label{dipA-modVill-gaugesym}
&\mathcal A_\tau \sim \mathcal A_\tau + \Delta_\tau \alpha + 2\pi k_\tau~,
\\
&\mathcal A_{xx} \sim \mathcal A_{xx} + \Delta_x^2 \alpha + 2\pi k_{xx}~,
\\
&n_{\tau xx} \sim n_{\tau xx} + \Delta_\tau k_{xx} - \Delta_x^2 k_\tau~,
\\
&\alpha\in {\mathbb R}~,
\\
&k_\tau,k_{xx}\in {\mathbb Z}~.
\fe
The gauge parameters have their own gauge symmetries
\ie
&\alpha\sim\alpha+c+\frac{2\pi m\hat x}{L_x}+2\pi k~,
\\
&k_{\tau}\sim k_{\tau} -\Delta_\tau k~,
\\
&k_{xx}\sim k_{xx}-\Delta_{xx}k-m(\delta_{\hat x,0}-\delta_{\hat x,L_x-1})~,
\\
&c\in {\mathbb R}~,\\
&m,k\in {\mathbb Z}~,
\fe
with $c$ and $m$ constants on the lattice.

The gauge configurations have a quantized $\mathbb{Z}$-valued electric flux
\ie\label{eleflui}
\frac{1}{2\pi}\sum_{\tau\text{-link}} \mathcal E_{xx} = - \sum_{\tau\text{-link}} n_{\tau xx}\in \mathbb{Z}~,
\fe
and a quantized $\mathbb{Z}_{L_x}$-valued electric dipole flux
\ie\label{ZLxefi}
-\sum_{\tau\text{-link}} \hat x n_{\tau xx}\text{ mod } L_x~.
\fe
Both \eqref{eleflui} and \eqref{ZLxefi} are gauge invariant.

In contrast to the dipole $\phi$-theory, the $U(1)$ dipole gauge theory has no ``vortices.'' Relatedly, there is no gauge invariant field strength of the integer gauge field $n_{\tau xx}$. So we do not modify the Villain action \eqref{dipA-modVill-action}.

We can add a theta-term to the action \eqref{dipA-modVill-action}:
\ie
\frac{i\theta}{2\pi} \sum_{\tau\text{-link}} \mathcal E_{xx} = - i\theta \sum_{\tau\text{-link}} n_{\tau xx}~,
\fe
where $\theta \sim \theta + 2\pi$.\footnote{We could also add a discrete theta-term associated with the $\mathbb{Z}_{L_x}$-valued dipole flux \eqref{ZLxefi}. However, the dipole flux is not invariant under the {time-like $\mathbb Z_{L_x}$ dipole symmetry} (see Section \ref{sec:dipA-modVill-sym}). Therefore, adding a nontrivial discrete theta-term makes the partition function vanish.}
The full action is
\ie\label{dipA-modVill-fullaction}
S = \frac{\Gamma}{2} \sum_{\tau\text{-link}} \mathcal E_{xx}^2 + \frac{i\theta}{2\pi} \sum_{\tau\text{-link}} \mathcal E_{xx}~.
\fe
Note that we could not add such a $\theta$-term to the continuum action \eqref{dipA-cont2-gaugesym} since the electric field $E_{xx}$ has mass dimension +3.  Below, this will be discussed further.

The Villain model \eqref{dipA-modVill-fullaction} has gauge invariant operators
\ie\label{dipA-modVill-operator}
\exp\left( in\sum_{\text{site: fixed }\hat \tau} \mathcal A_{xx} \right)~,\quad n\in\mathbb{Z}~.
\fe
Unlike the operators \eqref{eq:operator} in the continuum, these lattice operators are gauge invariant only after exponentiation due to the integer $k_{xx}$ symmetry.

The model also has defects that describe fractons
\ie\label{dipA-modVill-fracton}
\exp\left( in \sum_{\tau\text{-link: fixed }\hat x} \mathcal A_\tau \right)~,\quad n\in\mathbb{Z}~.
\fe
These are the lattice counterparts of the continuum defects \eqref{eq:defect}. Moreover, for $\gcd(n,L_x)\ne 1$, the particle can hop by $kL_x/\gcd(n,L_x)$ sites for any integer $k$. For $k=1$, this is captured by the defect
\ie\label{dipA-modVill-fractonM}
&\exp\left( in \sum_{\tau\text{-link: }\hat \tau' \ge \hat \tau} \mathcal A_\tau\left(\hat \tau',\hat x+\frac{L_x}{\gcd(n,L_x)}\right) \right)
\\
&\times \exp\left(-\frac{in}{\gcd(n,L_x)}\sum_{r=1}^{\gcd(n,L_x)-1} ~\sum_{\hat y = \hat x +1}^{\hat x+{L_x\over \gcd(n,L_x)}}~\sum_{\text{site: }\hat y \le \hat y' < \hat y + {rL_x\over \gcd(n,L_x)}}\mathcal A_{xx}(\hat \tau,\hat y')\right)
\\
&\times \exp\left( in \sum_{\tau\text{-link: }\hat \tau' < \hat \tau} \mathcal A_\tau(\hat \tau',\hat x) \right)~.
\fe
This is the lattice counterpart of the Euclidean version of the continuum defect \eqref{eq:defectLm}.  The first and the third line represent the motion of the particle in (Euclidean) time.  And the second line represents an operator moving the particle from $\hat x$ to $\hat x+{L_x\over \gcd(n,L_x)}$.

A dipole can move as long as its separation is fixed. This is described by
\ie\label{dipA-modVill-dipdefect}
\exp\left( in\sum_{\tau x\text{-plaq: }\hat \tau < \hat \tau_0} \Delta_x \mathcal A_\tau(\hat \tau,\hat x_1)  + i n\sum_{\text{site: }\hat x_1 < \hat x \le \hat x_2} \mathcal A_{xx}(\hat \tau_0,\hat x) + in\sum_{\tau x\text{-plaq: }\hat \tau \ge \hat \tau_0} \Delta_x \mathcal A_\tau(\hat \tau,\hat x_2)  \right)~.
\fe
The coefficients of these defects are quantized because of the integer gauge symmetry of $(k_\tau,k_{xx})$ in \eqref{dipA-modVill-gaugesym}.

Given the gauge invariant defects \eqref{dipA-modVill-fracton} and the gauge invariant field strength \eqref{calEdef} we can write additional gauge invariant defects
\ie\label{dipA-modVill-dipdefect-frac}
&\exp\left( in \sum_{\tau\text{-link}}  \Delta_x \mathcal A_\tau(\hat \tau,\hat x+\hat x_0) \right)\exp\left( \frac{in}{\hat x_0} \sum_{\tau\text{-link: }\hat x < \hat x' < \hat x + \hat x_0}  (\hat x' - \hat x)\mathcal E_{ xx}(\hat \tau,\hat x') \right)
\\
&=\exp\left( \frac{in}{\hat x_0} \sum_{\tau\text{-link}} [\mathcal A_\tau(\hat \tau,\hat x + \hat x_0)  - \mathcal A_\tau(\hat \tau,\hat x)] - \frac{2\pi in}{\hat x_0}\sum_{\tau\text{-link: }\hat x < \hat x' < \hat x + \hat x_0}  (\hat x' - \hat x) n_{\tau xx}(\hat \tau,\hat x') \right)~,
\fe
for any $\hat x_0$, where $n\in \mathbb Z$.

These defects can be interpreted as the worldlines of a dipole of fractional charges $\pm {n\over \hat x_0}$ at $\hat x$ and at $\hat x +\hat x_0$.  Surprisingly, these dipoles are mobile as long as their separation is fixed:
\ie\label{movingfd}
&\exp\left( \frac{in}{\hat x_0} \sum_{\tau\text{-link: }\hat \tau < \hat \tau_0} [\mathcal A_\tau(\hat \tau,\hat x_1 + \hat x_0)  - \mathcal A_\tau(\hat \tau,\hat x_1)] - \frac{2\pi in}{\hat x_0} \sum_{\tau\text{-link: }\hat \tau < \hat \tau_0 \atop \hat x_1 < \hat x' < \hat x_1 + \hat x_0}  (\hat x' - \hat x_1) n_{\tau xx}(\hat \tau,\hat x') \right)
\\
&\times \exp\left( \frac{in}{\hat x_0} \sum_{\hat x_1 < \hat x \le \hat x_2} ~~\sum_{\text{site: }\hat x \le \hat x' < \hat x + \hat x_0}\mathcal A_{xx}(\hat \tau_0,\hat x') \right)
\\
&\times \exp\left( \frac{in}{\hat x_0} \sum_{\tau\text{-link: }\hat \tau \ge \hat \tau_0} [\mathcal A_\tau(\hat \tau,\hat x_2 + \hat x_0)  - \mathcal A_\tau(\hat \tau,\hat x_2)] - \frac{2\pi in}{\hat x_0} \sum_{\tau\text{-link: }\hat \tau \ge \hat \tau_0 \atop \hat x_2 < \hat x' < \hat x_2 + \hat x_0}  (\hat x' - \hat x_2) n_{\tau xx}(\hat \tau,\hat x') \right)~.
\fe
This is the lattice version of \eqref{dipA-dipdefectid}.  Unlike the continuum problem, here the real charge $r$ is restricted to be the rational number $n/\hat x_0$.

\subsubsection{Relation to the 2+1d $U(1)$ tensor gauge theory}\label{sec:dipA-compactify}

Similar to the discussion in Section \ref{sec:dipphi-compactify}, here we will relate the 1+1d model \eqref{dipA-modVill-action} to a 2+1d model.

In \cite{Gorantla:2021svj}, the Villain version of the 2+1d $U(1)$ tensor gauge theory of \cite{paper1} was studied:
\ie\label{2+1dU1}
S = \frac{\Gamma}{2} \sum_{\tau\text{-link}} \mathcal E_{xy}^2 + \frac{i\theta}{2\pi} \sum_{\tau\text{-link}} \mathcal E_{xy}~.
\fe
Here  $\mathcal E_{xy} = \Delta_\tau \mathcal A_{xy}-\Delta_x \Delta_y \mathcal A_\tau - 2\pi n_{\tau xy}$ is the gauge-invariant electric field, and $\mathcal A_\tau,\mathcal A_{xy}$ are real-valued gauge fields and $n_{\tau x y}$ is the Villain integer gauge field.
The gauge transformations are
\ie
&\mathcal A_\tau \sim \mathcal A_\tau + \Delta_\tau \alpha + 2\pi k_\tau~,
\\
&\mathcal A_{xy} \sim \mathcal A_{xy} + \Delta_x\Delta_y \alpha + 2\pi k_{xy}~,
\\
&n_{\tau xy} \sim n_{\tau xy} + \Delta_\tau k_{xy} - \Delta_x\Delta_y k_\tau~,
\fe
where $\alpha$ is a real-valued gauge parameter and $k_\tau,k_{xy}$ are integer-valued gauge parameters.
We refer the readers to \cite{Gorantla:2021svj} for more detaila of this 2+1d lattice model.

We will now place this 2+1d model on the slanted torus \eqref{slanted}.
Following an identical discussion in Section \ref{sec:dipphi-compactify}, we find the exact equivalence between the 2+1d model \eqref{2+1dU1} and the 1+1d  model \eqref{dipA-modVill-action} under the identification
\ie
&\mathcal A_{xx}(\hat \tau,\hat x) = \mathcal A_{xy}(\hat \tau,\hat x-1,0)~, \qquad \mathcal A_\tau(\hat \tau,\hat x) = \mathcal A_\tau(\hat \tau,\hat x,0)~,
\\
&n_{\tau xx}(\hat \tau,\hat x) = n_{\tau xy}(\hat \tau,\hat x-1,0)~.
\fe
Due to this equivalence, the analysis in the rest of this subsection follows from the discussion of the 2+1d $U(1)$ tensor gauge theory on a slanted torus in \cite{Rudelius:2020kta}.

\subsubsection{Global symmetry} \label{sec:dipA-modVill-sym}

In the Villain model \eqref{dipA-modVill-fullaction}, the global electric symmetry acts as
\ie\label{eteranU}
\mathcal A_\tau \rightarrow \mathcal A_\tau + \Lambda_\tau ~, \qquad \mathcal A_{xx} \rightarrow \mathcal A_{xx} + \Lambda_{xx}~, \qquad n_{\tau xx} \rightarrow n_{\tau xx} + m_{\tau xx}~,
\fe
where $(\Lambda_\tau,\Lambda_{xx};m_{\tau xx})$ is a flat gauge field, i.e.,
\ie
\Delta_\tau \Lambda_{xx} - \Delta_x^2 \Lambda_\tau - 2\pi m_{\tau xx} = 0~.
\fe
The Noether current is\footnote{Recall our conventions, as discussed in footnote \ref{conventionsf}.}
\ie
&\mathcal J_\tau^{xx} = i\Gamma \mathcal E_{xx} - \frac{\theta}{2\pi}~,
\\
&\Delta_\tau \mathcal J_\tau^{xx} = 0~,\qquad \Delta_x^2 \mathcal J_\tau^{xx} = 0~,
\fe
where the equations in the second line are the conservation equation and the Gauss law respectively.

Using the freedom in $\alpha$ and $(k_\tau,k_{xx})$, we can set
\ie\label{dipA-modVill-elecsym}
&\Lambda_\tau = \frac{c_\tau}{L_\tau} + 2\pi m {\hat x \over L_x} \delta_{\hat \tau,0}~, \qquad 0\le \hat x< L_x~,
\\
&\Lambda_{xx} = \frac{c_{xx}}{L_x}~,
\\
&m_{\tau xx} = - m \left( \delta_{\hat x, 0} - \delta_{\hat x, L_x-1} \right)\delta_{\hat \tau,0}~,
\fe
with $m=0,1,\ldots,L_x-1$, and circle-valued $c_\tau\sim c_\tau +2\pi$ and $c_{xx}\sim c_{xx}+2\pi$. This will ultimately lead to \eqref{timelike-dip-sym}. As we will discuss below, the parameters $c_{xx}$ and  $(c_\tau, m)$ generate space-like and time-like global symmetries, respectively.  In the rest of this sub-subsection, we will discuss the space-like symmetry, and leave the time-like symmetry to the next one.

In terms of a Hilbert space interpretation, the transformation associated with $c_{xx}$ is a standard symmetry transformation, acting on states and operators such as \eqref{dipA-modVill-operator}. Since $c_{xx}$ is circle-valued, it is related to a $U(1)$ space-like symmetry. This is to be contrasted with the $\mathbb{R}$ space-like symmetry in the continuum theory discussed in Section \ref{sec:dipAcontinuum}.
The charge of the symmetry is
\ie\label{dipA-modVill-charge}
Q^{xx}(\hat x) = \mathcal J_\tau^{xx}~.
\fe
Using the Gauss law and the fact that it should be single valued, $Q^{xx}(\hat x)=\bar Q^{xx}$ is an integer constant, independent of $\hat x$.

\subsubsection{Restricted mobility of defects}

How should we interpret the symmetries associated with $c_\tau$ and $m$ in \eqref{dipA-modVill-elecsym}?

The circle-valued parameter $c_\tau$ does not correspond to a standard symmetry.  It does not act on states or operators.  Instead, it acts on defects, such as \eqref{dipA-modVill-fracton}, so it is a $U(1)$ time-like symmetry. The symmetry operator of this $U(1)$ time-like symmetry is the bilocal operator
\ie\label{dipA-modVill-U1timelikesymop}
U_{c_\tau}(\hat\tau;\hat x_1,\hat x_2)=\exp\left(ic_\tau [\Delta_x \mathcal J^{xx}_\tau(\hat \tau,\hat x_2) - \Delta_x \mathcal J^{xx}_\tau(\hat \tau,\hat x_1)]\right)~.
\fe
Because of the Gauss law and the conservation equation it is invariant under deformations of $\hat x_1$, $\hat x_2$ and $\hat \tau$ as long as they do not cross any defect.
In particular, when there is no defect, the $U(1)$ time-like symmetry operator is trivial because of the Gauss law:
\ie\label{Uisatri}
U_{c_\tau}(\hat \tau;\hat x_1,\hat x_2) = \exp \left( ic_\tau \sum_{\tau\text{-link: }\hat x_1 < \hat x \le \hat x_2} \Delta_x^2 \mathcal J^{xx}_\tau(\hat \tau,\hat x) \right) = 1~.
\fe

\begin{figure}[t]
\begin{center}
\raisebox{-1\height}{\includegraphics[scale=0.35]{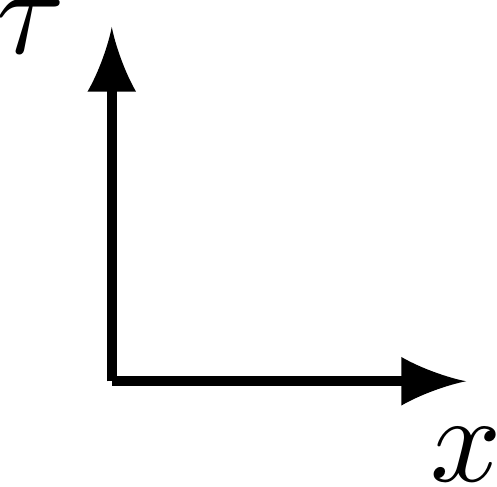}}~~~
\raisebox{-0.5\height}{\includegraphics[scale=0.25]{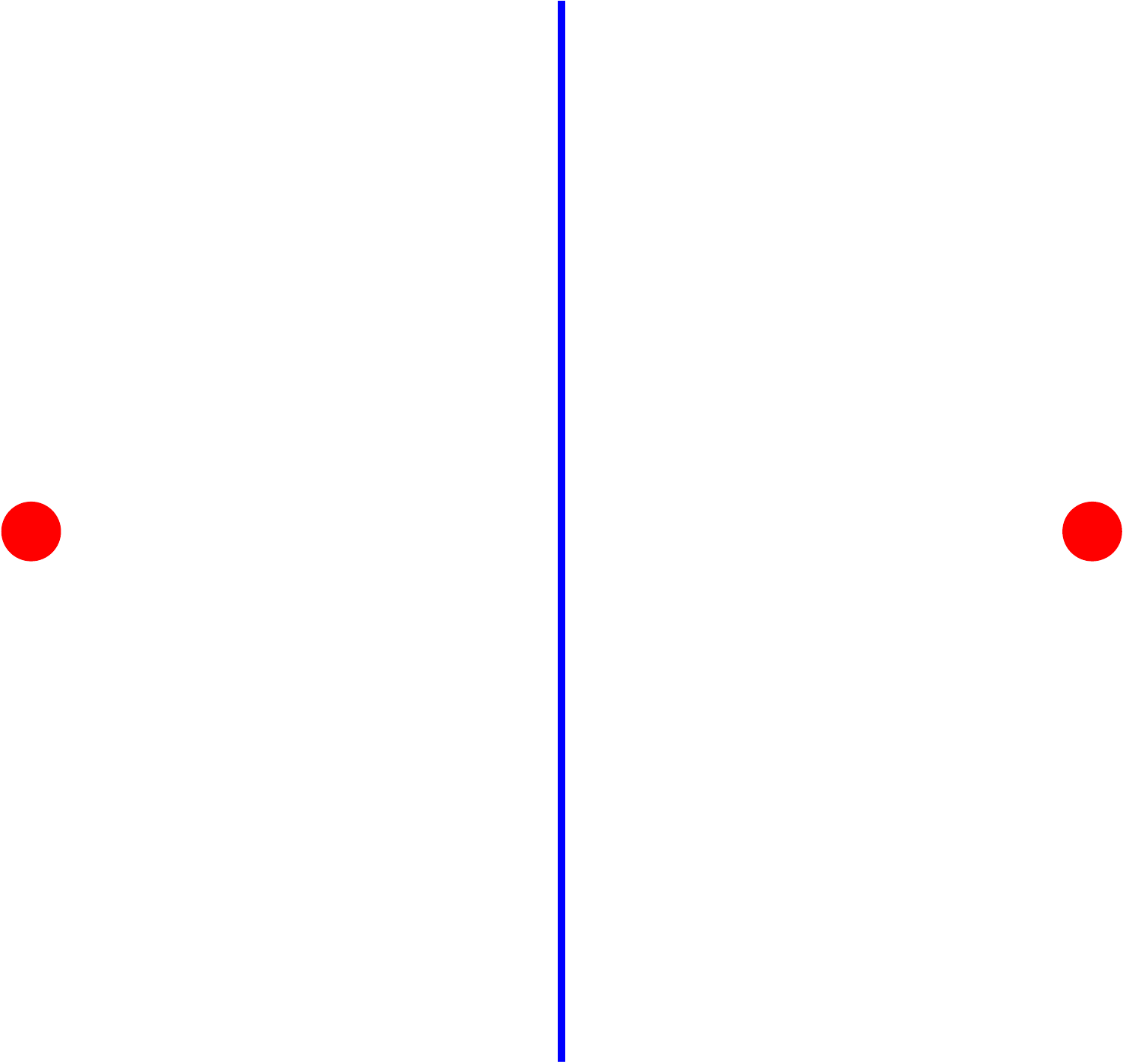}}~~~~~~~$\longrightarrow ~~~e^{i n c_\tau}$~~
\raisebox{-0.5\height}{\includegraphics[scale=0.25]{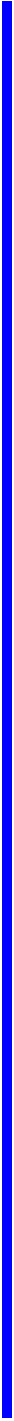}}
\caption{The Euclidean configuration for the action \eqref{dipA-modVill-timelikesymaction} of the $U(1)$ time-like symmetry operator (red dots) with a circle-valued parameter $c_\tau$ on the fracton defect (blue line) of charge $n$.}\label{fig:dipA-timelikesymaction}
\end{center}
\end{figure}

However, it is nontrivial in the presence of defects because the presence of the defect changes the Gauss law.  The action of this time-like symmetry on defects is
\ie\label{dipA-modVill-timelikesymaction}
U_{c_\tau}(\hat \tau;\hat x_1,\hat x_2) \exp\left(in \sum_{\tau\text{-link: fixed }\hat x} \mathcal A_\tau \right) = e^{in c_\tau}\exp\left(in \sum_{\tau\text{-link: fixed }\hat x} \mathcal A_\tau \right)~,\qquad \hat x_1<\hat x<\hat x_2~.
\fe
The action is trivial if $\hat x$ is not in between $\hat x_1$ and $\hat x_2$. See Figure \ref{fig:dipA-timelikesymaction}.

This $U(1)$ time-like symmetry leads to a selection rule stating that amplitudes like
\ie\label{defectco}
\left\langle \prod_i e^{iq_i\sum_{\hat \tau} \mathcal A_\tau (\hat \tau, \hat x_i)}\right\rangle
\fe
are nonzero only when
\ie
\sum_i q_i =0~.
\fe

In infinite volume, we can send one of these defects to infinity and then the sum of the charges $q_i$ of the remaining defects can be nonzero. In that case, this selection rule becomes the statement of total charge conservation.  The discussion here, using the time-like symmetry, gives a precise meaning to this charge conservation in compact space.

The $\mathbb{Z}_{L_x}$-valued parameter $m$ in \eqref{dipA-modVill-elecsym} also does not act on states and operators.  Instead, it acts on the defects \eqref{dipA-modVill-fracton}, \eqref{dipA-modVill-dipdefect} so it generates a $\mathbb{Z}_{L_x}$ time-like symmetry. The symmetry operator is the bilocal operator
\ie\label{dipA-modVill-ZLtimelikesymop}
\mathbf U_m(\hat \tau; \hat x_1,\hat x_2)&=
\exp\left( \frac{2\pi i m}{L_x} \left[\hat x_2 \Delta_x \mathcal J^{xx}_\tau(\hat\tau,\hat x_2 - 1) - \mathcal J^{xx}_\tau(\hat\tau,\hat x_2) \right] \right)
\\
&\quad \times \exp\left(- \frac{2\pi i m}{L_x} \left[\hat x_1 \Delta_x \mathcal J^{xx}_\tau(\hat\tau,\hat x_1 - 1) - \mathcal J^{xx}_\tau(\hat\tau,\hat x_1) \right] \right)~.
\fe

The exponent of $\mathbf U_m(\hat \tau; \hat x_1,\hat x_2)$ in \eqref{dipA-modVill-ZLtimelikesymop} is not well-defined because of the identification $\hat x \sim \hat x + L_x$. In contrast, $\mathbf U_m(\hat \tau; \hat x_1,\hat x_2)$ itself is well-defined because, under this identification, it changes by $\exp[2\pi i m \sum_{\tau\text{-link: }\hat x_1 \le \hat x < \hat x_2} \Delta_x^2 \mathcal J^{xx}_\tau(\hat \tau,\hat x)]$ which is trivial because $\Delta_x^2 \mathcal J^{xx}_\tau$ is an integer even in the presence of defects.
This $\mathbb{Z}_{L_x}$ time-like symmetry becomes the $\mathbb{Z}$ time-like symmetry of the continuum theory in Section \ref{sec:dipAcontinuum}.

Because of the Gauss law and the conservation equation, the operator $\mathbf U_m(\hat \tau; \hat x_1,\hat x_2)$ is invariant under  deformations of $\hat x_1$, $\hat x_2$ and $\hat \tau$ as long as they not cross any defects. In particular, when there is no defect, the $\mathbb Z_{L_x}$ time-like symmetry operator is trivial because of the Gauss law:
\ie
\mathbf U_m(\hat \tau;\hat x_1,\hat x_2) = \exp \left( \frac{2\pi i m}{L_x} \sum_{\tau\text{-link: }\hat x_1 \le \hat x < \hat x_2} \hat x\Delta_x^2 \mathcal J^{xx}_\tau(\hat \tau,\hat x) \right) = 1~.
\fe

However, it is nontrivial in the presence of defects because the presence of the defect changes the Gauss law.  The action of this time-like symmetry on defects is
\ie\label{dipA-modVill-timelikedipsymaction}
\mathbf U_m(\hat \tau;\hat x_1,\hat x_2) \exp\left(in \sum_{\tau\text{-link: fixed }\hat x} \mathcal A_\tau \right) = e^{\frac{2\pi i n m \hat x}{L_x}}\exp\left(in \sum_{\tau\text{-link: fixed }\hat x} \mathcal A_\tau \right)~,\qquad \hat x_1<\hat x<\hat x_2~.
\fe
The action is trivial if $\hat x$ is not in between $\hat x_1$ and $\hat x_2$.

The dipole defects \eqref{dipA-modVill-dipdefect} and \eqref{dipA-modVill-dipdefect-frac} (or \eqref{movingfd}) carry charge $n$ under the $\mathbb Z_{L_x}$ time-like dipole symmetry.  This is obvious in \eqref{dipA-modVill-dipdefect} and in the first line of \eqref{dipA-modVill-dipdefect-frac}.  The second line of \eqref{dipA-modVill-dipdefect-frac} can be interpreted as smearing this dipole charge over the interval $(\hat x,\hat x+\hat x_0)$.

As with the $U(1)$ time-like symmetry, the $\mathbb Z_{L_x}$ time-like symmetry leads to a selection rule on correlation functions of defects.  In particular, \eqref{defectco} is nonzero only when
\ie\label{dipA-ZLxselection}
\sum_i \hat x_i q_i =0\mod L_x~.
\fe

The selection rule \eqref{dipA-ZLxselection}  implies that two fractons of $U(1)$ time-like charge $n$ carry the same $\mathbb{Z}_{L_x}$ time-like symmetry charges only if their positions differ by a multiple of $L_x/{\gcd(n,L_x)}$.
This implies that a single fracton cannot move by itself arbitrarily but it can hop by $kL_x/{\gcd(n,L_x)}$ sites for any integer $k$. Comparing with \eqref{dipA-modVill-fractonM}, this explains the allowed mobility of the fracton defect in terms of global symmetries.

Again, in infinite volume, we can send some of these defects to infinity, and then the sum of the dipoles $\hat x_i q_i$ of the remaining defects can be nonzero.\footnote{Note that when we do that and the remaining defects in the interior of the space have nonzero $U(1)$ charge, their dipole moment depends on the origin of the coordinate.} In that case, this selection rule becomes the statement of total dipole charge conservation.

As for the ordinary charge conservation, our discussion using the time-like symmetry gives us a precise way to formulate the notion of conserved dipole charges in compact space.

\subsubsection{Gauge fixing the integers}\label{sec:dipA-modVill-gaugefix}

Following the same procedure as in \cite{Gorantla:2021svj}, after gauge fixing the integer gauge fields, the action \eqref{dipA-modVill-fullaction} can be written in terms of a new gauge field $(\bar{\mathcal  A}_\tau, \bar{\mathcal  A}_{xx})$ as
\ie\label{dipA-modVill-action-gaugefix}
S = \frac{\Gamma}{2} \sum_{\tau\text{-link}} \mathcal E_{xx}^2 + \frac{i\theta}{2\pi} \sum_{\tau\text{-link}} \mathcal E_{xx}~,
\fe
where $\mathcal E_{xx} = \Delta_\tau \bar{\mathcal A}_{xx} - \Delta_x^2 \bar{\mathcal A}_\tau$ is the electric field. The new gauge field $(\bar{\mathcal A}_\tau, \bar{\mathcal A}_{xx})$ is defined as
\ie\label{dipA-modVill-newfield}
\Delta_\tau \bar{\mathcal A}_{xx} - \Delta_x^2 \bar{\mathcal A}_\tau = \Delta_\tau \mathcal A_{xx} - \Delta_x^2 \mathcal A_\tau - 2\pi n_{\tau xx}~.
\fe
It has the gauge symmetry
\ie\label{eq:U(1)_barA_gauge_symmetry}
&\bar{\mathcal A}_\tau \sim \bar{\mathcal A}_\tau +\Delta_\tau \bar\alpha~,
\\
&\bar{\mathcal A}_{xx} \sim \bar{\mathcal A}_{xx} +\Delta_x^2 \bar\alpha~.
\fe
More generally, $\bar\alpha$ may not be single-valued in which case the above corresponds to a change of trivialization.

Unlike $(\mathcal A_\tau,\mathcal A_{xx})$, the new gauge fields $(\bar{\mathcal A}_\tau,\bar{\mathcal A}_{xx})$ need not be single-valued. Instead, they can have transition functions. Around the $\tau$-cycle, we have
\ie
&\bar{\mathcal A}_\tau(\hat\tau+L_\tau,\hat x)-\bar{\mathcal A}_\tau(\hat\tau,\hat x)=\Delta_\tau\bar\gamma_T(\hat\tau,\hat x)~,
\\
&\bar{\mathcal A}_{xx}(\hat\tau+L_\tau,\hat x)-\bar{\mathcal A}_{xx}(\hat\tau,\hat x)=\Delta_x^2\bar\gamma_T(\hat\tau,\hat x)~.
\fe
Around the $x$-cycle, we have
\ie
&\bar{\mathcal A}_\tau(\hat\tau,\hat x+L_x)-\bar{\mathcal A}_\tau(\hat\tau,\hat x)=\Delta_\tau\bar\gamma_X(\hat\tau,\hat x)~,
\\
&\bar{\mathcal A}_{xx}(\hat\tau,\hat x+L_x)-\bar{\mathcal A}_{xx}(\hat\tau,\hat x)=\Delta_x^2\bar\gamma_X(\hat\tau,\hat x)~.
\fe
These transition functions are subject to the cocycle condition
\ie\label{eq:cocycle_condtion}
\bar\gamma_T(\hat\tau,\hat x+L_x)-\bar\gamma_T(\hat\tau,\hat x)-\bar\gamma_X(\hat\tau+L_\tau,\hat x)+\bar\gamma_X(\hat\tau,\hat x)=2\pi n\hat x + 2\pi p~, \quad n,p\in\mathbb{Z}~.
\fe
They transform under the gauge transformation \eqref{eq:U(1)_barA_gauge_symmetry} as
\ie
&\bar\gamma_T(\hat \tau,\hat x) \sim \bar\gamma_T(\hat \tau,\hat x) + \bar{\alpha}(\hat \tau +L_\tau,\hat x)-\bar{\alpha}(\hat \tau ,\hat x)~,
\\
&\bar\gamma_X(\hat \tau,\hat x) \sim \bar\gamma_X(\hat \tau,\hat x) + \bar{\alpha}(\hat \tau,\hat x+L_x)-\bar{\alpha}(\hat \tau ,\hat x)~.
\fe
In addition, they are also subject to the same identifications \eqref{dipphi-modVill-gaugesym-barphi} as $\bar \phi$. It implies that $p\sim p+L_x$. The cocycle condition \eqref{eq:cocycle_condtion} is invariant under both the gauge transformation and the identifications.

$(\bar{\mathcal A}_\tau,\bar{\mathcal A}_{xx})$ can have nontrivial electric fluxes. For example, the configuration
\ie
\bar{\mathcal A}_\tau(\hat \tau,\hat x) = 0~,\qquad \bar{\mathcal A}_{xx}(\hat \tau, \hat x) = 2\pi n \frac{\hat \tau}{L_\tau L_x}~,
\fe
has a transition function
\ie
\bar \gamma_T(\hat \tau,\hat x) = 2\pi n \frac{\hat x(\hat x-L_x)}{2L_x}~,
\fe
in the $\tau$-direction. It gives rise to a nontrivial $\mathbb{Z}$-valued electric flux:
\ie
\frac{1}{2\pi}\sum_{\tau\text{-link}} \mathcal E_{xx} =\frac{1}{2\pi}\Delta_x\big[\bar\gamma_T(\hat\tau,\hat x+L_x)-\bar\gamma_T(\hat\tau,\hat x)-\bar\gamma_X(\hat\tau+L_\tau,\hat x)+\bar\gamma_X(\hat\tau,\hat x)\big] = n \in \mathbb Z~.
\fe
In terms of the original integer gauge fields, it is $-\sum_{\tau\text{-link}} n_{\tau x x}$ \eqref{eleflui}.

There is also another $\mathbb{Z}_{L_x}$-valued dipole electric flux. Consider the configuration
\ie
\bar{\mathcal A}_\tau(\hat \tau,\hat x) = 2\pi p \frac{\hat x}{L_xL_\tau}~,\qquad \bar{\mathcal A}_{xx}(\hat \tau, \hat x) = 0~.
\fe
It has a transition function
\ie
\bar \gamma_X(\hat \tau,\hat x) = 2\pi p \frac{\hat \tau}{L_\tau}~,
\fe
in the $x$-direction. This configuration carries a nontrivial $\mathbb Z_{L_x}$ dipole flux
\ie
-\frac{1}{2\pi}\big[\bar\gamma_T(\hat\tau, L_x)-\bar\gamma_T(\hat\tau,0)-\bar\gamma_X(\hat\tau+L_\tau,0)+\bar\gamma_X(\hat\tau,0) \big]= p \text{ mod } L_x~.
\fe
In terms of the original integer gauge fields, it is $-\sum_{\tau\text{-link}} \hat x n_{\tau x x}$ mod $L_x$.

\subsubsection{Spectrum}

We will now determine the spectrum of the theory. We will work with a continuous Lorentzian time, denoted by $t$, while keeping the space discrete. We do this by introducing a lattice spacing $a_\tau$ in the $\tau$-direction, taking the limit $a_\tau\rightarrow 0$, while keeping $\Gamma' = \Gamma a_\tau$ fixed, and then Wick rotating from Euclidean time to Lorentzian time. We pick the temporal gauge $\bar{\mathcal A}_0 =0$ and Gauss law tells us that
\ie
\Delta_x^2 \mathcal E_{xx}(t,\hat x)=0~.
\fe
It is solved by
\ie
\mathcal E_{xx}(t,\hat x)=\check{\mathcal E}_x(t) \hat x+\check{\mathcal E}_{xx}(t)~.
\fe
Since $\mathcal E_{xx}$ is single-valued, $\check{\mathcal E}_x$ has to vanish. Up to a time-independent gauge transformation, the solution is
\ie
\bar{\mathcal A}_{xx} =\frac{1}{L_x} f(t)~,
\fe
where $f(t)$ has periodicity $f(t)\sim f(t)+2\pi$.\footnote{This follows from the identification $e^{if} = \exp(i\sum_{\text{site: fixed }\hat \tau} \mathcal A_{xx})$ (see \eqref{dipA-modVill-operator}).}

The effective Lorentzian action is
\ie
S = \oint dt\left[\frac{\Gamma'}{2L_x}  \dot f(t)^2 - \frac{\theta}{2\pi}  \dot f(t)\right]~,
\fe
Let $\Pi$ be the conjugate momentum of $f(t)$. The periodicity of $f(t)$ implies that $\Pi$ is an integer. The Hamiltonian is
\ie
H =\frac{L_x}{2\Gamma'} \left(\Pi+\frac{\theta}{2\pi}\right)^2~.
\fe

This theory is reminiscent of the ordinary $U(1)$ gauge theory in 1+1d.  It has no local degrees of freedom.  All the gauge invariant information is summarized in the holonomy \eqref{dipA-modVill-operator}. And its dynamics is that of a quantum mechanical rotor.

\subsection{Continuum limit}\label{sec:dipA-cont}

\renewcommand{\arraystretch}{1.5}
\begin{table}[t]
\begin{center}
\begin{tabular}{|c|c|c|c|c|}
\hline
&Lattice gauge theory& Dipole $A$-theory & Dipole $\tilde A$-theory
\tabularnewline
\hline
Gauge parameter  & $\phi$ of Section \ref{Villain} & $\phi$ of Section \ref{sec:dipphi-cont2} & $\Phi$ of Section \ref{sec:dipphi-cont3}
\tabularnewline
\hline
Space-like  & \multirow{2}{*}{$U(1)$} & \multirow{2}{*}{$\mathbb R$} & \multirow{2}{*}{$U(1)$}
\tabularnewline
symmetry  & $$ & $$ & $$
\tabularnewline
\hline
Time-like & $U(1)$ & $U(1)$ & --
\tabularnewline
\cline{2-4}
symmetry & $\mathbb Z_{L_x}$ dipole & $\mathbb Z$ dipole & $U(1)$ dipole
\tabularnewline
\hline
\multirow{2}{*}{Fluxes} & $\mathbb{Z}$-valued flux & -- & $\mathbb{Z}$-valued flux
\tabularnewline
\cline{2-4}
& $\mathbb Z_{L_x}$-valued dipole flux
& $\mathbb{Z}$-valued dipole flux
& circle-valued dipole flux
\tabularnewline
\hline
Basic defect & $\exp\left( i\sum_{\tau\text{-link: fixed }\hat x} \mathcal A_\tau \right)$
& $\exp\left(i\oint d\tau~ A_\tau \right)$ & not present
\tabularnewline
\hline
Basic operator& $\exp\left( i\sum_{\text{site: fixed }\hat \tau} \mathcal A_{xx} \right)$
& $\oint dx~A_{xx}$ & $\exp\left(\oint dx~\tilde A_{xx}\right)$
\tabularnewline
\hline
\end{tabular}
\caption{Relation between symmetries and fluxes of the lattice theory of Section \ref{sec:dipA-modVill}, and its continuum limits in Section \ref{sec:dipA-cont}. All these symmetries are electric symmetries. There is another continuum theory, the $\hat A$-theory, whose gauge parameter has the same global properties as $\hat \phi$ of Section \ref{sec:dipphi-cont1}. All of its global symmetries are noncompact. We discuss it briefly in Section \ref{sec:dipA-cont1}.
}\label{tbl:dipA-lat-cont}
\end{center}
\end{table}

Below, we will consider three continuum limits. They have similar Lagrangians, but they are different in various global aspects, such as their global symmetries and fluxes, summarized in Table \ref{tbl:dipA-lat-cont}. The gauge parameters of these continuum gauge theories have the same global properties as the continuum scalar fields in Section \ref{sec:dipphi-cont}. One of these theories reproduces the continuum theory in Section \ref{sec:dipAcontinuum}.
In all these limits, we introduce the spatial and temporal lattice spacings $a,a_\tau$, and take the limit $a,a_\tau \rightarrow 0$ and $L_x,L_\tau \rightarrow \infty$ such that $\ell_x = a L_x$ and $\ell_\tau = a_\tau L_\tau$ are fixed.

\subsubsection{1+1d $U(1)$ dipole $A$-theory}\label{sec:dipA-cont2}
Following Section \ref{sec:dipphi-cont2}, we scale the lattice coupling constant as
\ie
\Gamma = \frac{1}{g^2 a_\tau a^3}~.
\fe
where $g$ is a fixed continuum coupling constant with mass dimension $2$. We define new continuum gauge fields,
\ie\label{dipA-cont1fields}
A_\tau = a_\tau^{-1} \bar{\mathcal A}_\tau~,\qquad A_{xx} = a^{-2} \bar{\mathcal A}_{xx}~,
\fe
with mass dimensions $1$ and $2$ respectively. Recall that $(\bar{\mathcal A}_\tau,\bar{\mathcal A}_{xx})$ are the gauge-fixed versions of the lattice gauge fields $(\mathcal A_\tau,\mathcal A_{xx})$. Then the theory reduces to the continuum theory discussed in Section \ref{sec:dipAcontinuum}, and it reproduces the defects and the operators discussed there. Recall that there is no theta-term in Section \ref{sec:dipAcontinuum}.

\begin{center}
\emph{Defects and operators}
\end{center}

Let us substitute \eqref{dipA-cont1fields} in the defects \eqref{dipA-modVill-fracton} and \eqref{dipA-modVill-dipdefect}, and take their continuum limit to find the defects in this continuum theory.

There are particles that cannot move continuously:
\ie\label{dipA-cont2-fractondefect}
\exp\left( in \oint d\tau~ A_\tau(\tau,x) \right)~,
\fe
where $n$ is quantized by the large gauge transformation $\alpha(\tau,x) = 2\pi \frac{\tau}{\ell_\tau}$. For $n\ne1$, while they cannot move continuously, they can hop from $x$ to $x+\frac{k\ell_x}{n}$ for any integer $k$.

We also have the observables
\ie\label{dipA-cont2-dipdefectd}
\oint_{\mathcal C} \left[d\tau~ \partial_x A_\tau(\tau,x) + dx~ A_{xx}(\tau,x) \right] ~,
\fe
where $\mathcal C$ is a closed curve in the spacetime. Note that \eqref{dipA-cont2-dipdefectd} does not have to be exponentiated.\footnote{Indeed, after using the limit \eqref{dipA-cont1fields} (and dropping the bar) in \eqref{dipA-modVill-dipdefect} with fixed lattice points $\hat x_1$ and $\hat x_2$, the coefficient in the exponent vanishes in the limit $a\rightarrow 0$, so we can expand it to find the continuum observable \eqref{dipA-cont2-dipdefectd}.} The integrated version of \eqref{dipA-cont2-dipdefectd} can be exponentiated
\ie\label{dipA-cont2-dipdefect}
\exp\left(ir \oint_{\mathcal C} \left[d\tau~ (A_\tau(\tau,x+x_0) - A_\tau(\tau,x)) + dx~ \int_x^{x+x_0}dx'~ A_{xx}(\tau,x') \right]\right) ~,
\fe
with any real $r$.  When $r$ is quantized, \eqref{dipA-cont2-dipdefect} represents a defect of a dipole of probe particles with charges $\pm r$ separated by fixed amount, $x_0$.  It is the continuum limit of \eqref{dipA-modVill-dipdefect-frac} with $x_0 = a\hat x_0$, and $r = na/x_0$ fixed, where $n,\hat x_0$ are scaled appropriately.

Finally, when $\mathcal C$ is purely space-like, \eqref{dipA-cont2-dipdefectd} is a gauge invariant operator.

\begin{center}
\emph{Global symmetry}
\end{center}

It is interesting that unlike the lattice theory, in this continuum theory, the quantization of the line defects and the line operators are very different. Below, we will discuss the space-like and time-like symmetries that act on them.

There is an $\mathbb R$ space-like symmetry $A_{xx}(\tau,x) \rightarrow A_{xx}(\tau,x) + c_{xx}$. Its charge is found by taking the continuum limit of the charge \eqref{dipA-modVill-charge} on the lattice:
\ie
\frac{i}{g^2} E_{xx}~.
\fe
It is independent of $x$. The scaling to the continuum limit turned the quantized $U(1) $ charge on the lattice to an $\mathbb R$ charge. This is the analog of the one-form global symmetry charge of the standard $U(1)$ 1+1d gauge theory. But unlike that case, here, this charge is not quantized.

There is a $U(1)$ time-like symmetry $A_\tau(\tau,x) \rightarrow A_\tau(\tau,x) + \frac{c_\tau}{\ell_\tau}$, where $c_\tau \sim c_\tau + 2\pi$. Its symmetry operator is the continuum version of the operator \eqref{dipA-modVill-U1timelikesymop} on the lattice:
\ie\label{Uonetls}
U_{c_\tau}(\tau;x_1,x_2)=\exp\left(-\frac{c_\tau}{g^2} [\partial_x E_{xx}(\tau,x_2) - \partial_x E_{xx}(\tau,x_1)]\right)~.
\fe
This is the analog of the $U(1)$ time-like symmetry of the 1+1d $U(1)$ gauge theory.
Similar to its lattice counterpart \eqref{dipA-modVill-U1timelikesymop}, it is invariant under deformations of $ x_1$, $ x_2$ and $\tau$ as long as they do not cross any defect, because of the Gauss law and the conservation equation. In particular, when there is no defect, the $U(1)$ time-like symmetry operator is trivial.

More explicitly, in the presence of the defect \eqref{dipA-cont2-fractondefect},
\ie
\exp\left( in \oint d\tau~ A_\tau(\tau,x_0) \right)~,
\fe
the equation of motion of $A_\tau$ (i.e., Gauss law) leads to
\ie
{1\over g^2}\partial_x^2 E_{xx}=-in \delta(x-x_0)
\fe
and therefore, for $x_1<x_0<x_2$, the value of \eqref{Uonetls} is $e^{inc_\tau}$.

There is also a $\mathbb Z$ dipole time-like symmetry $A_\tau \rightarrow A_\tau + 2\pi m\frac{x}{\ell_x\ell_\tau}$, where $m$ is an integer. Its symmetry operator is found by taking the continuum limit of the operator \eqref{dipA-modVill-ZLtimelikesymop} on the lattice:
\ie
\mathbf U_m(\tau;x_1,x_2) &= \exp \left(-\frac{2\pi m }{g^2\ell_x} [x_2\partial_x E_{xx}(\tau,x_2) - E_{xx}(\tau,x_2)] \right)
\\
&\quad \times \exp \left(\frac{2\pi m }{g^2\ell_x} [x_1\partial_x E_{xx}(\tau,x_1) - E_{xx}(\tau,x_1)] \right)~.
\fe
It is well-defined under $x \rightarrow x+\ell_x$ because it is shifted by $\exp \left( -\frac{2\pi m}{g^2} \int_{x_1}^{x_2}dx~ \partial_x^2 E_{xx}(\tau,x) \right)$, which is trivial because $\frac{i}{g^2}  \int_{x_1}^{x_2}dx~ \partial_x^2 E_{xx}(\tau,x)$ is an integer even in the presence of defects.
Similar to its lattice counterpart \eqref{dipA-modVill-ZLtimelikesymop}, it is invariant under deformations of $x_1$, $x_2$ and $ \tau$ as long as they do not cross any defect, because of the Gauss law and the conservation equation. In particular, when there is no defect, the $U(1)$ time-like symmetry operator is trivial.

This symmetry was ${\mathbb Z}_{L_x}$ on the lattice, and became $\mathbb Z$ in the continuum limit.  Note that this symmetry does not have an analog in the ordinary 1+1d $U(1)$ gauge theory.

\begin{center}
\emph{Fluxes}
\end{center}

In this continuum limit, there are no configurations with nontrivial electric flux and therefore, there is no $\theta$-term in the action.  However, there are configurations with nontrivial
dipole flux:
\ie
&A_\tau(\tau,x) = 2\pi p \frac{x}{\ell_x\ell_\tau}~,\qquad &&A_{xx}(\tau,x) = 0~,
\\
&\gamma_X(\tau,x) = 2\pi p \frac{\tau}{\ell_\tau}~,\qquad &&\gamma_T(\tau,x) = 0~, \qquad p\in \mathbb Z~.
\fe
Unlike on the lattice, the dipole flux in this continuum limit becomes $\mathbb{Z}$-valued:
\ie
-\frac{1}{2\pi}\left[ \gamma_T(0,\ell_x) - \gamma_T(0,0) - \gamma_X(\ell_\tau,0) + \gamma_X(0,0)\right] = p\in\mathbb{Z}~.
\fe

\begin{center}
\emph{Spectrum}
\end{center}

The spectrum consists of states charged under the $\mathbb R$ global symmetry. The energy of the state carrying charge $q\in \mathbb R$ is
\ie
E \sim \frac{g^2 \ell_x}{2}q^2~,
\fe
which is finite in the continuum limit $a\rightarrow 0$. Since $q$ is not quantized, the spectrum is continuous.

\subsubsection{1+1d $U(1)$ dipole $\tilde A$-theory}\label{sec:dipA-cont3}
Following Section \ref{sec:dipphi-cont3}, we scale the lattice coupling constants as
\ie
\Gamma = \frac{1}{\tilde g^2 a_\tau a}~,
\fe
where $\tilde g$ is a fixed continuum coupling constant with mass dimension $1$. We define new continuum gauge fields,
\ie\label{dipA-cont2fields}
\tilde A_\tau = a_\tau^{-1} a \bar{\mathcal A}_\tau~,\qquad \tilde A_{xx} = a^{-1} \bar{\mathcal A}_{xx}~,
\fe
with mass dimensions $0$ and $1$ respectively. Then the action \eqref{dipA-modVill-action-gaugefix} becomes
\ie
S = \oint d\tau dx~ \left[ \frac{1}{2\tilde g^2} \tilde E_{xx}^2 + \frac{i\theta}{2\pi} \tilde E_{xx} \right]~.
\fe
where $\tilde E_{xx} = \partial_\tau \tilde A_{xx} - \partial_x^2 \tilde A_\tau$ is the electric field with mass dimension $2$.  The gauge symmetry is
\ie
\tilde A_\tau \sim \tilde A_\tau + \partial_\tau \tilde \alpha~, \qquad \tilde A_{xx} \sim \tilde A_{xx} + \partial_x^2 \tilde \alpha~,
\fe
where $\tilde \alpha$ is the gauge parameter with mass dimension $-1$, which has its own gauge symmetry $\tilde \alpha \sim \tilde \alpha + \tilde c + 2\pi x$, where $\tilde c$ is a real constant. The global properties of $\tilde \alpha$ are the same as those of $\Phi$ of Section \ref{sec:dipphi-cont3}.

\begin{center}
\emph{Defects and operators}
\end{center}

Let us substitute \eqref{dipA-cont2fields} in the defects \eqref{dipA-modVill-fracton} and \eqref{dipA-modVill-dipdefect}, and take their continuum limit to find the defects in this continuum theory.

There are no fracton defects because, after substituting \eqref{dipA-cont2fields} in the defect \eqref{dipA-modVill-fracton}, the coefficient diverges in the limit $a\rightarrow 0$, unless $n=0$. We can also see this in the continuum: the would-be defect ``$\oint d\tau~\tilde A_\tau$'' is not invariant under the large gauge transformation $\tilde \alpha = \frac{\tilde c \tau}{\ell_\tau}$, where $\tilde c$ is a real constant. Related to that, $\oint d\tau~\tilde A_\tau$ cannot be exponentiated because it is dimensionful.

However, there are mobile dipole defects:
\ie\label{dipA-cont3-defect}
\exp\left( i n \oint_{\mathcal C} \left[ d\tau~ \partial_x \tilde A_\tau(\tau,x) + dx~ \tilde A_{xx}(\tau,x) \right] \right)~,
\fe
where $n$ is quantized.\footnote{When $\mathcal C$ is space-like, this can be seen by the large gauge transformation $\tilde \alpha(\tau,x) = 2\pi \frac{x(x-\ell_x)}{2\ell_x}$. Such a gauge transformation has its own transition functions consistent with its periodicities, $\tilde \alpha \sim \tilde \alpha + \tilde c + 2\pi x$. \label{globlsymgau}} When $\mathcal C$ is purely space-like, it is a gauge invariant operator with a quantized coefficient, whose lattice counterpart is \eqref{dipA-modVill-operator}.

 There are also gauge invariant mobile dipole defects derived from \eqref{dipA-modVill-dipdefect-frac}
\ie\label{dipA-cont3-defect2}
\exp\left( \frac{in}{x_0} \oint_{\mathcal C} \left[ d\tau~ (\tilde A_\tau(\tau,x+x_0) - \tilde A_\tau(\tau,x)) + dx \int_x^{x+x_0}dx'~\tilde A_{xx}(\tau,x') \right] \right)~,
\fe
where $x_0$ is the (fixed) separation of the dipole, and $n$ is quantized.

\begin{center}
\emph{Global symmetry}
\end{center}

There is a $U(1)$ global symmetry with charge
\ie
Q^{xx} = \frac{i}{\tilde g^2} \tilde E_{xx} - \frac{\theta}{2\pi}~.
\fe
It acts on the gauge fields as
\ie
\tilde A_{xx} \rightarrow \tilde A_{xx} + \frac{\tilde c_{xx}}{\ell_x}~,
\fe
where $\tilde c_{xx}$ is circle-valued, i.e., $\tilde c_{xx} \sim \tilde c_{xx} + 2\pi$. The charged operator is \eqref{dipA-cont3-defect} with $\mathcal C$ purely space-like, and its charge is $n$, which is quantized due to the large gauge transformation in footnote \ref{globlsymgau}.  Unlike the continuum theory of Section \ref{sec:dipA-cont2}, here the space-like $U(1)$ global symmetry of the lattice theory remains $U(1)$ in the continuum.

The $U(1)$ time-like symmetry of the lattice theory is absent in this continuum limit because the shift
\ie
\tilde A_\tau(\tau,x) \rightarrow \tilde A_\tau(\tau,x) + \frac{\tilde c_\tau}{\ell_\tau}~,
\fe
is a gauge transformation with gauge parameter $\tilde \alpha = \tilde c_\tau \frac{\tau}{\ell_\tau}$ for any $\tilde c_\tau \in \mathbb R$. So the would-be time-like symmetry operator $-\frac{1}{\tilde g^2} [\partial_x \tilde E_{xx}(\tau,x_2)-\partial_x \tilde E_{xx}(\tau,x_1)]$ is trivial.
This is consistent with the fact that there are no fracton defects.

Finally, the ${\mathbb Z}_{L_x}$ dipole time-like symmetry \eqref{dipA-modVill-ZLtimelikesymop} of the lattice theory becomes a $U(1)$ dipole time-like symmetry with symmetry operator
\ie
\mathbf U_{\rho}(\tau;x_1,x_2) &= \exp\left(-\frac{\rho}{\tilde g^2} [x_2 \partial_x \tilde E_{xx}(\tau,x_2) - \tilde E_{xx}(\tau,x_2)]\right)
\\
&\quad \times \exp\left(\frac{\rho}{\tilde g^2} [x_1 \partial_x \tilde E_{xx}(\tau,x_1) - \tilde E_{xx}(\tau,x_1)]\right)~,
\fe
where $\rho$ is a real parameter with $\rho \sim \rho + 2\pi$.
The exponent is well-defined under $x \rightarrow x+\ell_x$ because it is shifted by $-\frac{1}{\tilde g^2} [\partial_x \tilde E_{xx}(\tau,x_2)-\partial_x \tilde E_{xx}(\tau,x_1)]$, which is trivial.
Similar to its lattice counterpart \eqref{dipA-modVill-ZLtimelikesymop}, it is invariant under deformations of $x_1,x_2$, and $\tau$ as long as they do not cross any defects, because of Gauss law and the conservation equation.
In particular, it is trivial in the absence of any defect insertions.

It acts on the gauge fields, up to gauge transformations, as
\ie
\tilde A_\tau(\tau,x) \rightarrow \tilde A_\tau(\tau,x) + \frac{\rho x}{\ell_\tau}~.
\fe
The charged defects are \eqref{dipA-cont3-defect} and \eqref{dipA-cont3-defect2} with $\mathcal C$ that wraps around the $\tau$-direction once.  Both have charge $n$.

\begin{center}
\emph{Fluxes}
\end{center}

There are configurations, such as
\ie
&\tilde A_\tau(\tau,x) = 0~,\qquad && \tilde A_{xx}(\tau,x) = 2\pi n \frac{\tau}{\ell_x \ell_\tau}~,
\\
&\tilde \gamma_X(\tau,x) = 0~,\qquad && \tilde \gamma_T(\tau,x) = 2\pi n \frac{x(x-\ell_x)}{2\ell_x}~,\qquad k\in\mathbb Z~,
\fe
that realize a nontrivial $\mathbb{Z}$-valued electric flux
\ie
\frac{1}{2\pi}\oint d\tau dx~ \tilde E_{xx} = n~.
\fe
This flux allows a nontrivial $\theta$-term.

There are also configurations that realize a nontrivial dipole flux, such as
\ie
&\tilde A_\tau(\tau,x) = \vartheta x\frac{1}{\ell_\tau} ~,\qquad &&\tilde A_{xx}(\tau,x) = 0~,
\\
&\tilde \gamma_X(\tau,x) = \vartheta \ell_x\frac{\tau}{\ell_\tau}~,\qquad &&\tilde \gamma_T(\tau,x) = 0~.
\fe
Importantly, $\vartheta$ and $\vartheta+2\pi$ are related by a change of trivialization with $\tilde\alpha = 2\pi x\frac{\tau}{\ell_\tau}$ and the identification $\tilde\gamma_T(\tau,x)\sim \tilde\gamma_T(\tau,x)+2\pi x$.
Thus, they should be identified and  the parameter $\vartheta$ is circle-valued, i.e., $\vartheta\sim \vartheta+2\pi$.
The dipole flux of this configuration is
\ie
-\frac{1}{\ell_x}\left[ \tilde \gamma_T(0,\ell_x) - \tilde \gamma_T(0,0) - \tilde \gamma_X(\ell_\tau,0) + \tilde \gamma_X(0,0)\right]  = \vartheta\text{ mod } 2\pi~.
\fe
Unlike on the lattice, the dipole flux is circle-valued in this continuum limit.

\begin{center}
\emph{Spectrum}
\end{center}

The spectrum consists of states charged under the $U(1)$ global symmetry. The energy of the state with charge $n\in\mathbb Z$ is
\ie
E \sim \frac{\tilde g^2 \ell_x}{2}\left(n+\frac{\theta}{2\pi}\right)^2~,
\fe
which is finite in the continuum limit $a\rightarrow 0$. Since $n$ is quantized, the spectrum is discrete.

\subsubsection{1+1d dipole $\hat A$-theory}\label{sec:dipA-cont1}

Following Section \ref{sec:dipphi-cont1}, we scale the lattice coupling constant as
\ie
\Gamma = \frac{1}{\hat g^2 a_\tau a^2}~,
\fe
where $\hat g$ is a fixed continuum coupling constant with mass dimension $\frac32$. We define new continuum gauge fields, $\hat A_\tau = a_\tau^{-1} a^{\frac12} \bar{\mathcal A}_\tau$, and $\hat A_{xx} = a^{-\frac32} \bar{\mathcal A}_{xx}$, with mass dimensions $\frac12$ and $\frac32$, respectively. Then the action \eqref{dipA-modVill-action-gaugefix} becomes
\ie
S = \oint d\tau dx~ \frac{1}{2\hat g^2} \hat E_{xx}^2 ~,
\fe
where $\hat E_{xx} = \partial_\tau \hat A_{xx} - \partial_x^2 \hat A_\tau$ is the electric field with mass dimension $\frac52$. There is no $\theta$-term in this limit. The gauge symmetry is
\ie
\hat A_\tau \sim \hat A_\tau + \partial_\tau \hat\alpha~, \qquad \hat A_{xx} \sim \hat A_{xx} + \partial_x^2 \hat\alpha~,
\fe
where $\hat \alpha$ is the gauge parameter with mass dimension $-\frac12$. The global properties of $\hat \alpha$ are the same as those of $\hat \phi$ of Section \ref{sec:dipphi-cont1}.

There are no fracton defects but there are mobile dipole defects and line operators with real coefficients. There is no $U(1)$ time-like symmetry, whereas the electric symmetry and dipole time-like symmetry are noncompact.

\subsection{More comments}

We can study another continuum limit, in which we take the gauge coupling $g$ in the $A$-theory to zero.  We take $g=\epsilon g'\to 0$ with fixed $g'$.  This can be absorbed by rescaling $A_\tau = \epsilon A'_\tau$ and $A_{xx}= \epsilon A'_{xx}$.  Therefore, we should also take the gauge parameter $\alpha'= {1\over \epsilon} \alpha$.  This has the effect of decompactifying the underlying gauge group from $U(1)$ to $\mathbb R$.  Correspondingly, there are no identifications in the space of gauge parameters $\alpha'$.  In this case, the fractons are not quantized.  In fact, the observable $\oint d\tau~A'_\tau$ does not need to be exponentiated to a defect.  Similarly, the operators $\oint dx~A'_{xx}$ do not need to be exponentiated and all the global symmetries are noncompact.

In ordinary classical gauge theory with gauge algebra $\mathfrak u(1)$, the gauge group can be $U(1)$ or $\mathbb R$.  Here, we see that there are more options.  All of them arise from the same underlying lattice theory with $U(1)$ gauge symmetry (or as in the Villain formulation, $\mathbb R$ with another $\mathbb Z$ gauge field), but arise in different continuum limits.

Just as the ordinary $U(1) $ and $\mathbb R$ gauge theories differ in their fluxes, operators, defects, and global symmetries, the same is true in the various different continuum theories here.

We have not discussed the higher dimensional versions of this theory.  One difference from the 1+1d case we discussed here is that in order to preserve the magnetic symmetries, the lattice Villain model should be modified.  We expect that the subtleties we discussed here will still be present in the higher dimensional theory.

\section{1+1d $\mathbb Z_N$ dipole gauge theory}\label{sec:dipZN}

In this section, we will study the $\mathbb Z_N$ lattice dipole gauge theory, and its BF version. Surprisingly, while the $U(1)$ dipole gauge theory has immobile fracton defects, a particle in the $\mathbb{Z}_N$ theory in noncompact space can hop by $N$ sites on its own, and is, therefore, not fully immobile. As we will see, in a lattice with $L_x$ sites with periodic boundary conditions, the particle can hop by even smaller steps -- steps of $\gcd(N,L_x)$ sites.

\subsection{1+1d $\mathbb Z_N$ lattice dipole gauge theory}

The $\mathbb Z_N$ lattice dipole gauge theory is defined by the action
\ie\label{dipZN-action}
S = -\Gamma \sum_{\tau\text{-link}} \cos\left[ \frac{2\pi}{N} (\Delta_\tau m_{xx} - \Delta_x^2 m_\tau) \right]~,
\fe
where $\Gamma$ is the gauge coupling constant.
The integer fields $m_\tau$ and $m_{xx}$ are placed on the $\tau$-links and the sites, respectively.
The gauge symmetry is
\ie\label{dipZN-gaugesym}
m_\tau \sim m_\tau + \Delta_\tau k + N k_\tau~,\qquad m_{xx} \sim m_{xx} + \Delta_x^2 k + N k_{xx}~,
\fe
where $k,k_\tau,k_{xx}$ are integer gauge parameters. It has an electric global symmetry that shifts
\ie\label{ZNelecg}
m_\tau \rightarrow m_\tau + p_\tau~,\qquad m_{xx} \rightarrow m_{xx} + p_{xx}~,
\fe
where $(p_\tau,p_{xx})$ is a flat $\mathbb Z_N$ gauge field, i.e.,
\ie
\Delta_\tau p_{xx} - \Delta_x^2 p_\tau = 0 \mod N~.
\fe

Using the gauge freedom of $k$, we can set $p_\tau = 0$ at $\hat \tau \ne 0$. The flatness condition then implies that
\ie
&\Delta_\tau p_{xx}(\hat \tau,\hat x) = 0 \mod N~,
\\
&\Delta_x^2 p_\tau(0,\hat x) = 0 \mod N~.
\fe
Using the residual (time-independent) gauge freedom in $k$, we can set
\ie\label{eq:electric_ZN}
&p_\tau(\hat \tau,\hat x) = \left( \bar p_\tau + \bar r_\tau \frac{N}{\gcd(N,L_x)} \hat x \right) \delta_{\hat \tau,0}~,\qquad 0\le \hat x < L_x~,
\\
&p_{xx}(\hat \tau,\hat x) = (\bar p_{xx} + \bar r_{xx})\delta_{\hat x,0} - \bar r_{xx}\delta_{\hat x,L_x-1}~,
\fe
where $\bar p_\tau,\bar p_{xx}$ are integers modulo $N$, whereas $\bar r_\tau,\bar r_{xx}$ are integers modulo $\gcd(N,L_x)$.

Similar to the electric global symmetries in the dipole $U(1)$ gauge theory in the previous section, the parameters $\bar p_{xx}$ and $\bar r_{xx}$ are associated with  $\mathbb Z_N$ and $\mathbb Z_{\gcd(N,L_x)}$   space-like global symmetries, respectively. On the other hand, the parameters $\bar p_\tau$ and $\bar r_\tau$ correspond to $\mathbb Z_N$ and $\mathbb Z_{\gcd(N,L_x)}$ time-like symmetries, respectively.

\subsection{An integer BF lattice model}\label{sec:dipZN-modVill}

For $\Gamma \gg 1$, the partition function is dominated by configurations satisfying $\Delta_\tau m_{xx} - \Delta_x^2 m_\tau = 0 \mod N$.  Therefore, we can replace the action \eqref{dipZN-action} by  the BF-type action
\ie\label{dipZN-intBFaction}
S = \frac{2\pi i}{N} \sum_{\tau\text{-link}} \tilde m (\Delta_\tau m_{xx} - \Delta_x^2 m_\tau)~,
\fe
where $\tilde m$ is an integer Lagrange multiplier field. In addition to \eqref{dipZN-gaugesym}, there is another gauge symmetry making $\tilde m$ $\mathbb{Z}_N$-valued
\ie\label{dipZN-periodicity}
\tilde m \sim \tilde m + N\tilde k~,
\fe
where $\tilde k$ is an integer gauge parameter.

 The BF-type action \eqref{dipZN-intBFaction} is similar to the topological $\mathbb Z_N$ ordinary lattice gauge theory action in \cite{Dijkgraaf:1989pz,Kapustin:2014gua}.  Following steps similar to those in Appendix C.2 of \cite{Gorantla:2021svj}, the action \eqref{dipZN-action} and the effective action \eqref{dipZN-intBFaction} can be related to a number of other actions of the Villain form and of a modified Villain form.

There is a related lattice spin model given by the action
\ie\label{dipZN-clock-action}
S = -\tilde \Gamma_0 \sum_{\tau\text{-link}} \cos\left( \frac{2\pi}{N}\Delta_\tau \tilde m \right) - \tilde \Gamma \sum_{\text{site}} \cos\left( \frac{2\pi}{N}\Delta_x^2 \tilde m  \right)~,
\fe
where $\tilde m$ is an integer field at each site with identification \eqref{dipZN-periodicity}, and $\tilde \Gamma_0,\tilde \Gamma$ are coupling constants. It is natural to  refer to it as the dipole $\mathbb Z_N$ clock model. For $\tilde \Gamma_0,\tilde \Gamma \gg 1$, the partition function is dominated by configurations satisfying $\Delta_\tau \tilde m = \Delta_x^2 \tilde m = 0 \mod N$, so we can replace the action \eqref{dipZN-clock-action} by  the BF action  \eqref{dipZN-intBFaction}. Now, $m_\tau$ and $m_{xx}$ are interpreted as integer Lagrange multiplier fields.

\subsubsection{Relation to 2+1d $\mathbb Z_N$ tensor gauge theory}\label{sec:dipZN-compactify}
As in Sections \ref{sec:dipphi-compactify} and \ref{sec:dipA-compactify}, there is an exact equivalence between the action \eqref{dipZN-intBFaction} and the integer $BF$-action of the 2+1d $\mathbb Z_N$ tensor gauge theory \cite{Gorantla:2021svj}
\ie
S = \frac{2\pi i}{N} \sum_\text{cube} \tilde m^{xy} (\Delta_\tau m_{xy} - \Delta_x \Delta_y m_\tau)~,
\fe
on a slanted spatial torus with identifications \eqref{slanted}.\footnote{There is a similar relation between the original 1+1d model \eqref{dipZN-action}, and the 2+1d $\mathbb Z_N$ tensor gauge theory with action
\ie
S = -\Gamma \sum_\text{cube} \cos \left[ \frac{2\pi}{N} (\Delta_\tau m_{xy} - \Delta_x \Delta_y m_\tau) \right]~,
\fe
on a slanted spatial torus with identifications \eqref{slanted}.}
Here $\tilde m^{xy}, m_{xy}, m_\tau$ are  the  $\mathbb{Z}_N$-valued fields of the 2+1d model.
The equivalence  follows from
\ie
\Delta_y m_\tau(\hat x, \hat y) = \Delta_x m_\tau(\hat x, \hat y) \implies \Delta_x \Delta_y m_\tau(\hat x,\hat y) = \Delta_x^2 m_\tau(\hat x+1,\hat y)~.
\fe
The remaining fields are related as $m_{xx}(\hat x) = m_{xy}(\hat x-1,0)$, and $\tilde m(\hat x) = \tilde m^{xy}(\hat x-1, 0)$.

Due to this equivalence, all the analysis in the rest of this section follows from the 2+1d $\mathbb Z_N$ tensor gauge theory on this slanted torus \cite{Rudelius:2020kta}.

\subsubsection{Global symmetry}

The electric space-like global symmetries of the original model \eqref{dipZN-action} are also present in the BF model  \eqref{dipZN-intBFaction}. It is generated by the operator $e^{\frac{2\pi i}{N} \tilde m}$. More specifically,  the $\mathbb{Z}_N$ electric symmetry associated with $\bar p_{xx}$ in \eqref{eq:electric_ZN} is generated by $e^{\frac{2\pi i }{N} \bar p_{xx}\tilde m(\hat x=0)}$ and the $\mathbb{Z}_{\gcd(N,L_x)}$ electric dipole symmetry in \eqref{eq:electric_ZN} associated with $\bar r_{xx}$ is generated by $e^{-\frac{2\pi i}{N}\bar r_{xx} \Delta_x\tilde m(\hat x=L_x-1)}$.

In fact, there are additional space-like symmetries in the BF model that are not present in the original model \eqref{dipZN-action}.\footnote{This is a common property of BF models and  modified Villain versions of various system and one of the motivations to introduce them \cite{Gorantla:2021svj}.  The original systems or their Villain versions have various symmetries, like momentum symmetries and electric symmetries.  The  BF models and modified Villain versions of these theories, when they exist, have additional symmetries like winding symmetries and magnetic symmetries.  (The gauge theory in Section \ref{sec:dipA-modVill} does not have a modification of its Villain version and therefore all the symmetries are visible already in the Villain theory.)} It has magnetic symmetries that shift $\tilde m$ by
\ie
\tilde m(\hat \tau,\hat x) \rightarrow \tilde m(\hat \tau,\hat x)+ \tilde p + \tilde r \frac{N}{\gcd(N,L_x)} \hat x~, \qquad 0\le \hat x<L_x~,
\fe
where $\tilde p$ and $\tilde r$ are integers modulo $N$ and $\gcd(N,L_x)$, respectively.

These magnetic symmetries are manifest in the dipole $\mathbb Z_N$ clock model \eqref{dipZN-clock-action}, while the electric symmetries are not present there.

The $\mathbb{Z}_N$ magnetic symmetry associated with $\tilde p$ is implemented by  $W^{\tilde p}$, with the generator
\ie
W = \exp \left( \frac{2\pi i }{N}\sum_{\text{site: fixed }\hat \tau} m_{xx} \right)~.
\fe
The $\mathbb{Z}_{\gcd(N,L_x)}$ magnetic symmetry associated with $\tilde r$ is implemented by $\mathbf W^{\tilde r}$, with the generator
\ie
\mathbf W = \exp \left( \frac{2\pi i }{\gcd(N,L_x)}\sum_{\text{site: fixed }\hat \tau} \hat x m_{xx} \right)~.
\fe

\subsubsection{Ground state degeneracy}\label{sec:dipZN-gsd}

We will now count the number of ground states in the  BF lattice model. In this case, it is equivalent to counting the number of solutions to the ``equations of motion.''  (The quotation is because, strictly, for integer fields there is no equation of motion.) Summing over $m_\tau$ and $m_{xx}$ gives
\ie\label{eq:eqofmotion_tildem}
\Delta_x^2 \tilde m = 0 \mod N~, \qquad \Delta_\tau \tilde m = 0 \mod N~.
\fe
The most general solution to these equations is
\ie
\tilde m(\hat \tau,\hat x) = \tilde p + \tilde r \frac{N}{\gcd(N,L_x)} \hat x~, \qquad 0\le \hat x<L_x~,
\fe
where $\tilde p$ and $\tilde r$ are integers modulo $N$ and $\gcd(N,L_x)$, respectively. Therefore, the number of solutions is
\ie\label{dipZN-GSD}
N \gcd(N,L_x)~.
\fe

As discussed in Section \ref{sec:dipZN-compactify}, this ground state degeneracy can also be computed from the 2+1d $\mathbb{Z}_N$ tensor gauge theory of \cite{Gorantla:2021svj}  on the slanted torus \eqref{slanted}. Indeed, the ground state degeneracy \eqref{dipZN-GSD} agrees with (4.8) of  \cite{Rudelius:2020kta} (with $L_{x}^{\text{eff}}=L_{y}^{\text{eff}}=1$ and $M=L_x$).

Another way to count the ground states is to use the algebra generated by the electric and magnetic space-like symmetry operators
\ie
&e^{\frac{2\pi i}{N} \tilde m(\hat x)} W = e^{-2\pi i/N}W e^{\frac{2\pi i}{N} \tilde m(\hat x)}~,
\\
&e^{\frac{2\pi i}{N} \tilde m(\hat x)} \mathbf{W} = e^{-2\pi i \hat x/\text{gcd}(N,L_x)}\mathbf{W} e^{\frac{2\pi i}{N} \tilde m(\hat x)}~.
\fe
Using \eqref{eq:eqofmotion_tildem}, the $L_x$ operators $e^{\frac{2\pi i}{N}\tilde m(\hat x)}$ can be generated by two operators $e^{\frac{2\pi i}{N}\tilde m(0)}$ and $e^{\frac{2\pi i}{N}\Delta_x \tilde m(0)}$.
We then find that the minimal representation of the  algebra has dimension $N \gcd(N,L_x)$:
\ie
&e^{\frac{2\pi i}{N} \tilde m(\hat x)}|\tilde p,\tilde r\rangle=e^{\frac{2\pi i}{N}\left(\tilde p +\tilde r \frac{N}{\gcd(N,L_x)}\hat x\right)}|\tilde p,\tilde r\rangle~,
\\
&W|\tilde p,\tilde r\rangle=|\tilde p-1,\tilde r\rangle
\\
&\mathbf W|\tilde p,\tilde r\rangle=|\tilde p,\tilde r-1\rangle~.
\fe
This reproduces the same ground state degeneracy \eqref{dipZN-GSD}.  To conclude, the space-like global symmetries lead to the ground state degeneracy $N \gcd(N,L_x)$.

Observe that the ground state degeneracy (GSD) is always between $N$ and $N^2$, but its value depends sensitively on the number of lattice sites $L_x$.

One consequence of it is that the GSD does not have a good $L_x\rightarrow \infty$ limit. For example, if we take $L_x=sN+1$ with $s\to \infty$, then for every finite but large $L_x$ the GSD is $N$.  In the other extreme, we can take $L_x=sN$ with $s\to \infty$ to find that the GSD is $N^2$.  This sensitivity to how we take the limit is a manifestation of UV/IR mixing.

While the strange dependence of the GSD on $L_x$ might seem peculiar,
it is perhaps not as surprising as that in other exotic models in higher spacetime dimensions (such as the Haah code \cite{Haah:2011drr}).
It is well known that certain systems with frustration exhibit a ground state degeneracy that depends sensitively on the details of the lattice, such as the number of lattice sites.  It would be nice to understand whether a similar interpretation of the degeneracy exists in the 1+1d $\mathbb Z_N$ dipole gauge theory.

Since all these degenerate ground states can be distinguished by the local operators $e^{2\pi i \tilde m /N}$, the large ground state degeneracy \eqref{dipZN-GSD} is not robust and will be lifted when the system is perturbed by these local operators. This would be the case if we start with the lattice theory \eqref{dipZN-action}, which does not have the magnetic symmetries. On the other hand, since the magnetic symmetries are natural in the dipole $\mathbb Z_N$ clock model \eqref{dipZN-clock-action}, the ground state degeneracy is robust in this model.

\subsubsection{Defects and their restricted mobility}\label{sec:dipZN-def-mob}

The theory has a defect
\ie\label{dipZN-particle}
W_\tau(\hat x) = \exp\left( \frac{2\pi i}{N} \sum_{\tau\text{-link: fixed }\hat x} m_\tau \right)~,
\fe
that represents the world-line of a static particle. On a noncompact space, the particle can hop by $kN$ sites for any integer $k$. When $k=1$, this is described by the defect
\ie
\exp\left[\frac{2\pi i}{N} \left(\sum_{\tau \text{-link: }\hat \tau < 0} m_\tau(\hat \tau,0)  +  \sum_{\text{site: }0 < \hat x < N} \hat x m_{xx}(0,\hat x) + \sum_{\tau \text{-link: }\hat \tau \ge 0} m_\tau(\hat \tau,N) \right) \right]
\fe
More generally, $n$ particles can simultaneously hop by $kN/\gcd(n,N)$ sites for any integer $k$. When $k=1$, this is described by the defect
\ie\label{chargen-def-hop}
&\exp\left[\frac{2\pi i n}{N} \sum_{\tau \text{-link: }\hat \tau < 0} m_\tau(\hat \tau,0)\right] \exp\left[\frac{2\pi in}{N\gcd(n,N)}  \sum_{\text{site: }0 < \hat x < N} \hat x m_{xx}(0,\hat x) \right]
\\
&\times  \exp\left[ -\frac{2\pi i n}{N\gcd(n,N)}\sum_{r=1}^{\gcd(n,N)-1}~ \sum_{\hat x = 1}^{N\over \gcd(n,N)} \sum_{\text{site: }\hat x \le \hat x' < \hat x + {rN\over \gcd(n,N)}} m_{xx}(0,\hat x')\right]
\\
&\times \exp\left[ \frac{2\pi i n}{N} \sum_{\tau \text{-link: }\hat \tau \ge 0} m_\tau\left(\hat \tau,\frac{N}{\gcd(n,N)}\right) \right]~.
\fe

When space is a circle with $L_x$ sites, i.e., $\hat x\sim \hat x+L_x$, the particle can hop by any multiple of $\gcd(N,L_x)$ sites. More generally, $n$ particles can simultaneously hop by any multiple of $\gcd(N,L_x)/\gcd(n,N,L_x)$ sites. This, more complicated hopping is similar to the way the $U(1)$ fracton can hop by a large amount, as in \eqref{eq:defectLm}.

Let us clarify this fact.  For finite $L_x \gg N$, the hopping by $N$ sites is simple.  There is also a more complicated possibility of moving several times around the space always in jumps of $N$ sites (on the covering space) and ending up only $\gcd(N,L_x)$ sites away from the starting point in real space.

How should we interpret it for infinite $L_x$?

The simple hopping by $N$ sites is clearly possible.  And if we take a continuum limit, it is a microscopic hop and we conclude that the fracton is completely mobile.  In this case, there is no need to consider the more complicated process involving going around the space.

Alternatively, we can think of the limit $L_x\to \infty$ without taking a continuum limit.  Then, for any finite, but large $L_x$, the more complicated jumps that circle around the space, are also possible.   Then the possible hops on the lattice  depend on how we take $L_x\to \infty$. For example, if we take $L_x=sN+1$ with $s\to \infty$, then for every finite, but large $L_x$ the particle is completely mobile. In the other extreme, for $L_x=sN$ with $s\to \infty$, the particle can hop only by $N$ sites.

There is, however, an important difference between the simple hop by $N$ sites and the shorter hop by $\gcd(N,L_x)$ sites. The former is implemented by an operator stretching over the $N$ sites, while the latter is implemented by a more complicated operator affecting degrees of freedom all over the lattice.  This difference becomes more significant as $L_x$ gets larger.

The fact that the number of sites the particle can hop on the lattice depends on the long distance geometry of the lattice is another  manifestation of UV/IR mixing.

We conclude that in the $\mathbb{Z}_N$ dipole gauge theory, the particle represented by the defect \eqref{dipZN-particle} is not completely immobile. This is to be contrasted with the defect \eqref{dipA-modVill-fracton} in the $U(1)$ dipole gauge theory, which cannot be deformed and hence represents an immobile fracton. This difference in the particle mobility between the $U(1)$ and $\mathbb{Z}_N$ theories has already been discussed in higher-dimensional tensor gauge theories in \cite{Bulmash:2018lid,Ma:2018nhd,Oh2021}.

Having discussed the defect for a single particle, we will now consider that for a dipole of particles with opposite charges.  A dipole of particles can move by arbitrary number of sites as long as their separation is fixed. It is described by the defect
\ie
\exp\left[\frac{2\pi i}{N} \left(\sum_{\tau x\text{-plaq: }\hat \tau < 0} \Delta_x m_\tau(\hat \tau,\hat x_1)  +  \sum_{\text{site: }\hat x_1 < \hat x \le \hat x_2} m_{xx}(0,\hat x) + \sum_{\tau x\text{-plaq: }\hat \tau \ge 0} \Delta_x m_\tau(\hat \tau,\hat x_2) \right) \right]~.
\fe

The  restricted mobility of the particle can be understood as a result of the time-like symmetries that act on the defects.
Let us discuss the time-like symmetries.

The  BF lattice model  \eqref{dipZN-intBFaction} has the same time-like symmetries \eqref{eq:electric_ZN} as the $\mathbb{Z}_N$ lattice dipole gauge theory \eqref{dipZN-action}. The $\mathbb Z_N$ electric time-like symmetry associated with $\bar p_\tau$ is generated by the $\mathbb Z_N$ operator
\ie
U(\hat \tau;\hat x_1,\hat x_2) = \exp \left(-\frac{2\pi i}{N}  [\Delta_x \tilde m(\hat\tau, \hat x_2) - \Delta_x \tilde m(\hat\tau, \hat x_1)] \right)~.
\fe
The $\mathbb Z_{\gcd(N,L_x)}$ electric dipole time-like symmetry associated with $\bar r_\tau$ is generated by the $\mathbb Z_{\gcd(N,L_x)}$ operator
\ie
\mathbf U(\hat \tau;\hat x_1,\hat x_2) &= \exp \left( -\frac{2\pi i}{\gcd(N,L_x)}[ \hat x_2 \Delta_x \tilde m(\hat \tau, \hat x_2-1)- \tilde m(\hat\tau, \hat x_2) ] \right)
\\
&\quad \times \exp \left( \frac{2\pi i}{\gcd(N,L_x)}[ \hat x_1 \Delta_x \tilde m(\hat \tau, \hat x_1-1)- \tilde m(\hat\tau, \hat x_1) ] \right)~.
\fe
Both $U(\hat \tau;\hat x_1,\hat x_2)$ and $\mathbf U(\hat \tau;\hat x_1,\hat x_2)$ are invariant under deformations of $\hat x_1$, $\hat x_2$, and $\hat \tau$, because of \eqref{eq:eqofmotion_tildem}, as long as they do not cross any defect.
In particular, they are trivial operators in the absence of any defect insertions.

These time-like symmetries act on the defect $W_\tau(\hat x)$ as
\ie
&U(\hat \tau;\hat x_1,\hat x_2) W_\tau(\hat x)  = e^{\frac{2\pi i}{N}}W_\tau(\hat x)~,\quad \text{if}\quad\hat x_1<\hat x<\hat x_2\,,
\\
&\mathbf U(\hat \tau;\hat x_1,\hat x_2) W_\tau(\hat x) = e^{\frac{2\pi i}{\gcd(N,L_x)} \hat x}W_\tau(\hat x)~,\quad \text{if}\quad\hat x_1<\hat x<\hat x_2\,.
\fe
The action is trivial if $\hat x$ is not in between $\hat x_1, \hat x_2$, which follows from the Gauss law.

The defects $W_\tau(\hat x)^n$ and $W_\tau(\hat x+\frac{\gcd(N,L_x)}{\gcd(n,N,L_x)})^n$ carry the same $\mathbb Z_N$ and $\mathbb Z_{\gcd(N,L_x)}$ time-like charges. Therefore, the time-like global symmetry explains the allowed mobility of the particles. This is to be compared with the ground state degeneracy of $N\gcd(N,L_x)$, which is a consequence of space-like global symmetry.

\section{Conclusions}

We studied the continuum Lifshitz theory of a compact scalar $\phi$ with action \eqref{dipphi-action1} on a spatial circle in 1+1d. It has momentum and winding dipole global symmetries with a mixed 't Hooft anomaly between them.  This leads to an infinite degeneracy in the spectrum.

We also studied a continuum $U(1)$ dipole gauge theory of $(A_\tau,A_{xx})$, which is the 1+1d version of the symmetric tensor gauge theory of \cite{Gu:2006vw,Xu:2006,Pankov:2007,Xu2008,Gu:2009jh,rasmussen,Pretko:2016lgv,Slagle:2018kqf,Du:2021pbc,Pretko:2016kxt}. This theory has defects, $\exp(i\oint d\tau ~A_\tau)$, which describe the world-lines of immobile particles -- fractons. Curiously, the line operator $\oint dx~A_{xx}$ is gauge invariant without exponentiating.

To understand the subtleties of the continuum field theories of the scalar and the tensor gauge field, we studied their lattice models. We used the modified Villain model for the scalar theory and a Villain version for the tensor gauge theory.  The lattice models are unambiguous and exhibit most of global symmetries of the continuum theories. Surprisingly, for each lattice model, there are several continuum limits one can take. These continuum limits are described by the same action, but they differ in various global aspects including the field identifications. Correspondingly, the global symmetries of the continuum models and their observables (defects and operators) are different.

We discussed three continuum limits of the scalar field theory. Two of them, the $\phi$- and $\Phi$-theories, are dual to each other, while the third, the $\hat \phi$-theory, is self-dual. The latter is also scale-invariant under the Lifshitz scale transformation $x\rightarrow \lambda x$ and $\tau\rightarrow \lambda^2 \tau$. Their global symmetries are summarized in Table \ref{tbl:lat-cont-sym}.

While the modified Villain lattice model has the local operators $e^{i\phi}$ and $e^{i\Delta_x\phi}$, none of the three continuum theories has both of these operators at the same time: the $\phi$-theory has $e^{i\phi}$ but no $e^{i\partial_x\phi}$, the $\Phi$-theory has $e^{i\partial_x\Phi}$ but no $e^{i\Phi}$, and the $\hat\phi$-theory has neither $e^{i\hat\phi}$ nor $e^{i\partial_x\hat\phi}$.  This observation follows from the mass dimensions of these continuum fields.\footnote{More precisely, when we say that the operator $e^{i\partial_x\phi}$ is absent in the continuum theory, we mean the following.  The lattice operator $e^{i\Delta_x\phi}$ flows to a nontrivial operator in the continuum limit.  At leading order it flows to the identity operator.  (In this sense it is trivial in the continuum.)  However, $e^{i\Delta_x\phi}-1$ flows, up to wave function renormalization, to $\partial_x\phi$, which is nontrivial.  And at higher orders we find additional operators.  The fact that $e^{i\Delta_x\phi}$ transforms under the lattice momentum dipole symmetry, leads to the fact that in the continuum, $\partial_x \phi$ also transforms (albeit inhomogeneously) under this symmetry.}
We summarized the local operators on the lattice and in the continuum in Table \ref{tbl:lat-cont-chargedop}.

We also discussed three continuum limits of the $U(1)$ dipole gauge theory. These are the pure gauge theories associated with the momentum symmetries of the $\phi$-, $\Phi$- and $\hat \phi$-theories. As already mentioned above, the $A$-theory has quantized fracton defects, but the line operators are not quantized. In the $\tilde A$-theory, there are no fracton defects and the line operators are exponentiated with a quantized coefficient. In the $\hat A$-theory, there are no fracton defects and the line operators are not quantized. The global symmetries and fluxes in these theories are summarized in Table \ref{tbl:dipA-lat-cont}.

Finally, we discussed  the BF lattice  version of the $\mathbb Z_N$ dipole gauge theory in 1+1d. Unlike the $U(1)$ theory, there are no fractons in the $\mathbb Z_N$ theory. The would-be fractons can hop by $N$ sites on the lattice.  In fact, if the lattice has $L_x$ sites with periodic boundary conditions, the would-be fracton can hop even by smaller steps, steps of $\gcd(N,L_x)$.  Surprisingly, the amount by which the particle can hop locally depends on the total number of sites of the lattice.  Consequently, the $L_x\to \infty$ limit is not well defined.  This subtlety, which is related to phenomena observed in \cite{Haah:2011drr,Yoshida:2013sqa,Meng,Manoj:2020wwy,Rudelius:2020kta}, reflects the UV/IR mixing in these theories, as emphasized in \cite{Gorantla:2021bda}.

One of the tools we used was the notion of a time-like global symmetry. This is a global symmetry that acts trivially on the operators and the Hilbert space without defects, but it acts nontrivially on defects extended in the time direction.
The time-like global symmetry leads to selection rules and constrains the shapes and locations   of line defects.
When these defects represent the world-line of   particles, the time-like global symmetry  explains their mobility restrictions as a result of a global symmetry, rather than a gauge symmetry.
This discussion based on the time-like global symmetry makes precise the intuition about conservation of ``gauge charges."
Specifically, in the $U(1)$ dipole gauge theory, the time-like global symmetry completely restricts the mobility of the fracton defects, while in the $\mathbb Z_N$ theory, it explains the relaxed mobility of the would-be fractons.

As we said above, we can summarize the role of the global symmetries in these exotic systems as follows.  {\it The space-like global symmetries lead to the peculiar ground state degeneracy and the time-like global symmetries lead to the unusual restricted mobility of the defects.}

Throughout the paper, we focused on models in 1+1d, but we expect many of these subtleties to be present in the higher dimensional versions of these theories. For instance, the origin of the infinite degeneracy in our 1+1d compact Lifshitz theory  is almost the same as the degeneracy in the 2+1d quantum Lifshitz theory \cite{Henley1997,Moessner2001}.
Similarly, the absence of the fracton defects in  the $\mathbb{Z}_N$ symmetric tensor gauge theory (unlike its $U(1)$ counterpart) was also observed in the higher-dimensional models of  \cite{Bulmash:2018lid,Ma:2018nhd,Oh2021}.

\emph{Note added later:} After the completion of this work, \cite{Oh2022,Pace2022} published similar results in related models in 2+1d.

\section*{Acknowledgements}

We thank X.\ Chen, M.\ Hermele, E.\ Lake, Y.\ Oz, W.\ Shirley, and T.\ Senthil for helpful discussions.
PG was supported by Physics Department of Princeton University.
HTL was supported in part by a Croucher fellowship from the Croucher Foundation, the Packard Foundation and the Center for Theoretical Physics at MIT.
The work of NS was supported in part by DOE grant DE$-$SC0009988.
NS was also supported by the Simons Collaboration on Ultra-Quantum Matter, which is a grant from the Simons Foundation (651440, NS).
The authors of this paper were ordered alphabetically.
Opinions and conclusions expressed here are those of the authors and do not necessarily reflect the views of funding agencies.

\appendix
\section{Time-like symmetries}\label{app:timelike}

Global symmetries including higher-form global symmetries, dipole global symmetries, and subsystem symmetries are implemented by operators acting at a particular time.  They act on states in the Hilbert space and they act on operators by conjugation.

In this appendix, we will discuss symmetries that act trivially on the Hilbert space in the absence of defects, but they act nontrivially on time-like defects.\footnote{More precisely, the time-like symmetry operator does not act on the defect.  It acts on the Hilbert space in the presence of a defect.}  For this reason, we will refer to these symmetries as time-like symmetries.  (Recall that, as we said in the introduction, we abuse the terminology here.  Our discussion is mostly in Euclidean signature, but we still use the phrase ``time-like'' for operators acting on defects along the Euclidean time direction.)
In contrast, global symmetries that act on states in the Hilbert space without defects and on operators will be called space-like symmetries.

Although the defects these symmetries can act on could be of various dimensions, for simplicity of the presentation, we will focus on symmetries acting on time-like line defects.

In relativistic systems, a one-form symmetry acts on line operators and line defects \cite{Gaiotto:2014kfa}.  And depending on how the line is oriented, the symmetry can be thought of as space-like or time-like.  Relativistic invariance relates them and there is no reason to distinguish between space-like and time-like symmetries.
However, in   non-relativistic systems, the distinction between space-like and time-like symmetries is quite significant.
We have already encountered several examples of that  in  Sections \ref{sec:dipA} and \ref{sec:dipZN}.

Since the time-like symmetries act on line defects, they give rise to selection rules of correlation functions of line defects.
More specifically, invariance under the time-like symmetry constrains the possible configurations and shapes of the line defects.
When the line defects represent the world-line of particles in the microscopic systems,  these selection rules restrict the mobility of these particles.
In particular, certain particles might be immobile due to the time-like symmetry, i.e., they are fractons.
Thus, the time-like symmetry gives a global-symmetry-based explanation of the restricted mobility of the fracton defects.

Below, we will start by phrasing the known symmetry properties of some relativistic systems in the language of space-like and time-like symmetries.  This will allow us to test our new language and to practice it.  Then, we will discuss some nonrelativistic systems with a subsystem global symmetry, where this language will lead to new results.
In the main text, we have discussed the time-like symmetries in $U(1)$ and $\mathbb{Z}_N$ gauge theories with dipole global symmetries.

\subsection{2+1d $U(1)$ gauge theory}\label{app:U1gauge}
Consider the 2+1d Maxwell theory described by the Lagrangian
\ie
\mathcal L = \frac{1}{g^2} F_{\mu \nu}F^{\mu \nu}~,\qquad \mu,\nu = \tau,x,y~,
\fe
where $F = dA$ is the field strength of the $U(1)$ one-form gauge field $A$ with gauge symmetry
\ie
A \sim A + d\alpha~.
\fe
We place the theory on a Euclidean 3-torus with lengths $\ell_\tau$, $\ell_x$ and $\ell_y$.

There is a $U(1)$ electric one-form symmetry \cite{Gaiotto:2014kfa}, which acts on the gauge fields as
\ie
A \rightarrow A + \lambda~,
\fe
where $\lambda$ is a flat gauge field. Using the gauge freedom of $\alpha$, we can set
\ie\label{U1-oneform-sym}
\lambda_\tau = \frac{c_\tau}{\ell_\tau}  ~,\qquad \lambda_i = \frac{c_i}{\ell_i}~, \qquad i = x,y~,
\fe
where $c_\tau$ and $c_i$ are circle-valued constants: $c_\tau \sim c_\tau + 2\pi$, and $c_i \sim c_i + 2\pi$.

The Noether current for the $U(1)$ electric one-form symmetry is
\ie
J_\mu = \frac{2i}{g^2} \epsilon_{\mu \nu \rho} F^{\nu \rho}~,\qquad dJ = 0~.
\fe
The charge is
\ie
Q(C) = \oint_{C} J~,
\fe
which is independent of small deformations of the closed curve $C$. The charged objects are the Wilson lines of $A$:
\ie
W(C) = \exp\left( i\oint_C A\right)~.
\fe

The circle-valued parameter $c_x$ generates a standard $U(1)$ global symmetry, \emph{i.e.}, a space-like symmetry. The action of this symmetry on the Wilson line operator $W(X)$, where $X$ is the $x$-cycle at a fixed time, is
\ie\label{symcono}
e^{ic_x Q(Y)} W(X)^n e^{-ic_xQ(Y)} = e^{inc_x}W(X)^n~.
\fe
Here $Q(Y)$ is the charge operator with the curve $C$ being the $Y$-cycle.
(There is a similar action of $c_y$ on $W(Y)^n$.)

\begin{figure}[t]
\begin{center}
\hfill\raisebox{-0.5\height}{\includegraphics[scale=0.15]{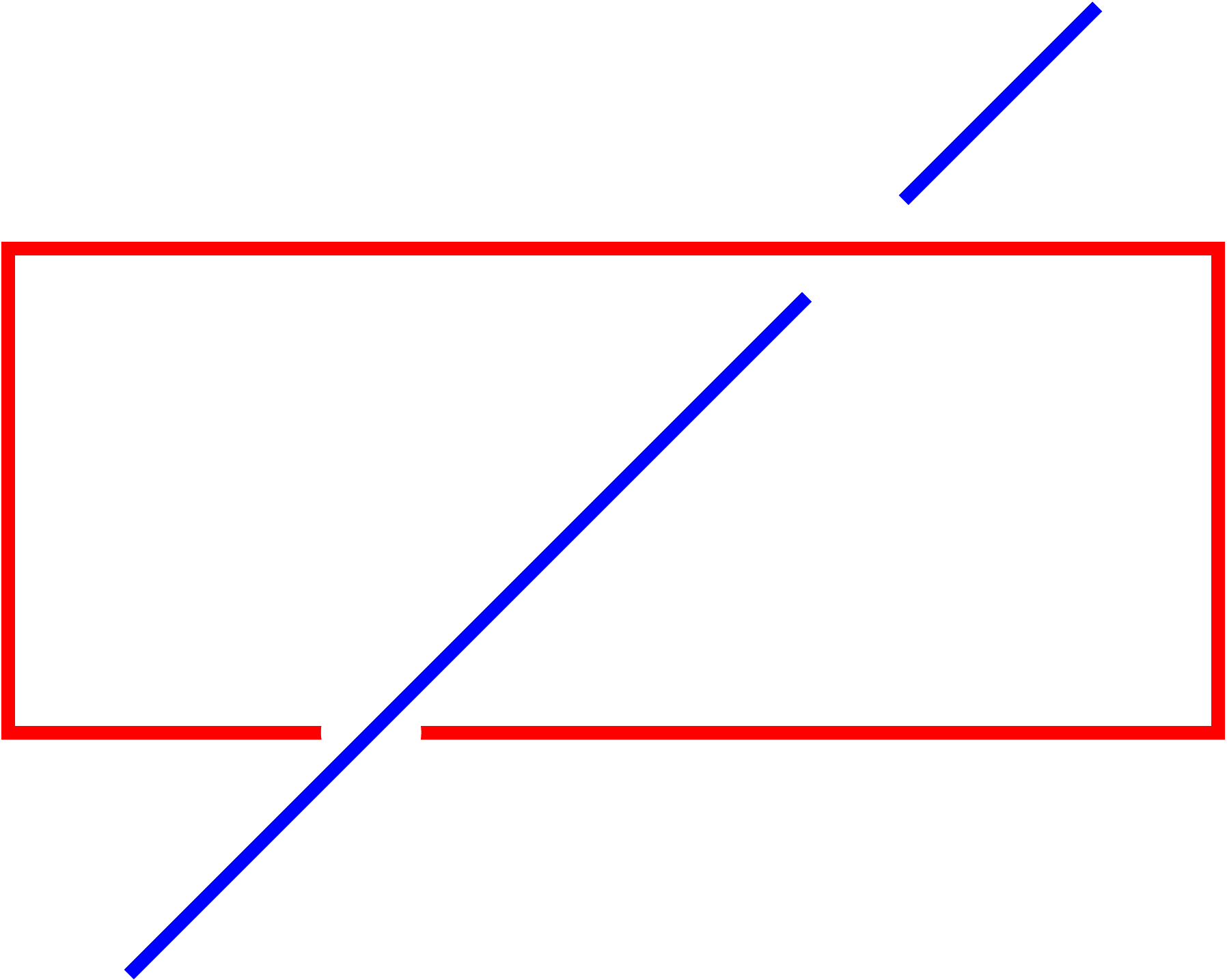}}~~~~$\longrightarrow ~~ e^{i c}$
\hspace*{-1cm}\raisebox{-0.5\height}{\includegraphics[scale=0.15]{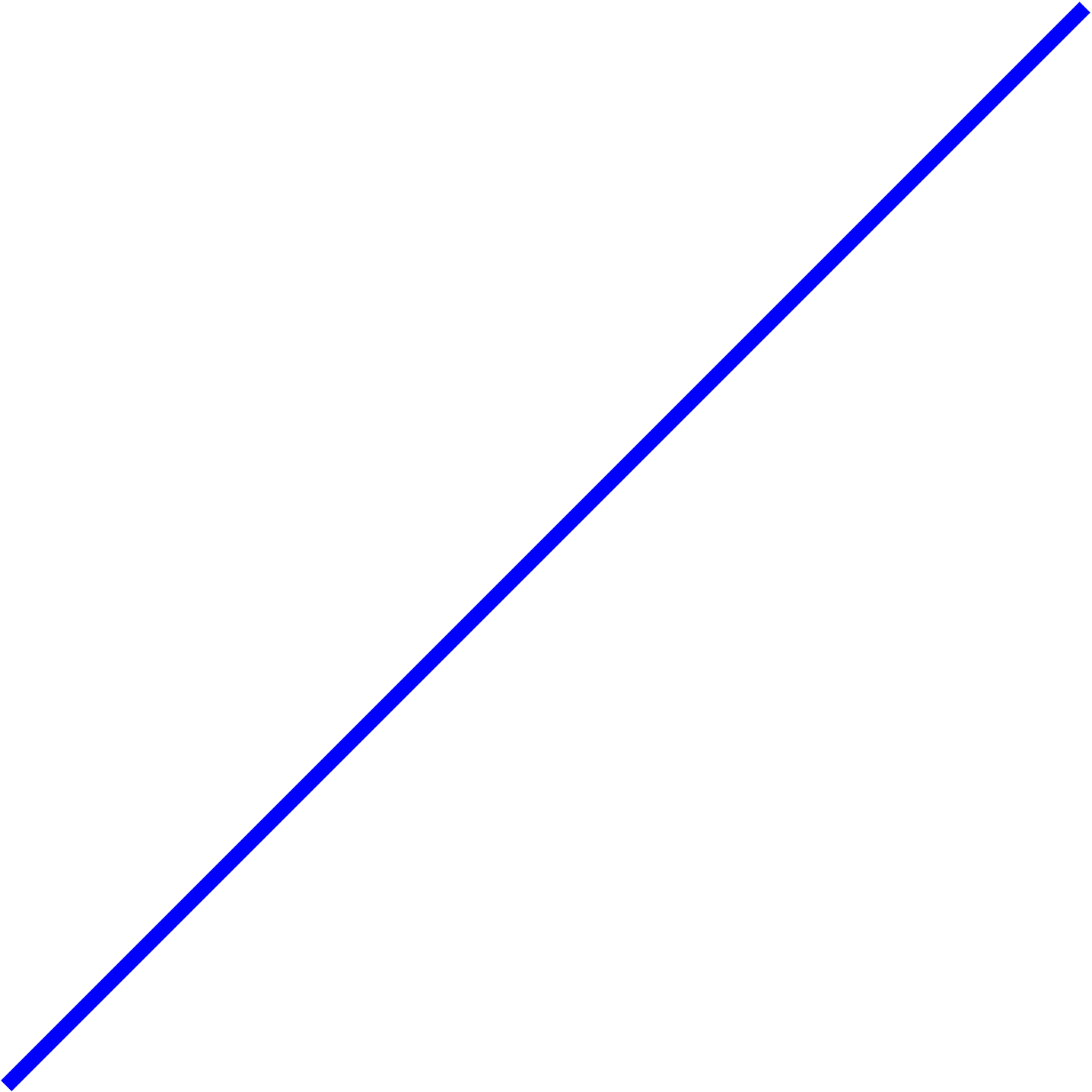}}~~~~~~~~~~~~~~~~~~~~~~~
\raisebox{-0.5\height}{\includegraphics[scale=0.15]{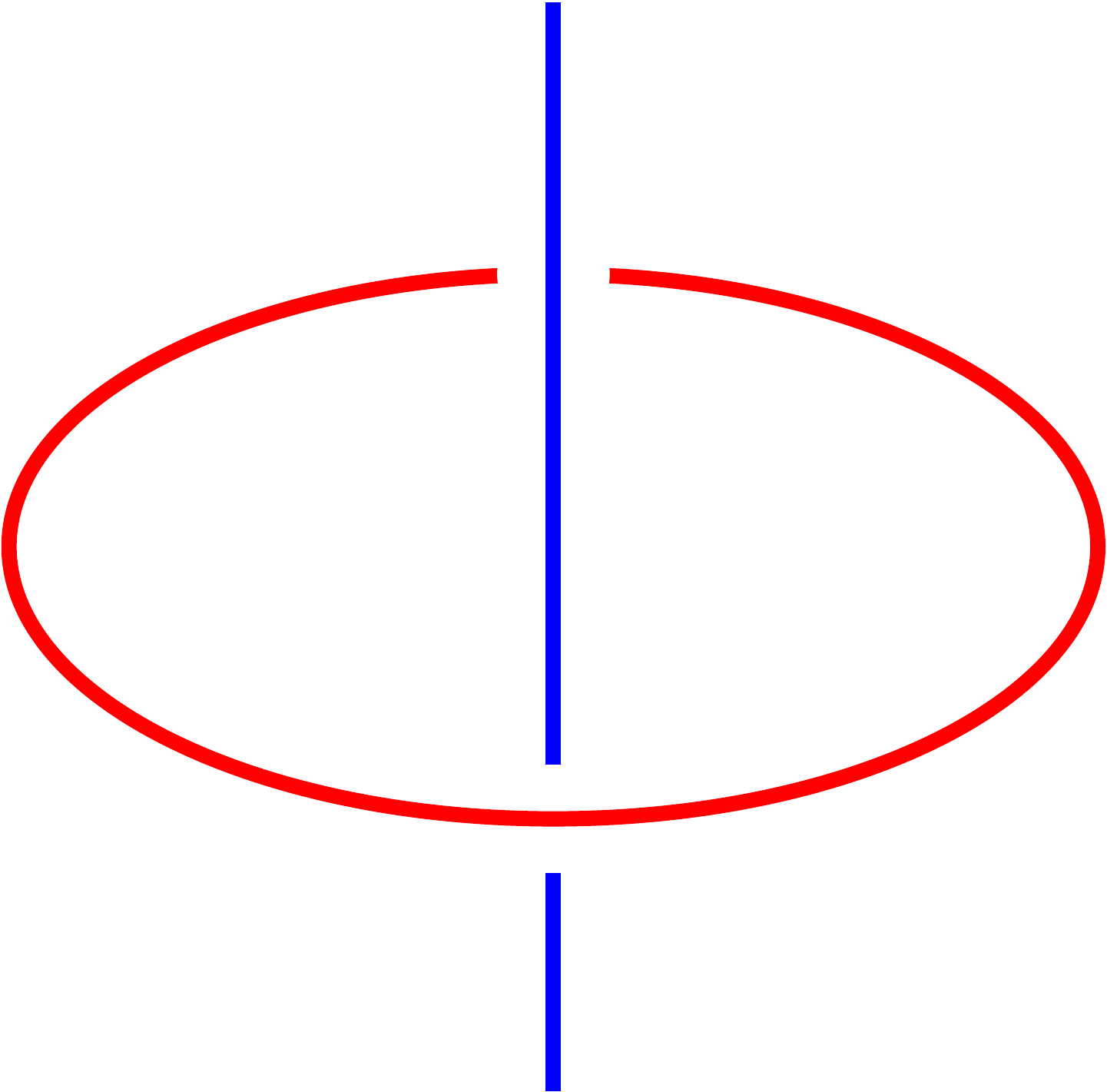}}~~~~$\longrightarrow ~~ e^{i c}$
\raisebox{-0.5\height}{\includegraphics[scale=0.15]{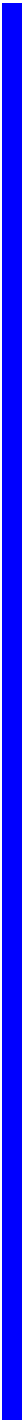}}~~~~~~~~~~~~~
\\
\hfill (a) Space-like symmetry action \hfill \raisebox{-0.5\height}{\includegraphics[scale=0.3]{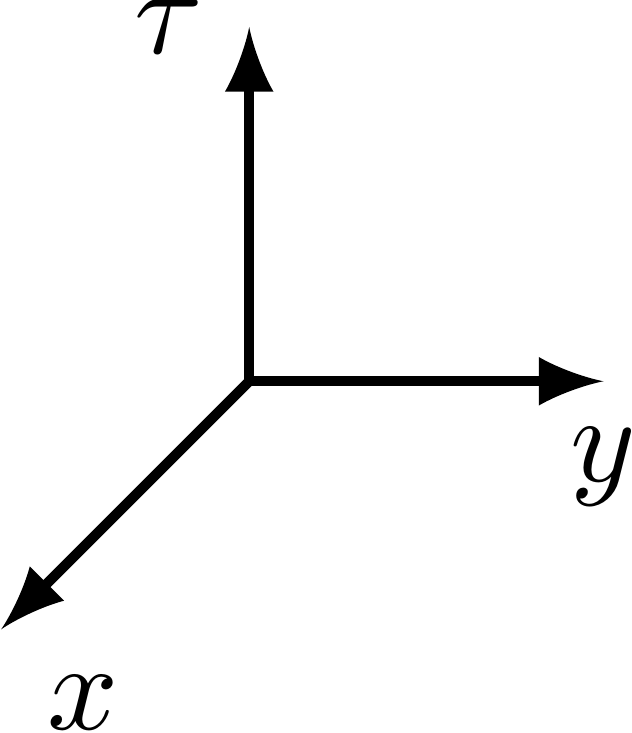}} \hfill (b) Time-like symmetry action \hfill \hfill
\caption{The Euclidean configurations for the action of a one-form symmetry on a line defect/operator. In (a), the space-like symmetry operator (red curve) acts on the line operator (blue line) giving the phase $e^{ic}$.  The two horizontal red lines are at fixed Euclidean time and the two vertical red lines are along the Euclidean time direction.  In (b), the time-like symmetry operator (red curve) acts on the line defect (blue line) giving the phase $e^{ic}$.  Here, the red curve is at fixed Euclidean time.   These phases are $U(1)$-valued in the 2+1d $U(1)$ gauge theory discussed in Appendix \ref{app:U1gauge}, while they are $\mathbb Z_N$-valued in the 2+1d $\mathbb Z_N$ gauge theory discussed in Appendix \ref{app:ZNgauge}.}\label{fig:one-form-symaction}
\end{center}
\end{figure}

In a standard way, this symmetry action by conjugation is represented as two operators inserted at different Euclidean times, before and after $W(X)^n$.  Then, the curves in the two lines in $Q(Y)$ can be deformed  to a closed curve in Euclidean space. See Figure \ref{fig:one-form-symaction}(a) for the Euclidean configuration of the above space-like symmetry action.\footnote{The action of the space-like symmetry operator on the charged operator is gauge equivalent to the shift \eqref{U1-oneform-sym}. The same is true for the action of the time-like symmetry, and also the other symmetries in the rest of this appendix.}

On the other hand, the circle-valued parameter $c_\tau$ generates a $U(1)$ time-like symmetry. It acts on the Wilson line defect $W(x,y) \equiv \exp[i\oint d\tau~A_\tau(\tau,x,y)]$ as
\ie
e^{ic_\tau Q(C)} W(x,y)^n = e^{in c_\tau} W(x,y)^n~,
\fe
where $C$ is a closed space-like curve that encloses the point $(x,y)$. Note that the action of the time-like symmetry on the defect is like an action on a state.  It does not involve conjugation. See Figure \ref{fig:one-form-symaction}(b) for the Euclidean configuration of this time-like symmetry action.

In finite volume, because of the time-like symmetry, the correlation function of line defects
\ie
\left\langle \prod_a W(x_a,y_a)^{q_a} \right\rangle~, \qquad q_a \in \mathbb Z~,
\fe
is trivial, unless
\ie\label{sumqaiz}
\sum_a q_a = 0~.
\fe

This constraint can also be obtained by nucleating a small time-like symmetry operator $e^{ic_\tau Q(C)}$ away from the defects, and then enlarging $C$ to wrap it around the entire compact space. During this process, whenever the symmetry operator crosses a defect, it picks up a phase. At the end of this process, the symmetry operator can be contracted to a trivial operator on the other side, i.e., it is as if we never inserted it, so we should get the original correlation function back, up to an overall phase. So, unless this phase is trivial, the correlation function vanishes.

In infinite volume, we can send one of the defects to infinity, so that the sum of the charges of the rest of them is conserved in time.

Since the $U(1)$ gauge theory is relativistic, both the space-like and the time-like symmetries (generated by $c_i$ and $c_\tau$, respectively) can be traced back to the same underlying $U(1)$ one-form global symmetry.

The discussion above is a complicated way to state well-known facts.  We can think of the defects as the world-lines of probe charged particles with charges $q_a$ and then \eqref{sumqaiz} states that the total charge in compact space must vanish.  And the process of removing one of the charges to infinity is the known fact of charge conservation.

However, one might not be satisfied with this low-brow presentation because $q_a$ are gauge charges and their definition needs more care.  For example, if the system includes dynamical charged particles, then $q_a$ can be screened and then the conclusions are slightly different.  Also, a gauge symmetry is not an intrinsic property of the system and it is preferable to use well-defined, gauge-invariant operators to derives such constraints.

The formulation of the discussion in terms  of generalized global symmetries \cite{Gaiotto:2014kfa} addresses these concerns.  The discussion we presented in this subsection, in terms of space-like and time-like symmetries is an adaptation of the generalized global symmetry framework to non-relativistic systems like the ones we study in this note.

\subsection{Time-like symmetry without gauge fields}
Time-like symmetries can also exist in theories without gauge fields. For example, consider the 2+1d compact boson described by the Lagrangian
\ie
\mathcal L = \frac{f}{2} (\partial_\mu \phi)^2~,\qquad \mu = \tau,x,y~,
\fe
with identifications $\phi(\tau,x,y) \sim \phi(\tau,x,y) + 2\pi$. Here, $f$ is a coefficient with mass dimension $1$.
We place the theory on a Euclidean 3-torus with lengths $\ell_\tau$, $\ell_x$ and $\ell_y$.

There is a $U(1)$ winding one-form symmetry \cite{Gaiotto:2014kfa} with current
\ie
J = \frac{1}{2\pi} d\phi~, \qquad d J = 0~.
\fe
The charge is
\ie
 Q( C) = \oint_{ C}  J~,
\fe
which is independent of small deformations of the closed curve $ C$. When $ C = X$ or $Y$, this is the usual winding charge that measures the winding number of $\phi$ in $x$ or $y$ directions, respectively.

Similar to the discussion in Section \ref{app:U1gauge},
there is  a $U(1)$ winding time-like symmetry, which is  part of the $U(1)$ winding one-form symmetry.
The objects charged under this time-like symmetry are vortex line defects in spacetime, which are defined in terms of the winding of $\phi$ around them.  Concretely, consider the line defect obtained by removing a nontrivial curve $C$ that wraps once around the $\tau$-direction. This creates another nontrivial spatial cycle $ C_0$ around $C$.  Part of the definition of the defect is that the scalar $\phi$ winds $q$ times along $ C_0$.

We can interpret this to mean that this line defect carries charge $q$ under the $U(1)$ winding time-like symmetry similar to the configuration in Figure \ref{fig:one-form-symaction}(b).  As in Section \ref{app:U1gauge}, the time-like symmetry constrains the configurations. When we have several such time-like cycles with charges $q_a$, and the space is compact, they should satisfy $\sum_a q_a=0$.  In particular, the case with a single vortex with nonzero $q$ cannot exist in a compact space.

Note that the 2+1d compact boson is exactly dual to the continuum 2+1d $U(1)$ gauge theory of Section \ref{app:U1gauge}. In the gauge theory description,  the vortex line defects are the Wilson lines of the gauge field along the time direction. In this sense, the discussion here does not add information to the discussion in Section \ref{app:U1gauge}. However, the analysis here applies without any significant change in more general systems.  First, we can add to the Lagrangian various terms, such that it is not dual to the free gauge theory.  Second, we can replace this theory with any non-linear sigma model whose target space $\mathcal M$ has nontrivial cycles.  Finally, this discussion underscores the fact that higher-form global symmetries (and therefore also time-like symmetries) are not specific to gauge theories.

\subsection{2+1d $\mathbb Z_N$ gauge theory}\label{app:ZNgauge}

Consider the ordinary 2+1d $\mathbb{Z}_N$ gauge theory, which can be described by the continuum Chern-Simons Lagrangian \cite{Maldacena:2001ss,Banks:2010zn,Kapustin:2014gua}:
\ie
\mathcal L = \frac{iN}{2\pi} A d\hat A~,
\fe
where $A$ and $\hat A$ are one-form gauge fields with gauge symmetry
\ie
A \sim A + d\alpha~,\qquad \hat A \sim \hat A + d\hat \alpha~.
\fe
We place the theory on a Euclidean 3-torus with lengths $\ell_\tau$, $\ell_x$ and $\ell_y$.

There is a $\mathbb Z_N \times \mathbb Z_N$ one-form global symmetry. The symmetry operators/defects are
\ie
W(C) = \exp\left( i\oint_C A \right)~, \qquad \hat W(C) = \exp\left( i\oint_C \hat A \right)~,
\fe
where $C$ is a closed curve in spacetime. They satisfy
\ie\label{ZNgauge-commrel}
&W(C)^N = \hat W(C)^N = 1~,
\\
&\langle W(C_1)\hat W(C_2) \rangle = e^{2\pi i \ell(C_1,C_2)/N} \langle 1\rangle~,
\fe
where $\ell(C_1,C_2)$ is the linking number of the closed curves $C_1,C_2$. The second line of \eqref{ZNgauge-commrel} implies that the symmetry operators are charged under each other. This signals a mixed 't Hooft anomaly between the two $\mathbb Z_N$ one-form symmetries, which results in a  ground state degeneracy of $N^2$.

\bigskip\centerline{\it Space-like and time-like global symmetries}\bigskip

The $\bZ_N\times \bZ_N$ one-form global symmetry acts on the gauge fields as
\ie
A \rightarrow A + \lambda~, \qquad \hat A \rightarrow \hat A + \hat \lambda~,
\fe
where $\lambda$ and $\hat \lambda$ are flat one-form $\mathbb Z_N$ gauge fields. (They have to be $\mathbb Z_N$ gauge fields for the exponential of the action to be invariant.) Using the gauge freedom of $\alpha$ and $\hat \alpha$, we can set
\ie
&\lambda_\tau = \frac{2\pi n_\tau}{N\ell_\tau}  ~,\qquad \lambda_i = \frac{2\pi n_i}{N\ell_i}~, \qquad i = x,y~,
\\
&\hat \lambda_\tau = \frac{2\pi \hat n_\tau}{N \ell_\tau}  ~, \qquad \hat \lambda_i = \frac{2\pi \hat n_i}{N\ell_i}~,
\fe
where $n_\tau$, $n_i$, $\hat n_\tau$, and $\hat n_i$ are integers modulo $N$.

The parameters $n_i$ and $\hat n_i$ are associated with space-like symmetries. They act on states in the Hilbert space, and on operators $W(C)$ and $\hat W(C)$ when $C$ is purely space-like.

On the other hand, $n_\tau$ and $\hat n_\tau$ are associated with   time-like symmetries. Instead of acting on operators and states, they act on the defects $W(C)$ and $\hat W(C)$ when $C$ is purely time-like. Consequently, in finite volume, the correlation function of defects,
\ie
\left\langle \prod_a W(x_a,y_a)^{q_a}\right\rangle\equiv \left\langle \prod_a \exp\left( iq_a \oint d\tau~ A_\tau(\tau,x_a,y_a) \right) \right\rangle~,\qquad q_a\in\mathbb Z~,
\fe
vanishes unless
\ie
\sum_a q_a = 0 \mod N~.
\fe

This can also be derived by nucleating a time-like symmetry operator and wrapping it around the space, similar to Section \ref{app:U1gauge}.

In infinite volume, we can move one defect to infinity, and then the rest of the defects are such that the sum of their charges is conserved modulo $N$.

Since the $\mathbb{Z}_N$ gauge theory is relativistic, both the space-like and the time-like symmetries (generated by   $n_i,\hat n_i$ and $n_\tau,\hat n_\tau$, respectively) can be traced back to the same underlying $\mathbb{Z}_N\times \mathbb{Z}_N$ one-form global symmetry.

\bigskip\centerline{\it Symmetry operators}\bigskip

We will now discuss the symmetry operators of the space-like and time-like symmetries.
The symmetry operator for the space-like symmetry associated with $n_x$ is $\hat W(Y)^{n_x}$, where $Y$ is the $y$-cycle at a fixed time.
The action of the space-like symmetry on a space-like defect $W(X)$ wrapping the $x$-cycle is given by the conjugation:
\ie
\hat W(Y) W(X) \hat W(Y)^{-1} = e^{-2\pi i /N}W(X) \,.
\fe
The discussions for the space-like symmetry along the other directions and for $\hat n_i$ are similar.  See Figure \ref{fig:one-form-symaction}(a) for the Euclidean configuration of the above space-like symmetry action.

Next, the symmetry operator for the time-like symmetry associated with $n_\tau$ is $\hat W(C)^{n_\tau}$. Here $C$ is a space-like curve encircling a line defect $W(x_0,y_0)$,  which extends in time at a point $(x_0,y_0)$ in space.  In equation, we denote this time-like symmetry action by:
\ie\label{ZNgauge-time-likesym-action}
\hat W(C) W(x_0,y_0) =e^{-2\pi i /N} W(x_0,y_0)~.
\fe
See Figure \ref{fig:one-form-symaction}(b) for the Euclidean configuration of this time-like symmetry action.

\subsection{2+1d $\mathbb Z_N$ tensor gauge theory}\label{app:ZNtensor}

We will now start considering  non-relativistic theories with subsystem global symmetries.

Our first example is the 2+1d $\mathbb{Z}_N$ Ising-plaquette model. We  can study it on the lattice, but we might as well use the continuum formulation of \cite{paper1} in terms of the 2+1d $\mathbb Z_N$ tensor gauge theory.  The Lagrangian is\footnote{The $x$ and the $y$ directions in space are distinguished. They define a foliation in space.
}
\ie
\mathcal L = \frac{iN}{2\pi} \phi^{xy} (\partial_\tau A_{xy} - \partial_x \partial_y A_\tau)~,
\fe
with gauge symmetry
\ie
&\phi^{xy} \sim \phi^{xy} + 2\pi n_x^{xy}(x) + 2\pi n_y^{xy}(y)~,
\\
&A_\tau \sim A_\tau + \partial_\tau \alpha~,\qquad A_{xy} \sim A_{xy} + \partial_x \partial_y \alpha~,
\fe
where $n_i^{xy}(x^i)\in \mathbb Z$.

We place the theory on a Euclidean 3-torus with lengths $\ell_\tau$, $\ell_x$, and $\ell_y$.  Later, we will generalize the discussion to other spatial manifolds.

We will start with the space-like global symmetries of this model \cite{paper1}.
There is $\mathbb Z_N$ electric subsystem symmetry generated by $e^{i\phi^{xy}}$. Due to Gauss law $\partial_x\partial_y\phi^{xy}=0$, it factorizes into
\ie
e^{i\phi^{xy}(x,y)} = e^{i\phi_x(x)}  e^{i\phi_y(y)}~,
\fe
where only the sum of zero modes of $\phi_i(x^i)$ is physical. There is also a $\mathbb Z_N$ magnetic subsystem  symmetry generated by
\ie
&W_x(x_1,x_2) = \exp\left( i \int_{x_1}^{x_2} dx \oint dy~ A_{xy} \right)~,
\\
&W_y(y_1,y_2) = \exp\left( i \oint dx \int_{y_1}^{y_2} dy~ A_{xy} \right)~,
\fe
which satisfy
\ie
W_x(0,\ell_x) = W_y(0,\ell_y)~.
\fe
These operators are $\mathbb Z_N$ operators:
\ie
e^{iN\phi^{xy}} = W_x(x_1,x_2)^N = W_y(y_1,y_2)^N = 1~.
\fe
They satisfy the commutation relations
\ie
&e^{i\phi^{xy}(x,y)} W_x(x_1,x_2) = e^{-2\pi i/N} W_x(x_1,x_2) e^{i\phi^{xy}(x,y)}~,\quad \text{if}\quad x_1 < x < x_2~,
\\
&e^{i\phi^{xy}(x,y)} W_y(y_1,y_2) = e^{-2\pi i/N} W_y(y_1,y_2) e^{i\phi^{xy}(x,y)}~,\quad \text{if}\quad y_1 < y < y_2~.
\fe
This implies that the symmetry operators are charged under each other, and signals a mixed 't Hooft anomaly between the two $\mathbb Z_N$ symmetries. This results in an infinite ground state degeneracy, which is regularized to $N^{L_x+L_y-1}$ on a lattice with $L_i$ sites in the $i$-direction.

In addition to these global symmetries, there is also a $\mathbb Z_N$ tensor time-like symmetry generated by the \emph{quadrupole operator}:
\ie\label{quad-op}
U(\tau;x_1,x_2;y_1,y_2) = e^{i\phi^{xy}(\tau,x_2,y_2)} e^{-i\phi^{xy}(\tau,x_2,y_1)} e^{-i\phi^{xy}(\tau,x_1,y_2)} e^{i\phi^{xy}(\tau,x_1,y_1)}~.
\fe
This time-like symmetry operator $U$ is  supported at  a collection of four points (which are the vertices of a rectangle) on the $xy$-plane at a fixed time.
This operator satisfies
\ie\label{xoxty}
\partial_\tau U = 0~,\qquad \partial_{x_{1,2}} U = 0~,\qquad \partial_{y_{1,2}} U = 0~.
\fe

The last two equations imply that we can deform the rectangle of $U$ along the $x$ or the $y$ directions without affecting any correlation functions, as long as the deformation does not cross any defects.   It is important that here we mean that the edges of the rectangle and not only its corners do not cross any defects.
This is reminiscent of the topological nature of the time-like symmetry arising from a one-form symmetry in relativistic systems, but here $U$ can be deformed only  along the foliated directions in space.

\begin{figure}[t]
\begin{center}
\raisebox{-0.8\height}{\includegraphics[scale=0.25]{axes2d.pdf}}~~~~
\raisebox{-0.5\height}{\includegraphics[scale=0.25]{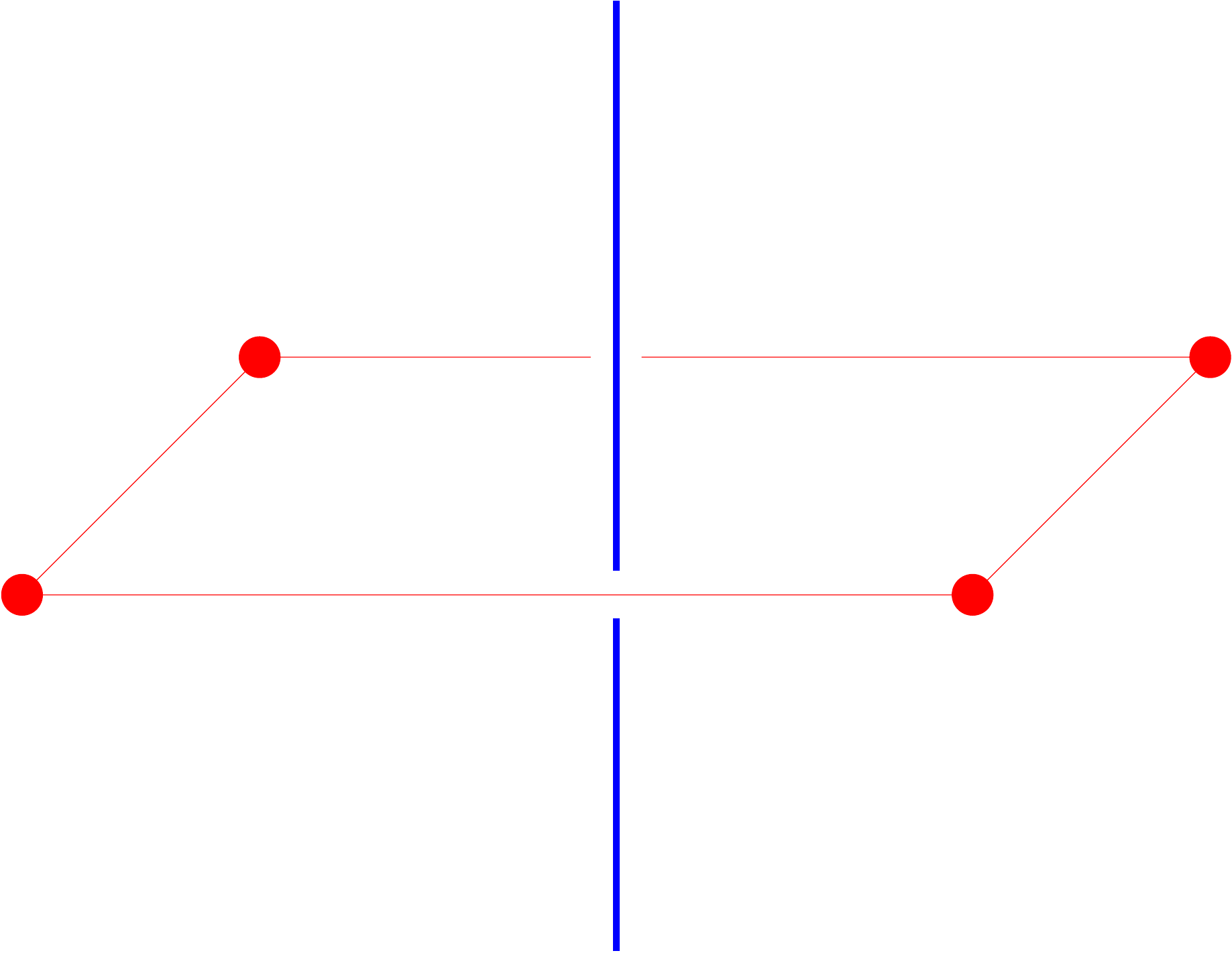}}~~~~$\longrightarrow ~~ e^{2\pi i/N}$
\raisebox{-0.5\height}{\includegraphics[scale=0.25]{defect.pdf}}
\caption{The Euclidean configuration for the action of the quadrupole operator $U(\tau_0;x_1,x_2;y_1,y_2)$ (red dots) on the fracton defect $W_\tau(x,y)$ (blue line).}\label{fig:quad-symaction}
\end{center}
\end{figure}

The defects charged under this time-like symmetry are the fracton defects,
\ie
W_\tau(x,y) = \exp\left( i\oint d\tau~ A_\tau(\tau,x,y) \right)~.
\fe
Their $N$'th powers are trivial, and they satisfy
\ie\label{2+1dZNA-timelikesym}
 U(\tau_0;x_1,x_2;y_1,y_2) W_\tau(x,y) = e^{-2\pi i/N} W_\tau(x,y)  ~,\quad \text{if}\quad x^i_1 < x^i < x^i_2~.
\fe
See Figure \ref{fig:quad-symaction}.

The $\mathbb Z_N$ tensor time-like symmetry acts as
\ie
A_\tau(\tau,x,y) \rightarrow A_\tau(\tau,x,y) + \frac{2\pi}{N\ell_\tau}\left[ n_x(x) + n_y(y) \right]  ~,
\fe
where $n_i(x^i)\in \mathbb Z$. In finite volume, this implies that the correlation function of defects,
\ie
\left\langle \prod_a W_\tau(x_a,y_a)^{q_a}\right\rangle\equiv \left\langle \prod_a \exp\left( iq_a \oint d\tau~ A_\tau(\tau,x_a,y_a) \right) \right\rangle~,\qquad q_a\in\mathbb Z~,
\fe
vanishes unless
\ie\label{trivial-phase}
\sum_a q_a [n_x(x_a) + n_y(y_a)] = 0 \mod N~.
\fe
This should be satisfied for all integer valued functions $n_x(x)$ and $n_y(y)$.

This constraint can also be obtained by nucleating a small quadrupole operator $U(\tau; x_1, x_2; y_1, y_2)$ \eqref{quad-op} away from the defects, and then enlarging it in, say the $x$ direction to wrap it around the strip bounded by $y_1$ and $y_2$ (similar argument works in the $y$ direction). During this process, whenever the symmetry operator crosses a defect, it picks up a phase. At the end of this process, the four corners of the quadrupole operator annihilate each other on the other side, i.e., it is as if we never inserted the quadrupole operator, so we should get the original correlation function back, up to an overall phase. So, unless this phase is trivial,
\ie\label{trivial-phase2}
\frac{2\pi}{N}\sum_{a: ~y_1<y_a<y_2} q_a \in 2\pi \mathbb Z~,
\fe
the correlation function vanishes. The condition \eqref{trivial-phase2}, supplemented with a similar condition in the other direction, is equivalent to \eqref{trivial-phase}.

Note the difference between this enlargement and annihilation operation and the analogous one in ordinary 2+1d gauge theories in Sections \ref{app:U1gauge} and \ref{app:ZNgauge}.  Here, the time-like charge operator is a quadrupole operator, supported at four points, which are annihilated at the end of this process.  In ordinary 2+1d gauge theories, the time-like charge operator is supported on a line.  And therefore, in order to annihilate it at the end of an analogous process, we need to move the line around the whole space, rather than to cover a strip in the $x$ or $y$ directions.  Therefore, the time-like symmetry in the tensor gauge theory case leads to stronger constraints than in the ordinary gauge theories.

In infinite volume, we can send some of the defects to infinity, so that, for the remaining defects,
\ie
\left(\sum_a q_a [n_x(x_a) + n_y(y_a)]\right)  \mod N~,
\fe
is conserved for all integer valued functions $n_x(x)$ and $n_y(y)$. In particular, a single fracton cannot move at all, whereas a dipole of fractons separated in the $x$-direction can move in the $y$-direction, and vice versa.
Thus,  this provides an explanation of the restricted mobility of the fractons in the 2+1d $\mathbb{Z}_N$ tensor gauge theory based on global symmetries.

\subsubsection{Twisted torus}\label{app:2+1dtwisted}

Above we have discussed the time-like symmetry on a rectangular spatial torus.
Next, we will place the 2+1d $\mathbb Z_N$ tensor gauge theory on a spatial torus with a more general complex structure.
Our discussion here follows \cite{Rudelius:2020kta} closely.

As shown in \cite{Rudelius:2020kta}, the space-like symmetry depends on the foliation.  Similarly, we will see that the time-like symmetry also depends on the foliation and the  complex structure of the torus. Consequently, the restricted mobility of the fractons are relaxed on a twisted torus.

Consider a
twisted spatial torus with identifications
\ie\label{twistedtorus-cont}
(x, y) \sim (x + m \ell_x^\text{eff}, y) \sim (x + k \ell_x^\text{eff}, y + \ell_y^\text{eff})~, \qquad \gcd(m,k)=1 ~.
\fe
We can combine these identifications to generate another identification
\ie\label{twistedtorus-cont2}
(x, y) \sim (x + \ell_x^\text{eff}, y + \tilde k \ell_y^\text{eff})~, \qquad \gcd(m,\tilde k) = 1 ~.
\fe

The coordinate system $(x,y)$ is associated with a choice of foliation of the torus.  Using the identifications \eqref{twistedtorus-cont}, we see that going along the constant $x$ or along the constant $y$ leaves, we return to the starting point after a shift $y\to y+m\ell_y^\text{eff}$ or $x \to x+m\ell_x^\text{eff}$, respectively.  Each pair of a constant $x$ and constant $y$ curves cross each other $m$ times within the fundamental domain of the torus \cite{Rudelius:2020kta}.

Let us start with the space-like symmetry.
In addition to the electric $\mathbb Z_N $ and the magnetic $ \mathbb Z_N$ global symmetries as before,  there is also a $\mathbb Z_{\gcd(N,m)}\times \mathbb Z_{\gcd(N,m)}$ global symmetry \cite{Rudelius:2020kta}.

Next, we will consider the time-like symmetry.
In addition to the $\mathbb Z_N$ time-like symmetry as before, there is also a $\mathbb Z_{\gcd(N,m)}$ time-like symmetry generated by the symmetry operator
\ie
\mathbf U(\tau) = \exp\left( \frac{iN}{\gcd(N,m)} \oint_F dx dy \left[ \Theta^P(x, 0; \ell_x^\text{eff}) - k \Theta^P(y, 0; \ell_y^\text{eff}) \right] \partial_x \partial_y \phi^{xy}(\tau, x,y) \right)~,
\fe
where $F$ is any fundamental domain on the spatial twisted torus, and $\Theta^P(x, x_0; \ell_x^\text{eff})$ is a suitable step function that satisfies
\ie
\Theta^P(0, x_0; \ell_x^\text{eff}) = 0~,\qquad \partial_x \Theta^P(x, x_0; \ell_x^\text{eff}) = \sum_{I \in \mathbb Z} \delta(x- x_0 - I \ell_x^\text{eff})~.
\fe
It satisfies
\ie
\partial_\tau \mathbf U(\tau) = 0~, \qquad \mathbf U(\tau)^{\gcd(N,m)} = 1~.
\fe

The time-like symmetries act on the gauge fields as
\ie
A_\tau(\tau,x,y) &\rightarrow A_\tau(\tau,x,y) + \frac{2\pi}{N\ell_\tau} \left[ n_x(x) + n_y(y) \right]
\\
&\qquad \qquad + \frac{2\pi r}{\gcd(N,m)\ell_\tau}\left[ \Theta^P(x, 0; \ell_x^\text{eff}) - k \Theta^P(y, 0; \ell_y^\text{eff}) \right]~,
\fe
where $n_i(x^i)$ is an integer-valued function with  $n_i(x^i + \ell_i^\text{eff}) = n_i(x^i)$, and $r=0,\ldots,\gcd(N,m)-1$.

On a twisted torus, the immobility of fractons is relaxed. Consider two charged defects $W_\tau(x, y)$, and $W_\tau(x',y')$. If $(x,y)$ and $(x',y')$ are related by the identifications \eqref{twistedtorus-cont}, these are two different labels of the same point in space and therefore these two defects are trivially the same. More interestingly, consider the case where there is an integer $I$ such that
\ie\label{Nmove-cont}
(x', y') = (x + I N \ell_x^\text{eff}, y)~, \qquad I \in {\mathbb Z}~.
\fe
Now, the two points $(x,y)$ and $(x',y')$ can label two different points in space and therefore the two defects can be different.  When \eqref{Nmove-cont} is satisfied, these two different defects carry the same $\mathbb Z_N$ and $\mathbb Z_{\gcd(N,m)}$ time-like charges.  Since they carry the same charges, we could ask whether this means that a fracton at $(x,y)$ can move to $(x',y')$ of \eqref{Nmove-cont}.

Let us examine it in the special case $I=1$.  Here we can see that indeed a fracton at $(x, y)$ can move to $(x+N\ell_x^\text{eff},y)$.  The defect\footnote{Here we omit the terms involving transition functions.}
\ie
&\exp\left( i\int_{-\infty}^{\tau_0} d\tau~ A_\tau(\tau, x, y)\right) \times \prod_{J=1}^N \exp\left( i \int_{x + (J-1) \ell_x^\text{eff}}^{x + J\ell_x^\text{eff}}d\tilde x \int_{y}^{y + J \tilde k \ell_y^\text{eff}}d\tilde y~ A_{xy}(\tau_0,\tilde x, \tilde y) \right)
\\
&\times \prod_{J=1}^N\exp\left( i\int_{\tau_0}^\infty d\tau  \left[ A_\tau(\tau, x + (J-1)\ell_x^\text{eff}, y + J \tilde k \ell_y^\text{eff}) - A_\tau(\tau, x + J\ell_x^\text{eff}, y + J \tilde k \ell_y^\text{eff}) \right] \right)
\\
&\times \exp\left( i\int_{\tau_0}^\infty d\tau~ A_\tau(\tau, x + N\ell_x^\text{eff}, y) \right)~
\fe
describes the motion of the fracton between these two different points at the time $\tau_0$, where $\tilde k$ is defined in \eqref{twistedtorus-cont2}.  (In reaching this conclusion, we used the fact that the defects in the second line are trivial because of the identification \eqref{twistedtorus-cont2}.)

This discussion of $I=1$ can be repeated for larger values of $I$.  But only  $I=1,\ldots, {m\over \gcd(N,m)}$ lead to distinct points on the torus.  Using the identifications \eqref{twistedtorus-cont}, the points corresponding to other values of $I$ are the same as for $I=1,\ldots, {m\over \gcd(N,m)}$.  In conclusion, a fracton at $(x, y)$ can move between ${m\over \gcd(N,m)}$ different points (with the last one being the same as $(x,y)$).  They can be labeled as
\ie\label{fracton-mob}
(x', y') = (x + J \gcd(N,m)\ell_x^\text{eff}, y )~, \qquad J=1,\ldots, {m\over \gcd(N,m)}~.
\fe

This result has the following interpretation.  As we said, the system depends on a choice of foliation with leaves at constant $x$ and constant $y$.  These leaves cross each other at $m$ points \cite{Rudelius:2020kta} within the fundamental domain.  The points in the set \eqref{fracton-mob} are on the same leaves.  But not all the $m$ intersection points between these leaves are included in \eqref{fracton-mob}.   As can be seen from \eqref{Nmove-cont}, we include only points that differ by a multiple of $N$ intersections in the covering space. They differ by an integer multiple of $\gcd(N,m)$ in the fundamental domain. The facts that these points are on the same leaves of the foliation and that the difference between them is a multiple of $N$ intersections in the covering space guarantee that a fracton at $(x,y)$ and a fracton at $(x',y')$ carry the same time-like symmetry charges.

This continuum discussion has a counterpart in its modified Villain and its BF lattice versions \cite{Gorantla:2021svj}.  Here, the identifications are
\ie\label{twistedtorus-lat}
(\hat x, \hat y) \sim (\hat x+M L_x^\text{eff},\hat y) \sim (\hat x + K L_x^\text{eff}, \hat y + L_y^\text{eff})~, \qquad  \gcd(M,K)=1
\fe
and the allowed mobility \eqref{fracton-mob} becomes an allowed mobility on the lattice from $(\hat x,\hat y)$ to
\ie\label{fracton-mobL}
(\hat x', \hat y') = (\hat x + J \gcd(N,M)L_x^\text{eff}, \hat y )~, \qquad J=1,\ldots, {M\over \gcd(N,M)}~.
\fe
As discussed in Section \ref{sec:dipZN-compactify}, in the special case $M=L_x$, $K=-1$, and $L_x^\text{eff} = L_y^\text{eff} = 1$, this 2+1d problem is equivalent to the 1+1d $\mathbb Z_N$ dipole gauge theory of Section \ref{sec:dipZN}.  Indeed, the allowed mobility \eqref{fracton-mobL} translates to the allowed mobility derived in Section \ref{sec:dipZN-def-mob}.

\subsection{3+1d $\mathbb Z_N$ anisotropic gauge theory}

Next, we will discuss an anisotropic model in 3+1d with lineons \cite{Shirley:2018nhn,Gorantla:2020jpy}.
The continuum  Lagrangian is
\ie
\mathcal L = \frac{iN}{2\pi} \left[ A_\tau (\partial_z \tilde A_{xy} - \partial_x \partial_y \tilde A_z) - A_z (\partial_\tau \tilde A_{xy} - \partial_x \partial_y \tilde A_\tau) - A_{xy} (\partial_\tau \tilde A_z - \partial_z \tilde A_\tau) \right]~,
\fe
with gauge symmetry
\ie
&A_\tau \sim A_\tau + \partial_\tau \alpha~,&&\tilde A_\tau \sim \tilde A_\tau + \partial_\tau \tilde \alpha~,
\\
&A_z \sim A_z + \partial_z \alpha~,&& \tilde A_z \sim \tilde A_z + \partial_z \tilde \alpha~,
\\
&A_{xy} \sim A_{xy} + \partial_x \partial_y \alpha~,\qquad&& \tilde A_{xy} \sim \tilde A_{xy} + \partial_x \partial_y \tilde \alpha~.
\fe

The space-like global symmetries were discussed in \cite{Gorantla:2020jpy}.
There is a $\mathbb Z_N$ electric global symmetry generated by
\ie\label{ZNaniso-ops}
&\tilde W_z(x,y) = \exp \left( i\oint dz~ \tilde A_z \right)~,
\\
&\tilde W_x(x_1,x_2;z) = \exp \left( i\int_{x_1}^{x_2}dx\oint dy~ \tilde A_{xy} \right)~,
\\
&\tilde W_y(y_1,y_2;z) = \exp \left( i\oint dx\int_{y_1}^{y_2}dy~ \tilde A_{xy} \right)~,
\fe
and a $\mathbb Z_N$ magnetic global symmetry generated by similar Wilson operators of $A$. Due to the Gauss law, the operator in the first line of \eqref{ZNaniso-ops} factorizes as
\ie
\tilde W_z(x,y) = \tilde W_z^x(x) \tilde W_z^y(y)~,
\fe
and the operators in the last two lines of \eqref{ZNaniso-ops} satisfy
\ie
\tilde W_x(0,\ell_x;z) = \tilde W_y(0,\ell_y;z)~.
\fe
Similar relations hold for the $\tilde A$ operators. They are all $\mathbb Z_N$ operators, and satisfy the commutation relations
\ie
&\tilde W_x(x_1,x_2;z) W_z(x,y) = e^{-2\pi i/N} W_z(x,y) \tilde W_x(x_1,x_2;z)~,\quad \text{if} \quad x_1 < x < x_2~,
\\
&\tilde W_y(y_1,y_2;z) W_z(x,y) = e^{-2\pi i/N} W_z(x,y) \tilde W_y(y_1,y_2;z)~,\quad \text{if} \quad y_1 < y < y_2~,
\fe
and vice versa. This implies that the symmetry operators are charged under each other, and signals a mixed 't Hooft anomaly between the two $\mathbb Z_N$ global symmetries. As a result, there is an infinite ground state degeneracy, which is regularized on a lattice with $L_i$ sites in the $i$-direction to $N^{L_x + L_y - 1}$.

\begin{figure}[t]
\begin{center}
\includegraphics[scale=0.25]{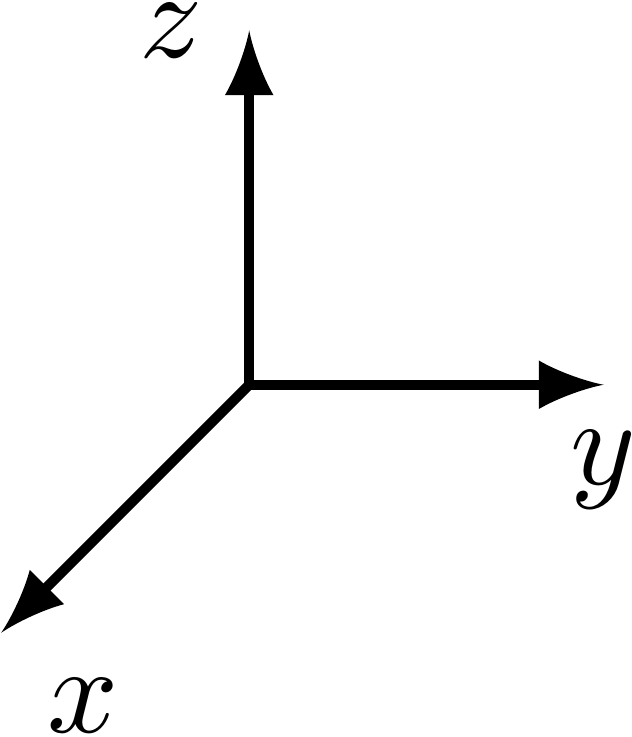}~~~~~~~~~~~
\includegraphics[scale=0.25]{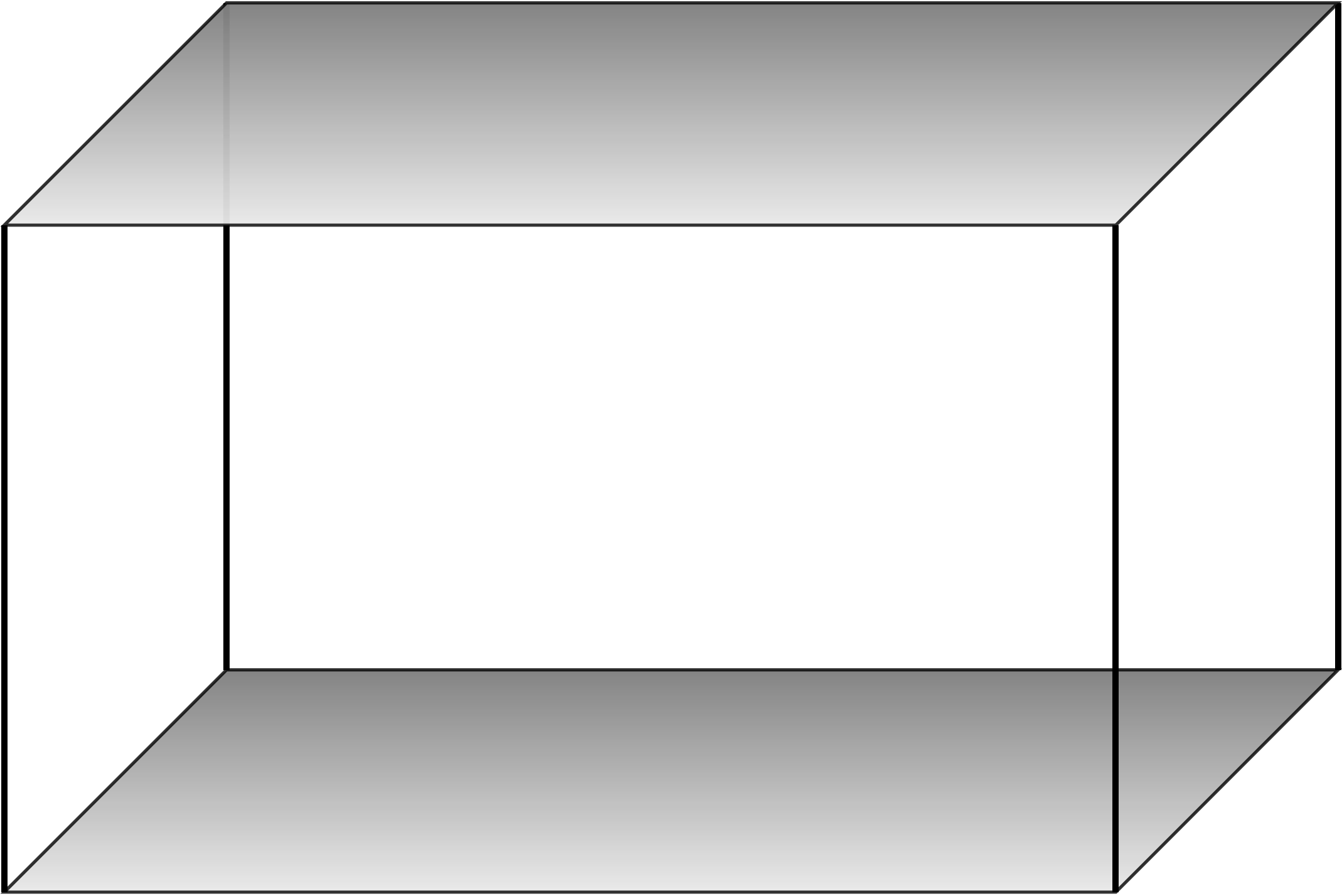}
\caption{The pillar operator $U(\tau;\mathcal B)$ made of $\tilde A$ gauge fields. Here, $\mathcal B = [x_1,x_2] \times [y_1,y_2] \times [z_1,z_2]$ is a spatial box. The pillar operator generates the $\mathbb Z_N$ electric time-like symmetry of the $\mathbb Z_N$ anisotropic gauge theory. A similar pillar operator made of $A$ gauge fields generates the $\mathbb Z_N$ magnetic time-like symmetry. These operators are reminiscent of terms in the lattice Hamiltonian of the anisotropic model.}\label{fig:pillar-op}
\end{center}
\end{figure}

Next, we will turn to the time-like symmetries. There are two $\mathbb Z_N$ time-like symmetries. The $\mathbb Z_N$ electric time-like symmetry is generated by the \emph{pillar operator}, which is reminiscent of similar terms in the lattice Hamiltonian of the anisotropic model (see Figure \ref{fig:pillar-op}):
\ie
&U(\tau;\mathcal B) =\exp\left( i\int_{x_1}^{x_2}dx\int_{y_1}^{y_2}dy~\left[ \tilde A_{xy}(\tau,x,y,z_2) - \tilde A_{xy}(\tau,x,y,z_1) \right] \right)
\\
&\times \exp\left( -i\int_{z_1}^{z_2}dz~\left[ \tilde A_z(\tau,x_2,y_2,z) - \tilde A_z(\tau,x_1,y_2,z) - \tilde A_z(\tau,x_2,y_1,z) + \tilde A_z(\tau,x_1,y_1,z)\right] \right)~,
\fe
where $\mathcal B = [x_1,x_2] \times [y_1,y_2] \times [z_1,z_2]$ is a spatial box. This operator satisfies
\ie
\partial_\tau U(\tau;\mathcal B) = 0~,\qquad \partial_{x^i_{1,2}} U(\tau;\mathcal B) = 0~.
\fe
The above equation implies that while $U$ is not completely topological, we can deform it along the $x$, $y$, or $z$ direction without changing any correlation functions, as long as the deformation does not cross any defects.

The defects charged under this time-like symmetry are the $z$-lineon defects,
\ie
W_\tau(x,y,z) = \exp \left( i\oint d\tau~A_\tau(\tau,x,y,z) \right)~.
\fe
Their $N$'th powers are trivial.
The action of the time-like symmetry on the $z$-lineon defect is:
\ie\label{ZNaniso-electimelikesym}
 U(\tau_0;\mathcal B) W_\tau(x,y,z)  = e^{2\pi i/N}  W_\tau(x,y,z)  ~, \quad \text{if} \quad (x,y,z)\in \mathcal B~.
\fe

The $\mathbb Z_N$ electric time-like symmetry acts as
\ie
A_\tau \rightarrow A_\tau + \frac{2\pi}{N\ell_\tau} [ n_x(x) + n_y(y) ]~,
\fe
where $n_i(x^i) \in \mathbb Z$ for $i=x,y$. In finite volume, this implies that the correlation function of defects,
\ie
\left\langle \prod_a \exp\left( iq_a \oint d\tau~ A_\tau(\tau,x_a,y_a,z_a) \right) \right\rangle~,\qquad q_a\in\mathbb Z~,
\fe
vanishes unless
\ie
\sum_a q_a [n_x(x_a) + n_y(y_a)] = 0 \mod N~.
\fe
This should be satisfied for all integer valued functions $n_x(x)$ and $n_y(y)$.
In infinite volume, we can send some of the defects to infinity, so that, for the remaining defects,
\ie
\left(\sum_a q_a [n_x(x_a) + n_y(y_a)]\right)  \mod N~,
\fe
is conserved for all integer valued functions $n_x(x)$ and $n_y(y)$. In particular, a single $z$-lineon can move only in $z$-direction, whereas a dipole of $z$-lineons separated in $x$-direction can move in the $yz$-plane, and vice versa.

 As above, we can derive these selection rules by nucleating the pillar operator, deforming it around the torus and annihilating it.

The $\mathbb Z_N$ magnetic time-like symmetry is obtained by exchanging $A$ and $\tilde A$.

To conclude, the restricted mobility of the lineons in this anisotropic model is explained by the two $\mathbb{Z}_N$ time-like global symmetries.

\subsection{X-cube model}
Here we will study the X-cube model \cite{Vijay:2016phm} using the continuum formulation of \cite{Slagle:2017wrc,paper3}.
We place the theory on a Euclidean 4-torus with a choice of foliation characterized by the spatial directions $x,y,z$ and the Euclidean time direction $\tau$.  We start by considering the case with simple periodicity in these directions given by the lengths $\ell_x$, $\ell_y$, $\ell_z$, and $\ell_\tau$. The Lagrangian is
\ie
\mathcal L = \frac{iN}{2\pi} \left[ \sum_{i,j,k \atop \text{cyclic}} A_{ij} (\partial_\tau \hat A^{ij} - \partial_k \hat A_\tau^{k(ij)}) + A_\tau \sum_{i<j}\partial_i \partial_j \hat A^{ij} \right]~,
\fe
with gauge symmetry
\ie
&A_\tau \sim A_\tau + \partial_\tau \alpha~,&& \hat A_\tau^{k(ij)} \sim \hat A_\tau^{k(ij)} + \partial_\tau \hat \alpha^{k(ij)}~,
\\
&A_{ij} \sim A_{ij} + \partial_i \partial_j \alpha~,\qquad && \hat A^{ij} \sim \hat A^{ij} + \partial_k \hat \alpha^{k(ij)}~.
\fe
We refer  the readers to \cite{paper3} for details of our notations.

There is a $\mathbb Z_N$ electric  global symmetry generated by \cite{paper3}
\ie
\hat W^z(x,y) = \exp\left( i\oint dz~\hat A^{xy} \right)~,
\fe
and similar operators in the other directions. Due to equation of motion of $A_\tau$ (i.e., Gauss law), we have
\ie
\hat W^z(x,y) = \hat W^z_x(x) \hat W^z_y(y)~.
\fe
There is also a $\mathbb Z_N$ magnetic   global symmetry generated by \cite{paper3}
\ie
W_{xy}(z_1,z_2;\mathcal C^{xy}) = \exp\left( i\int_{z_1}^{z_2} dz \oint_{\mathcal C^{xy}} (dx~A_{xz} + dy~A_{yz} ) \right)~,
\fe
where $\mathcal C^{xy}$ is a closed curve in the $xy$-plane,\footnote{Note that $W_{xy}(z_1,z_2;\mathcal C^{xy})$ depends only on the topology of $\mathcal C^{xy}$, \emph{i.e.}, it is invariant under small deformations of $\mathcal C^{xy}$ \cite{paper3}.} and similar operators in the other directions. They satisfy
\ie
W_{xy}(0,\ell_z;\mathcal C^{xy}_x) = W_{yz}(0,\ell_x;\mathcal C^{yz}_z)~,
\fe
and similarly in other directions. Here, $\mathcal C^{ij}_i$ is a closed curve in the $ij$-plane that wraps once in the $i$-direction but not in the $j$-direction. These operators are $\mathbb Z_N$ operators, \emph{i.e.}, their $N$'th powers are trivial, and they are the low energy limits of the logical operators of the X-cube model \cite{Vijay:2016phm}. They satisfy the commutation relations
\ie
\hat W^x(y_0,z_0) W_{xy}(z_1,z_2;\mathcal C) = e^{2\pi i I(\mathcal C,y_0)/N} W_{xy}(z_1,z_2;\mathcal C) \hat W^x(y_0,z_0)~,\quad \text{if}\quad z_1<z_0<z_2~,
\fe
where $I(\mathcal C,y_0)$ is the intersection number of the curve $\mathcal C$ and the line $y=y_0$ in the $xy$-plane. There are similar commutation relations in the other directions. These commutation relations imply that the symmetry operators are charged under each other. This means that there is a mixed 't Hooft anomaly between the two $\mathbb Z_N$ symmetries, and leads to an infinite ground state degeneracy, which is regularized on a lattice with $L_i$ sites in the $i$-direction to $N^{2L_x + 2L_y + 2L_z - 3}$.

\begin{figure}[t]
\begin{center}
\includegraphics[scale=0.25]{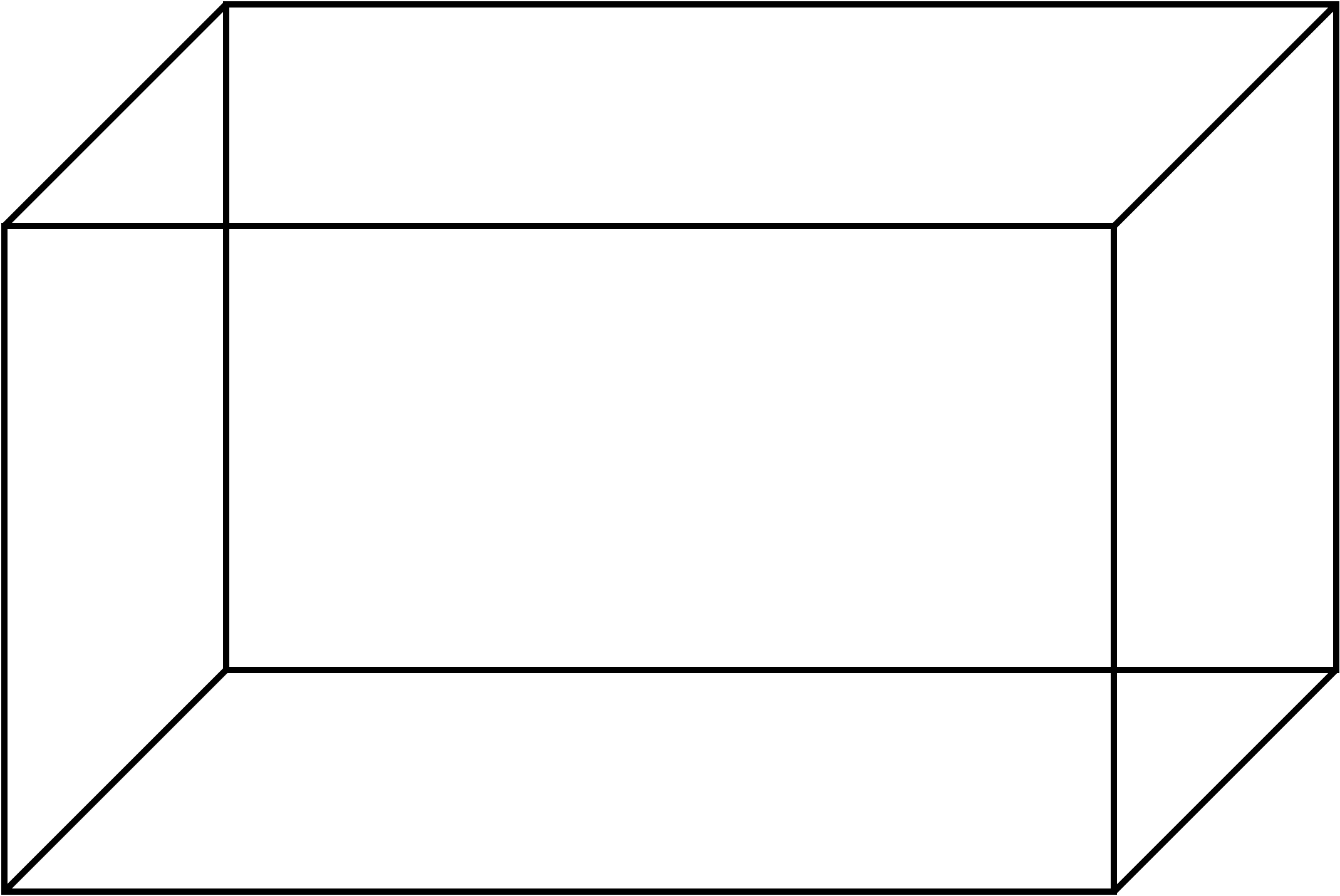}~~~~~~~~~~~~~~
\includegraphics[scale=0.25]{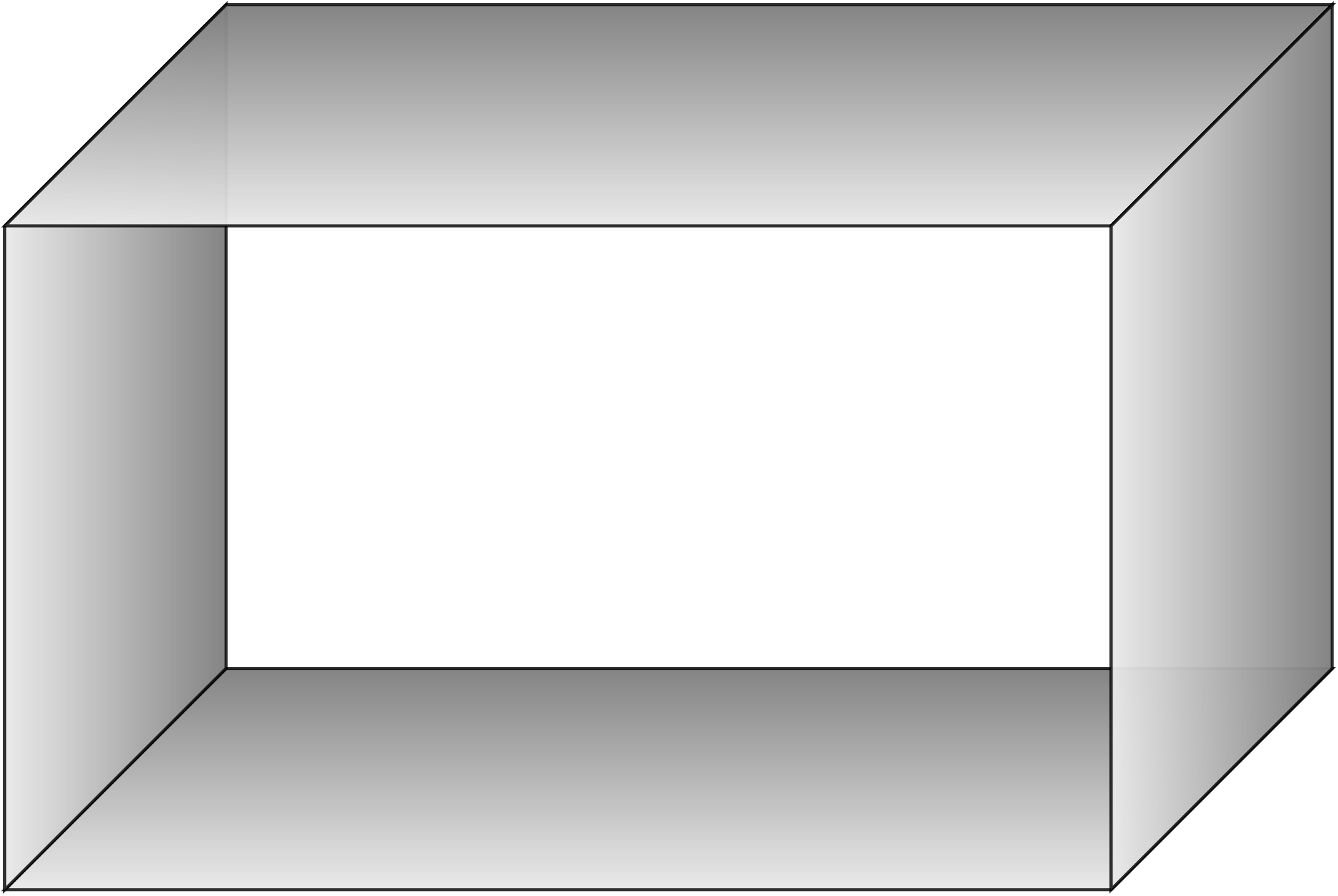}
\\
\includegraphics[scale=0.25]{axes.pdf}
\\
\includegraphics[scale=0.25]{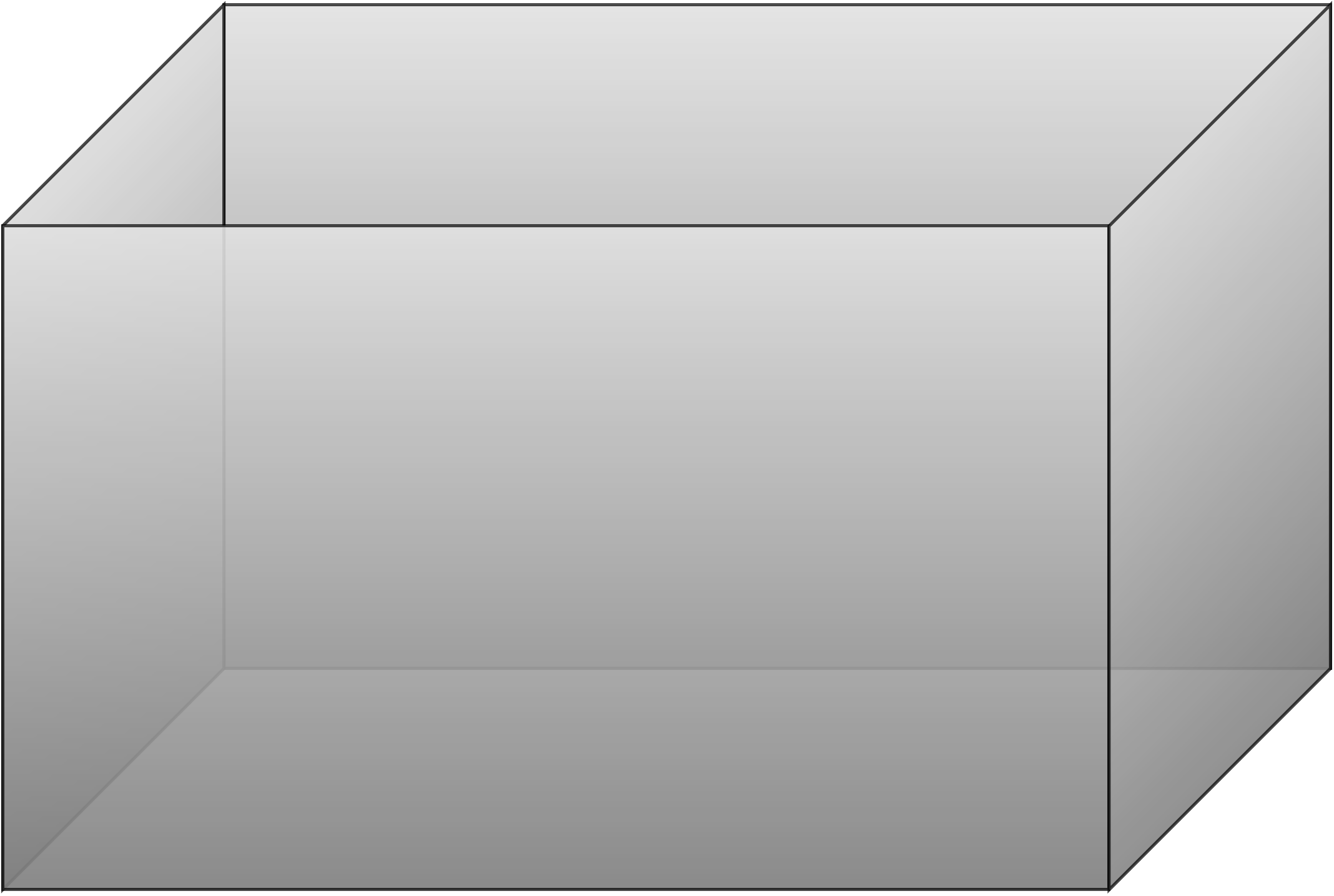}~~~~~~~~~~~~~~
\includegraphics[scale=0.25]{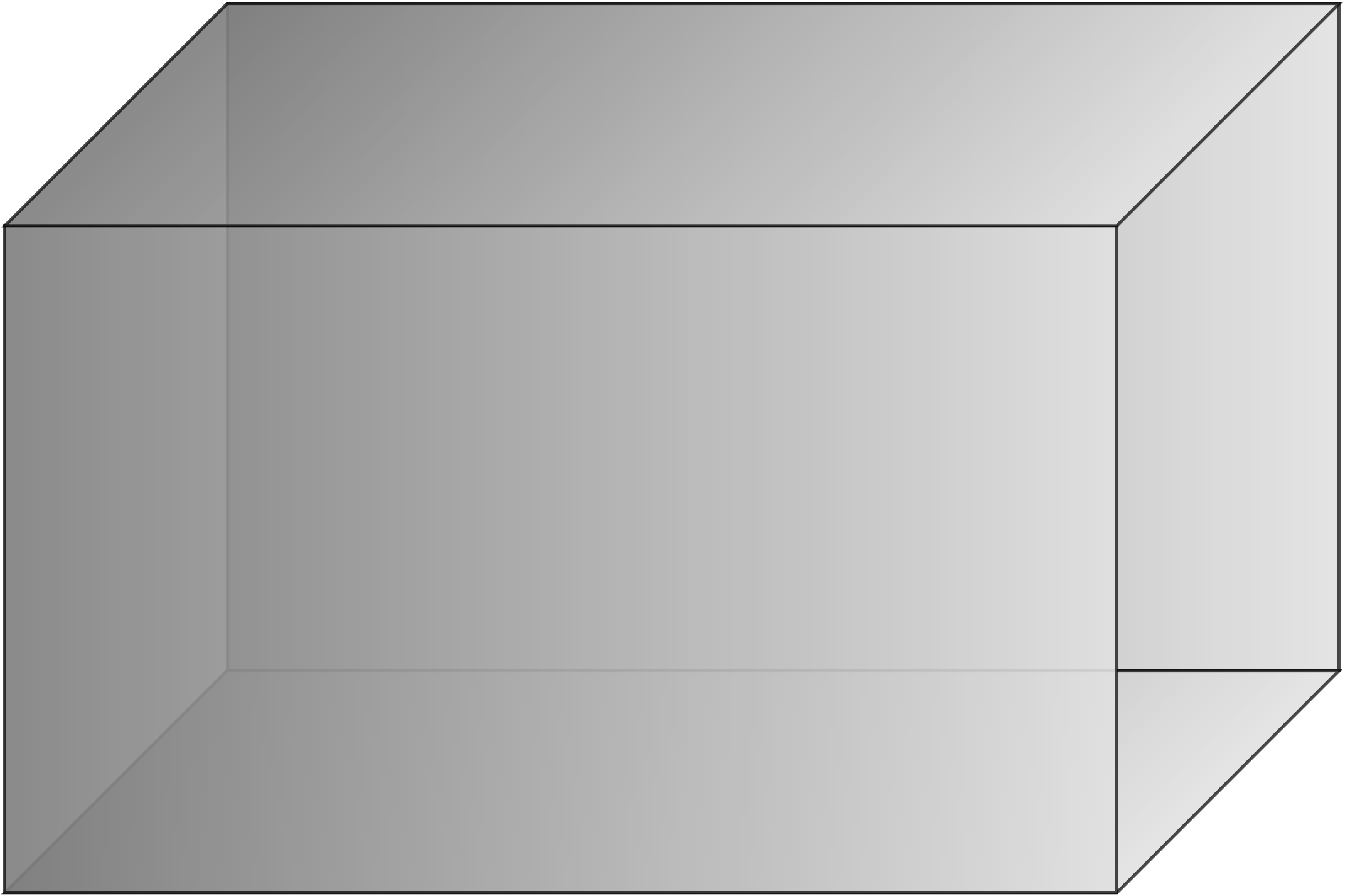}
\caption{The cage operator $U(\tau;\mathcal B)$, and the three belt operators $\hat U_{[yz]x}(\tau;\mathcal B)$, $\hat U_{[zx]y}(\tau;\mathcal B)$ and $\hat U_{[xy]z}(\tau;\mathcal B)$, respectively, in clockwise order. Here, $\mathcal B = [x_1,x_2] \times [y_1,y_2] \times [z_1,z_2]$ is a spatial box. The cage operator generates the $\mathbb Z_N$ electric time-like symmetry of the X-cube field theory, whereas the belt operators generate the $\mathbb Z_N$ magnetic time-like symmetry. These operators are reminiscent of the ``X-cube'' and ``vertex'' terms in the lattice Hamiltonian of the X-cube model.}\label{fig:cage-belt-op}
\end{center}
\end{figure}

There are also $\mathbb Z_N$ time-like global symmetries. The $\mathbb Z_N$ electric time-like symmetry is generated by the \emph{cage operator}, which is reminiscent of the ``X-cube'' term in the lattice Hamiltonian of the X-cube model (see Figure \ref{fig:cage-belt-op}):
\ie
&U(\tau;\mathcal B)
\\
&=\exp \left( i\int_{z_1}^{z_2} dz \left[ \hat A^{xy}(\tau,x_2,y_2,z) - \hat A^{xy}(\tau,x_1,y_2,z) - \hat A^{xy}(\tau,x_2,y_1,z) + \hat A^{xy}(\tau,x_1,y_1,z)\right] \right)
\\
&\times \exp \left( i\int_{x_1}^{x_2} dx \left[ \hat A^{yz}(\tau,x,y_2,z_2) - \hat A^{yz}(\tau,x,y_2,z_1) - \hat A^{yz}(\tau,x,y_1,z_2) + \hat A^{yz}(\tau,x,y_1,z_1)\right] \right)
\\
&\times \exp \left( i\int_{y_1}^{y_2} dy \left[ \hat A^{zx}(\tau,x_2,y,z_2) - \hat A^{zx}(\tau,x_1,y,z_2) - \hat A^{zx}(\tau,x_2,y,z_1) + \hat A^{zx}(\tau,x_1,y,z_1)\right] \right)~.
\fe
where $\mathcal B = [x_1,x_2] \times [y_1,y_2] \times [z_1,z_2]$ is a spatial box. This operator satisfies
\ie
\partial_\tau U(\tau;\mathcal B) = 0~,\qquad \partial_{x^i_{1,2}} U(\tau;\mathcal B) = 0~.
\fe
The above equation implies that while $U$ is not completely topological, we can deform it along the $x$, $y$, or $z$ direction without changing any correlation functions, as long as the deformation does not cross any defects.  By that we mean that the faces of the cube, rather than the edges, of the cube do not cross any defect.

The defects charged under this time-like symmetry are the fracton defects,
\ie
W_\tau(x,y,z) = \exp \left( i\oint d\tau~A_\tau(\tau,x,y,z) \right)~.
\fe
Their $N$'th powers are trivial, and they satisfy
\ie\label{XC-electimelikesym}
 U(\tau_0;\mathcal B) W_\tau(x,y,z)   = e^{2\pi i/N}  W_\tau(x,y,z)  ~, \quad \text{if} \quad (x,y,z)\in \mathcal B~.
\fe

The $\mathbb Z_N$ electric time-like symmetry acts as
\ie
A_\tau \rightarrow A_\tau + \frac{2\pi}{N\ell_\tau} [ n_x(x) + n_y(y) + n_z(z) ]~,
\fe
where $n_i(x^i) \in \mathbb Z$. In finite volume, this implies that the correlation function of defects,
\ie
\left\langle \prod_a W_\tau(x_a,y_a,z_a)^{q_a}\right\rangle\equiv \left\langle \prod_a \exp\left( iq_a \oint d\tau~ A_\tau(\tau,x_a,y_a,z_a) \right) \right\rangle~,\qquad q_a\in\mathbb Z~,
\fe
vanishes unless
\ie
\sum_a q_a [n_x(x_a) + n_y(y_a) + n_z(z_a)] = 0 \mod N~.
\fe
This should be satisfied for all integer valued functions $n_x(x)$, $n_y(y)$, and $n_z(z)$.

In infinite volume, we can send some of the defects to infinity, so that, for the remaining defects,
\ie
\left(\sum_a q_a [n_x(x_a) + n_y(y_a) + n_z(z_a)]\right)  \mod N~,
\fe
is conserved for all integer valued functions $n_x(x)$, $n_y(y)$, and $n_z(z)$.

In particular, a single fracton cannot move at all, whereas a dipole of fractons separated in $x$-direction can move in the $yz$-plane, and similarly in the other directions.
We have thus provided an explanation of the restricted mobility of fractons using global symmetries.

The authors of \cite{Slagle:2017wrc,Shirley:2018vtc} derived this restricted mobility using the conservation of the ``subsystem symmetry gauge charge." As we discussed at the end of Section
\ref{app:U1gauge}, it is preferable to avoid the use of conservation of gauge charges. Our presentation
here, recasts the discussion of \cite{Slagle:2017wrc,Shirley:2018vtc} using gauge invariant time-like symmetries rather than
gauge charges, thus making it more precise.

There is also a $\mathbb Z_N$ magnetic time-like symmetry generated by the \emph{belt operators}, which are reminiscent of the ``vertex'' terms in the lattice Hamiltonian of the X-cube model (see Figure \ref{fig:cage-belt-op}):
\ie
\hat U_{[xy]z}(\tau;\mathcal B) &=\exp \left( i\int_{y_1}^{y_2} dy \int_{z_1}^{z_2} dz~ \left[ A_{yz}(\tau,x_2,y,z) - A_{yz}(\tau,x_1,y,z) \right] \right)
\\
&\times \exp \left( i\int_{x_1}^{x_2} dx \int_{z_1}^{z_2} dz~ \left[ A_{zx}(\tau,x,y_1,z) - A_{zx}(\tau,x,y_2,z) \right] \right)~,
\fe
and similar operators in the other directions. The three operators are related by the constraint
\ie
&\hat U_{[xy]z}(\tau;\mathcal B)~ \hat U_{[yz]x}(\tau;\mathcal B)~ \hat U_{[zx]y}(\tau;\mathcal B) = 1~,
\fe
They also satisfy
\ie
\partial_\tau \hat U_{[xy]z}(\tau;\mathcal B) = 0~,\qquad \partial_{x^i_{1,2}}\hat U_{[xy]z}(\tau;\mathcal B) = 0~,
\fe
and similarly in the other directions.
Similar to the cage operator, the belt operator $\hat U$ can be deformed along the $x,y,z$ directions without changing any correlation functions, as long as the deformation does not cross any defects.

The defects charged under this time-like symmetry are the lineon defects, \emph{e.g.}, the $z$-lineon defect
\ie
\hat W_\tau^z(x,y,z) = \exp\left( i \oint d\tau~ \hat A_\tau^{z(xy)}(\tau,x,y,z) \right)~,
\fe
and similar defects in the other directions. Their $N$'th powers are trivial, and they satisfy
\ie\label{XC-magtimelikesym}
&  \hat U_{[zx]y}(\tau_0;\mathcal B) \hat W_\tau^z(x,y,z)   = e^{2\pi i/N}\hat W_\tau^z(x,y,z)  ~,\quad \text{if} \quad (x,y,z)\in \mathcal B~,
\\
&  \hat U_{[yz]x}(\tau_0;\mathcal B) \hat W_\tau^z(x,y,z)   = e^{-2\pi i/N}  \hat W_\tau^z(x,y,z)  ~,\quad \text{if} \quad (x,y,z)\in \mathcal B~,
\\
& \hat U_{[xy]z}(\tau_0;\mathcal B) \hat W_\tau^z(x,y,z)   =  \hat W_\tau^z(x,y,z)  ~,
\fe
and similarly for the other defects.

The $\mathbb Z_N$ magnetic time-like symmetry acts as
\ie
&\hat A_\tau^{x(yz)} \rightarrow \hat A_\tau^{x(yz)} + \frac{2\pi}{N\ell_\tau} [ \hat n_z(z) - \hat n_y(y) ]~,
\\
&\hat A_\tau^{y(zx)} \rightarrow \hat A_\tau^{y(zx)} + \frac{2\pi}{N\ell_\tau} [ \hat n_x(x) - \hat n_z(z) ]~,
\\
&\hat A_\tau^{z(xy)} \rightarrow \hat A_\tau^{z(xy)} + \frac{2\pi}{N\ell_\tau} [ \hat n_y(y) - \hat n_x(x) ]~,
\fe
where $\hat n_i(x^i) \in \mathbb Z$. In finite volume, this implies that the correlation function of defects,
\ie
\left\langle \prod_a \prod_{i,j,k\atop \text{cyclic}} \exp\left( i q^a_{k(ij)} \oint d\tau~ \hat A^{k(ij)}_\tau(\tau,x_a,y_a,z_a) \right) \right\rangle~,\qquad q^a_{k(ij)} \in\mathbb Z~,
\fe
vanishes unless
\ie
\sum_a \sum_{i,j,k\atop \text{cyclic}} q^a_{k(ij)}  [\hat n_j(x^j_a) - \hat n_i(x^i_a)] = 0 \mod N~.
\fe
Note that the sum constraint on $\hat A_\tau^{k(ij)}$ implies that the charges have a gauge symmetry
\ie
q^a_{k(ij)} \sim q^a_{k(ij)} + q^a~,\qquad q^a\in \mathbb Z~.
\fe
Indeed, the above condition for nonvanishing correlation function can be written as
\ie
\sum_a \left[ q^a_{[yz]x} \hat n_x(x_a) + q^a_{[zx]y} \hat n_y(y_a) + q^a_{[xy]z} \hat n_z(z_a) \right] = 0 \mod N~,
\fe
where $q^a_{[ij]k} = q^a_{i(jk)} - q^a_{j(ki)}$. This should be satisfied for all integer valued functions $\hat n_x(x)$, $\hat n_y(y)$, and $\hat n_z(z)$. In infinite volume, we can send some of the defects to infinity, so that, for the remaining defects,
\ie
\left( \sum_a \left[ q^a_{[yz]x} \hat n_x(x_a) + q^a_{[zx]y} \hat n_y(y_a) + q^a_{[xy]z} \hat n_z(z_a) \right] \right) \mod N~,
\fe
is conserved for all integer valued functions $\hat n_x(x)$, $\hat n_y(y)$, and $\hat n_z(z)$.

In particular, a single $z$-lineon can move only in the $z$-direction, whereas a dipole of $z$-lineons separated in the $x$-direction can move in the $yz$-plane. Similar restrictions apply to other lineons.
We have thus explained the restricted mobility of the lineons using the time-like global symmetries.

\subsubsection{Twisted torus}\label{Xcubetw}

Similar to Appendix \ref{app:2+1dtwisted},
we will now place the 3+1d $\mathbb Z_N$ tensor gauge theory on a spatial torus with a twist in the $xy$-plane given by the identifications \eqref{twistedtorus-cont}.  Our discussion will follow \cite{Rudelius:2020kta}.

As shown in \cite{Rudelius:2020kta}, the space-like symmetry depends on the foliation.  Now, as in Appendix \ref{app:2+1dtwisted}, we will see that the time-like symmetry also depends on the foliation and the complex structure of the torus.
Consequently, the restricted mobilities of the fractons and lineons are relaxed on a twisted torus.

Let us start with the space-like symmetries.
In addition to the electric  $\mathbb Z_N$ and the magnetic $ \mathbb Z_N$ global symmetries as before, there is also a $\mathbb Z_{\gcd(N,m)}\times \mathbb Z_{\gcd(N,m)}$ global symmetry \cite{Rudelius:2020kta}.

Next, we will move on to the time-like global symmetries.
In addition to the  $\mathbb Z_N$ electric time-like symmetry as in the untwisted case,  there is also a $\mathbb Z_{\gcd(N,m)}$ electric time-like symmetry generated by the symmetry operator
\ie
\mathbf U(\tau) = \exp\left( \frac{iN}{\gcd(N,m)} \oint_F dx dy dz \left[ \Theta^P(x, 0; \ell_x^\text{eff}) - k \Theta^P(y, 0; \ell_y^\text{eff}) \right] \sum_{i<j} \partial_i \partial_j \hat A^{ij} \right)~,
\fe
where $F$ is any fundamental domain of the spatial twisted torus. It satisfies
\ie
\partial_\tau \mathbf U(\tau) = 0~, \qquad \mathbf U(\tau)^{\gcd(N,m)} = 1~.
\fe

The electric time-like symmetries act on the gauge fields as
\ie
A_\tau(\tau,x,y,z) &\rightarrow A_\tau(\tau,x,y,z) + \frac{2\pi}{N\ell_\tau} \left[ n_x(x) + n_y(y) + n_z(z) \right]
\\
&\qquad \qquad + \frac{2\pi r}{\gcd(N,m)\ell_\tau}\left[ \Theta^P(x, 0; \ell_x^\text{eff}) - k \Theta^P(y, 0; \ell_y^\text{eff}) \right]~,
\fe
where $n_i(x^i) , r$ are integers,  $n_i(x^i + \ell_i^\text{eff}) = n_i(x^i)$. Here, $\ell_z^\text{eff} = \ell_z$.

As in the 2+1d $\mathbb Z_N$ tensor gauge theory in Appendix \ref{app:2+1dtwisted}, a fracton at $(x, y,z)$ can move between ${m\over \gcd(N,m)}$ different points (with the last one being the same as $(x,y,z)$).  They can be labeled as
\ie\label{fracton-mobx}
(x', y',z') = (x + J \gcd(N,m)\ell_x^\text{eff}, y ,z)~, \qquad J=1,\ldots, {m\over \gcd(N,m)}~.
\fe
These two points are on the same leaves of the foliation, and their difference is a multiple of $N$ intersections in the covering space. This guarantees that a fracton at $(x,y,z)$ and a fracton at $(x',y',z')$ carry the same time-like symmetry charges. We see that the restricted mobility of the fractons on a regular torus is relaxed due to the twisted boundary conditions.  This is consistent with the change in the electric time-like symmetries due to the twist.

As in the untwisted case, there is a $\mathbb Z_N$ magnetic time-like symmetry.  Since it is a subsystem symmetry, the twist reduces it significantly.  However, unlike the electric time-like symmetry, there is no additional magnetic time-like symmetry.
This means:

\begin{enumerate}
\item
The $z$-lineon defects $\hat W^z_\tau(x,y,z)$ and $\hat W^z_\tau(x',y',z')$ have the same magnetic time-like charges if and only if
\ie\label{zlineon-move}
(x',y',z') = (x+I \ell_x^\text{eff}, y, z + c_z)~,
\fe
for some $I=1,\ldots, m$ and $c_z \sim c_z + \ell_z$. In other words, a $z$-lineon at $(x,y,z)$ can move to $(x',y',z')$ if and only if $(x',y',z')$ is a point on the intersection of the leaves at constant $x$ and constant $y$.\footnote{For example, let us set $I=1$ and $c_z=0$. A $z$-lineon at $(x,y,z)$ can be thought of as product of an $x$-lineon and a $y$-lineon. Move the $x$-lineon to $(x+\ell_x^\text{eff},y,z)$, and the $y$-lineon to $(x,y-\tilde k \ell_y^\text{eff},z)$. Due to the identification \eqref{twistedtorus-cont2}, this is equivalant to a $z$-lineon at $(x + \ell_x^\text{eff},y,z)$.}

\item
The $x$-lineon defects $\hat W^x_\tau(x,y,z)$ and $\hat W^x_\tau(x',y',z')$ have the same magnetic time-like charges if and only if
\ie\label{xlineon-move}
(x',y',z') = (x+c_x, y+I\ell_y^\text{eff}, z)~,
\fe
for some $I=1,\ldots,m$ and $c_x\sim c_x +m\ell_x^\text{eff}$. In other words, an $x$-lineon at $(x,y,z)$ can move to $(x',y',z')$ if and only if $(x',y',z')$ is a point on the intersection of the leaves at constant $y$ and constant $z$.\footnote{For example, let us set $J=1$ and $c_x=0$. Move the $x$-lineon to $(x-k \ell_x^\text{eff},y,z)$. Due to the identification \eqref{twistedtorus-cont}, this is equivalant to an $x$-lineon at $(x,y+ \ell_y^\text{eff},z)$.}

\item
Similarly, the $y$-lineon defects $\hat W^y_\tau(x,y,z)$ and $\hat W^y_\tau(x',y',z')$ have the same magnetic time-like charges if and only if
\ie\label{ylineon-move}
(x',y',z') = (x+I \ell_x^\text{eff}, y+c_y, z)~,
\fe
for some $I = 1,\ldots,m$ and $c_y\sim c_y + m\ell_y^\text{eff}$. In other words, a $y$-lineon at $(x,y,z)$ can move to $(x',y',z')$ if and only if $(x',y',z')$ is a point on the intersection of the leaves at constant $x$ and constant $z$.

\end{enumerate}
Therefore, the restricted mobility in the untwisted case is relaxed due to the twist.  This new restricted mobility reflects the magnetic time-like symmetries.

\bibliographystyle{JHEP}
\bibliography{nonhollow_draft}

\end{document}